\documentclass[prb,superscriptaddress,amsfonts,amssymb,amsmath,floats,twocolumn,aps,footinbib,shownopacs,longbibliography]{revtex4-2}

\usepackage{amsmath}
\usepackage{amsfonts}
\usepackage{graphicx}
\usepackage{color}
\usepackage{xcolor}
\usepackage{float}
\usepackage{hyperref}
\usepackage{soul}
\usepackage{braket}

\begin{document}
\title {Cavity-control of Majorana bound states in superconductor-semiconductor heterostructures} 

\author{Francesco Buonemani}
\author{Massimo Balmelli}
\author{Olesia Dmytruk}
\affiliation{CPHT, CNRS, École polytechnique, Institut Polytechnique de Paris, Palaiseau, France}

\date{\today}

\begin{abstract}
 We theoretically study a hybrid superconductor-semiconductor platform hosting Majorana bound states coupled to a single mode photonic cavity. Starting with a one-dimensional wire coupled to a bulk $s$-wave superconductor embedded in a photonic cavity, we derive an effective light-matter Hamiltonian for such a platform. Assuming that the photonic vector potential is aligned along the tunneling between a wire and a superconductor, we find that the cavity coupling enters only in the effective superconducting pairing term. By solving the coupled electron-photon Hamiltonian using different approaches, such as exact diagonalization in case of zero or large number of photons, high-frequency expansion, and mean-field decoupling, we find that the phase boundary between the topological trivial phases is shifted to smaller values of the Zeeman energy compared to the uncoupled case. Cavity embedding has the strongest effect on the phase diagram in the semiclassical regime, corresponding to a large number of photons. In all cases, we find that at large values of the light-matter coupling strength the effective superconducting pairing is suppressed, driving the system into the gapless phase. We demonstrate that even small light-matter coupling strength allows for entering the topological phase at values of the Zeeman energy compared to the uncoupled platform.

\end{abstract}

\maketitle

\section{Introduction}

Majorana bound states (MBSs) are zero-energy excitations that appear at the boundaries of topological superconductors and are considered to be promising building blocks for topological quantum computation~\cite{alicea2012new,Beenakker_search_2013,Stanescu_2013,aguado_majorana_2017,lutchyn2018realizing,prada2020from,laubscher2021,Flensberg2021} The paradigmatic model for one-dimensional topological superconductivity is the Kitaev chain model, which supports a topological phase with unpaired MBSs at its ends~\cite{kitaev2001unpaired}. Two- and three-sites Kitaev chains based on quantum-dot platforms were realised in recent experiments~\cite{Dvir2023exp2site,tenHaaf2024exp2site,Zatelli2024exp2site,vanLoo2026single,Bordin2025exp3site,ten2025exp3site}. Although poor man’s MBSs that appear at sweet spot reproduce many features of MBSs, they lack the topological protection~\cite{Leijnse2012Parityqubits,SeoaneSouto2024,luethi2024from,luethi2025fate}. Among the proposed platforms for realizing MBSs in topological superconductors~\cite{graphene2014,graphene2015,magneticatom1,magneticatom2,magneticatom3}, proximitized semiconductor nanowire with strong spin–orbit coupling and subject to a magnetic field~
\cite{lutchyn_majorana_2010,Oreg2010helical} remains one of the most experimentally investigated systems~\cite{Mourik_2012,Das_2012,Deng_2012,Churchill_2013,Finck_2013,deng2016,deMoor2018,MicrosoftQ2023,MicrosoftQ2025,aghaee2025distinctlifetimesxz}.

One of the challenges in realizing the topological phase in nanowires is the relatively large magnetic field required to exceed the critical topological threshold~\cite{schrade2017low}. To overcome this constraint, several theoretical works have proposed using a finite supercurrent flowing through the proximitizing superconductor as an additional tuning parameter~\cite{romito2012manipulating,dmytruk2019majorana}.
At the same time, the unambiguous identification of MBSs in nanowires remains challenging. Smooth confinement potentials~\cite{kells2012,prada2012transport,penaranda2018,vuik2019}, disorder~\cite{liu2012zero}, and near-zero-energy Andreev bound states~\cite{prada2020from,liu2017andreev,reeg2018zero,hess2021local,hess2023trivial,sahu2023effect} can give rise to features that mimic MBSs. This ambiguity could be addressed using alternative probes, such as microwave spectroscopy in nanowires coupled to cavities~\cite{Cottet2013Squeezing,dmytruk2015cavity,Dartiailh2017Cavity,Cottet2017cavity,Dmytruk2023microwave,Prem2026Distinguishing}. In this approach, cavity response can reveal the parity and non-local properties of MBSs offering a promising route to distinguish them from trivial ABSs~\cite{Prem2026Distinguishing}.
Moreover, in the strong light-matter coupling regime, MBSs in topological superconductor hybridize with cavity photons, giving rise to Majorana polaritons~\cite{Trif2012Resonantly,bacciconi2024,dmytruk2024hybrid}.

More broadly, the combination of quantum materials with cavity electrodynamics are within the field of cavity quantum materials~\cite{Francisco2021manipulating,Schlawin2022quantummaterials,Lu2025cavity,bretscher2026fluctuationengineeringcavityquantum}. Recent experimental advances have demonstrated cavity-enhanced superconductivity~\cite{Keren2026Cavitysuperconductivity,montanaro2026cavity,zhang2026cavity,wang2026vacuum}, control of metal-to-insulator transition~\cite{Jarc2023metalinsulator}, and modification of quantum Hall effect~\cite{Appugliese2022Breakdown,enkner2025tunable}. On the theoretical side, cavity coupling has been explored in contexts of superconductivity~\cite{sentef2018,schlawin2019cavity,Curtis2019Eliashberg,alloca2019,kozin2025cavity}, correlated matter~\cite{passetti2023cavity,Fadler2024},  quantum transport~\cite{borici2025Cavity,ciuti2021Cavity,Winter2025Fractional}, and topology~\cite{,cordoba2020entropy,Vlasiuk2023cavity,bacciconi2024,becerra2026fermion,yang2026emergence}. In particular, it was demonstrated that cavity coupling could modify the topological phase boundaries in the Su-Schrieffer-Heeg (SSH) chains~\cite{dmytruk2023controlling,PerezGonzalez2025lightmatter,Nguyen2024Electron,Shaffer2024Entanglement,sueiro2025floquet,ritzzwilling2026topologicalmarkersonedimensionalfermionic} and nanowires~\cite{kobialka2026topology}.

\begin{figure}[t]
    \centering
\includegraphics[width=\linewidth]{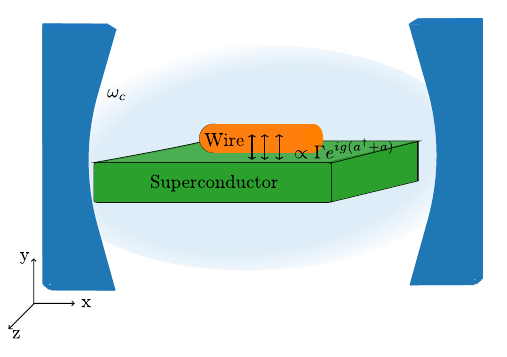}
    \caption{ 
Scheme of a setup: Quantum wire (in orange) aligned along $\bf{x}$ tunnel coupled with amplitude $\Gamma$ to a bulk superconductor (in green) embedded in a single mode photonic cavity (in blue) with frequency $\omega_c$. Tunneling amplitude $\Gamma$ is dressed by a phase $\Gamma  e^{ig(a+a^\dag)}$ that arises from photonic vector potential aligned along $\bf{y}$ direction.
}
    \label{fig:Scheme}
\end{figure}

In this work, we theoretically study a topological superconducting wire coupled to a single-mode cavity (see Fig.~\ref{fig:Scheme}).
We first derive an effective action for a one-dimensional semiconducting wire tunnel coupled to a three-dimensional superconductor embedded in a cavity. The cavity is described by a spatially homogeneous photonic vector potential aligned with the tunneling direction between the wire and the superconductor. 
By integrating out the bulk superconductor, we obtain an effective Hamiltonian in which the cavity field enters \textit{only} as a phase factor in the effective superconducting pairing term. 
We then analyze the resulting hybrid light–matter Hamiltonian in several regimes. In the dark cavity limit, corresponding to the zero-photon state, vacuum fluctuations are found to suppress the effective superconducting pairing, thereby reducing the critical Zeeman energy required to drive the system into the topological phase. A similar suppression occurs in the semiclassical limit, in which number of photons tends to infinity. Moreover, we also show that superconducting pairing can be completely suppressed at a specific value of the light–matter coupling strength, because of the renomalization due to the zeros of the Bessel function. In addition, we study the electron–photon system using a high-frequency expansion, which yields a purely electronic effective Hamiltonian with cavity-mediated electron interactions. In this framework, the renormalization of the chemical potential leads to results that are qualitatively similar to those obtained in the dark cavity regime. Finally, we use the mean-field decoupling between electrons and photons, and compute the topological phase diagram numerically. Comparing the results obtained within the different approaches, we find that the cavity reduces the critical Zeeman energy required to drive the system into the topological phase, with the largest reduction occurring in the semiclassical limit. However, in all cases considered, a sufficiently strong light-matter coupling suppresses the superconducting pairing and drives the system into a gapless phase.

The paper is organized as follows. In Section~\ref{Derivation} we introduce the model and derive the effective Hamiltonian for a quantum wire tunnel coupled to a bulk superconductor embedded in a cavity. We solve the light-matter coupled Hamiltonian in different regimes in Section~\ref{PhaseDiagram}. The dark cavity regime corresponding to the limit of zero photons in a cavity is considered in Section~\ref{Sec:DarkCavity}, while the semi-classical limit is addressed in Section~\ref{Sec:SemiClassical}. The high-frequency expansion approach for the electron-photon system is presented in Section~\ref{Sec:highfrequencyapproach}, followed by the mean-field solution of the photon-coupled proximitized quantum wire in Section~\ref{Sec:MeanField}. 
Finally,  Section~\ref{Conclusions} summarizes our main findings and concludes the paper.

\section{Effective Hamiltonian for the Majorana platform coupled to photons}
\label{Derivation}

In this section, we derive an effective Hamiltonian for a topological superconductor coupled to a single mode cavity.
We consider a microscopic platform for a topological superconductor that consists of a one-dimensional quantum wire tunnel coupled to a three-dimensional superconductor~\cite{stanescu2011majorana,alicea2012new,zyuzin2013correlations,chevallier2013from}.
The whole platform is embedded in a single mode photonic cavity (see Fig.~\ref{fig:Scheme}). 

The Hamiltonian of the electronic system describing a quantum wire aligned along the $\mathbf{x}$ direction and proximitized with a bulk superconductor consists of three terms, $H_{el} = H_{w} + H_{sc}+H_t$.  A one-dimensional quantum wire Hamiltonian $H_{w}$ consisting of $N_x$ lattice sites is given by~\cite{stanescu2011majorana,Stanescu_2013,chevallier2013from,cole2016proximity,reeg2018zero}
\begin{align}
    H_{w} &= \sum_{x=1}^{N_x} \sum_{\sigma\sigma'}\left[b_{x,\sigma}^{\dagger}(2t_w - \mu - V_Z \sigma_x)_{\sigma,\sigma'}\,b_{x,\sigma'}\right] \notag\\
    &-\sum_{x=1}^{N_x-1} \sum_{\sigma,\sigma'}\left[b_{x,\sigma}^{\dagger} (t_w + i\alpha \sigma_y)_{\sigma,\sigma'}\, b_{x+1,\sigma'} + \text{h.c.}\right],
    \label{eq:Hqwire}
\end{align}
where $b_{x,\sigma}^\dag$ ($b_{x,\sigma}$) creates (annihilates) an electron of spin $\sigma = \uparrow,\downarrow$ at the lattice site $x$, $\mu$ is the chemical potential of the quantum wire, and $t_w = \hbar^2/(2m_w a_l^2)$ is the hopping amplitude of the quantum wire, with $m_w$ being the effective mass of the quantum wire and $a_l$ the lattice constant. Here, $\alpha$ is the Rashba spin-orbit interaction (SOI) constant and $V_Z = g \mu_B B/2$ is the Zeeman energy arising from an applied magnetic field $B$ parallel to the quantum wire, with $g$ being the $g$-factor and $\mu_B$ the Bohr magneton.

The Hamiltonian of a three-dimensional  $s$-wave superconductor defined on a  lattice $N_x a_l \times N_y a_l \times N_z a_l$ reads~\cite{reeg2018zero}
\begin{align}
        H_{sc} &= \sum_{x,y,z}\sum_{\sigma} \left[c^{\dagger}_{x,y,z,\sigma}(6t_{sc} - \mu_{sc})c_{x,y,z,\sigma}\right] \notag \\
    &- t_{sc} \sum_{x,y,z}\sum_{\sigma}\Big[c^{\dagger}_{x,y,z,\sigma}c_{x+1, y,z,\sigma} + c^{\dagger}_{x,y,z,\sigma}c_{x, y+1,z,\sigma} \notag\\
    &+ c^{\dagger}_{x,y,z,\sigma}c_{x, y,z+1,\sigma} + \text{h.c.} \Big]\notag\\
    &-\sum_{x,y,z}\left[\Delta c^{\dagger}_{x,y,z,\downarrow} c^{\dagger}_{x,y,z,\uparrow} + \text{h.c.}\right],
    \label{eq:Hsc}
\end{align}
where $x\in \{1,N_x\}$, $y\in \{1,N_y\}$, $z\in \{1,N_z\}$  and $c^{\dagger}_{x,y,z,\sigma}$ ($c_{x,y,z,\sigma}$) is 
the creation (annihilation) operator acting on electrons with spin $\sigma$ located at site $(x,y,z)$. Here, $t_{sc} = \hbar^2/(2m_{sc} a_l^2)$ is the hopping amplitude of the superconductor, with $m_{sc}$ being the effective mass of the superconductor, $\mu_{sc}$ is the chemical potential of the superconductor,  and $\Delta$ is the pairing potential of the superconductor.

We assume that a one-dimensional quantum wire is placed on top of a bulk superconductor
(see Fig.~\ref{fig:Scheme}). The quantum wire and superconductor are coupled via a spin-conserving tunneling term that has the following form~\cite{alicea2012new,stanescu2011majorana,nakosai2013majorana,Stanescu_2013,zyuzin2013correlations,chevallier2013from,cole2016proximity,reeg2018zero}
\begin{align}
    H_t = -\sum_{x=1}^{N_x}\sum_\sigma  \left(\Gamma  c_{x,1,1,\sigma}^{\dagger} b_{x,\sigma} + \text{h.c.}\right),
    \label{eq:Htun}
\end{align}
where $\Gamma$ is the tunneling strength. 

Next, we couple the quantum wire-superconductor platform to a single mode cavity given by the Hamiltonian $H_{ph} = \omega_c a^\dag a$, where $a^\dag$ ($a$) is the photonic creation (annihilation) operator and $\omega_c$ is the cavity frequency. We consider a homogeneous photonic vector potential $\mathbf{A} =  \mathbf{u}_y \left(g/e\right)(a + a^{\dagger})$ pointing along $\mathbf{y}$, where $g$ is the light-matter coupling strength, and write down the light-matter coupling Hamiltonian using the Peierls substitution~\cite{dmytruk2021gauge,passetti2023cavity,dmytruk2023controlling,bacciconi2024,dmytruk2024hybrid,PerezGonzalez2025lightmatter,Nguyen2024Electron,becerra2026fermion}. We note that since the photonic vector potential is along $\mathbf{y}$ direction, the Peierls phase only dresses the hopping in $\mathbf{y}$ direction, $\Gamma \rightarrow \Gamma e^{i g (a+a^\dag)y_w}$. 
Therefore, the tunneling Hamiltonian~\eqref{eq:Htun} becomes
 \begin{align}
    \tilde{H}_t = -\sum_{x=1}^{N_x}\sum_\sigma  \left(\Gamma e^{i g (a+a^\dag)y_w} c_{x,y_1,1,\sigma}^{\dagger} b_{x,\sigma} + \text{h.c.}\right),
    \label{eq:HtunCoupl}
\end{align}
and the total light-matter coupled Hamiltonian reads
\begin{align}
    H_{C} = H_{ph} + H_{w} + H_{sc} + \tilde{H}_t.
    \label{eq:HCoulombMain}
\end{align}

Integrating out the bulk superconductor and performing a Fourier transformation (see Appendix~\ref{sec:Effective Electron-Photon Hamiltonian} for more details), we arrive at the effective wire-cavity Hamiltonian in momentum space 
\begin{align}
    &H_{\text{eff}} = \sum_{k_x} \Big[\sum_{\sigma,\sigma'}  b_{k_x,\sigma}^{\dagger} h_w(k_x) b_{k_x,\sigma'} \notag\\
    &- \Delta_{\text{eff}} \left(e^{2ig (a + a^{\dagger})} b_{k_x, \uparrow} b_{-k_x, \downarrow} + \text{h.c.}\right)\Big]+\omega_c a^\dagger a,
    \label{eq:HeffCoulomb}
 \end{align}
with
\begin{equation}
h_w(k_x) = 2t - \mu - V_Z\sigma_x - 2t\cos{k_x}+  2\alpha \sin{k_x}\,\sigma_y,
\end{equation}
that depends on the wire and cavity operators and in which coupling to the cavity enters \textit{only} as a phase $\propto (a+a^\dag)$ in the superconducting pairing term. 
This leads to a distinct model from one based on the standard Peierls substitution in an effective proximity-induced wire that would dress the hopping amplitudes~\cite{dmytruk2024hybrid, kobialka2026topology}.

\section{Topological phase diagram in presence of cavity}\label{PhaseDiagram}
In this section, we solve an effective Hamiltonian~\eqref{eq:HeffCoulomb}  to compute the topological phase diagram in the presence of the cavity coupling. We use three different approaches to address the problem, such as exact diagonalization of the BdG Hamiltonian that we arrive at in zero-photon or semiclassical regimes, high-frequency expansion, and mean-field decoupling between electrons and photons. In all cases, we find that cavity embedding reduces the value of the Zeeman energy required to reach the topological phase hosting MBSs. This result could be understood from suppression of the effective superconducting pairing that also leads to the appearance of the gapless phase at large values of the light-matter coupling strength. 

\subsection{Exact diagonalization}

\subsubsection{Dark cavity: $n=0$}\label{Sec:DarkCavity}
We start by considering the dark cavity limit that corresponds to having zero photons in a cavity~\cite{sentef2020quantum}. To solve Eq.~\eqref{eq:HeffCoulomb}, we project the $H_\text{eff}$ into the photonic Fock states $\ket{n},\ket{m}$ and take into account only the state with $n = m = 0$ photons. One can see that the effective superconducting pairing $\Delta_\text{eff}$ in Eq.~\eqref{eq:HeffCoulomb} is renormalized  as $\Delta_\text{eff} \to \braket{0|\Delta_\text{eff}\,e^{2i\,g(a^\dagger+a)}|0} = \Delta_\text{eff}e^{-2g^2}$ 
(see Appendix~\ref{sec:PhotonicOperators} for more details). Consequently, we arrive at an electron only Hamiltonian with a renormalized superconducting pairing that is suppressed due to cavity coupling. From now on, to lighten the notation, we drop the subscript $x$ on the momentum, since  electrons in the wire only propagate along this direction. The resulting Hamiltonian reads
\begin{align}
    H_{\text{eff}}^{n_{\text{ph}}=0} &= \sum_{k} \Big[\sum_{\sigma,\sigma'}  b_{k,\sigma}^{\dagger} h_w(k) b_{k,\sigma'} \notag\\
    &- \Delta_{\text{eff}}e^{-2g^2} \left( b_{k, \uparrow} b_{-k, \downarrow} + \text{h.c.}\right)\Big].    \label{eq:HeffCoulombDarkCavity}
\end{align}

 In the superconductor-semiconductor nanowire model~\cite{lutchyn_majorana_2010,Oreg2010helical}, the topological phase transition point is obtained by finding the parameters at which the bulk gap at $k = 0$ vanishes. Calculating the bulk energy spectrum of the Hamiltonian~\eqref{eq:HeffCoulombDarkCavity}, we find 
 the critical value of the Zeeman energy corresponding to the transition point between trivial and topological phases
\begin{align}
    V_Z^{t}(g) = \sqrt{\mu^2 + \left(e^{-2g^2}\Delta_{\text{eff}}\right)^2}
     \label{eq:DarkCavitytransition}
\end{align}
that now depends on the light-matter coupling $g$. We can also introduce the value of $g\equiv g_t$ at which the system enters the topological phase for a given value of $V_Z$.
The finite light-matter coupling $g$ that renormalizes the effective superconducting pairing $\Delta_{\text{eff}}$ modifies the topological criterion such that the Majorana bound states emerge at lower values of the Zeeman energy compared to the isolated nanowire, $V_Z^{t}(g)< \sqrt{\mu^2 + \Delta_{\text{eff}}^2}$. 
\begin{figure}[t!]
    \centering
    \includegraphics[width=\linewidth]{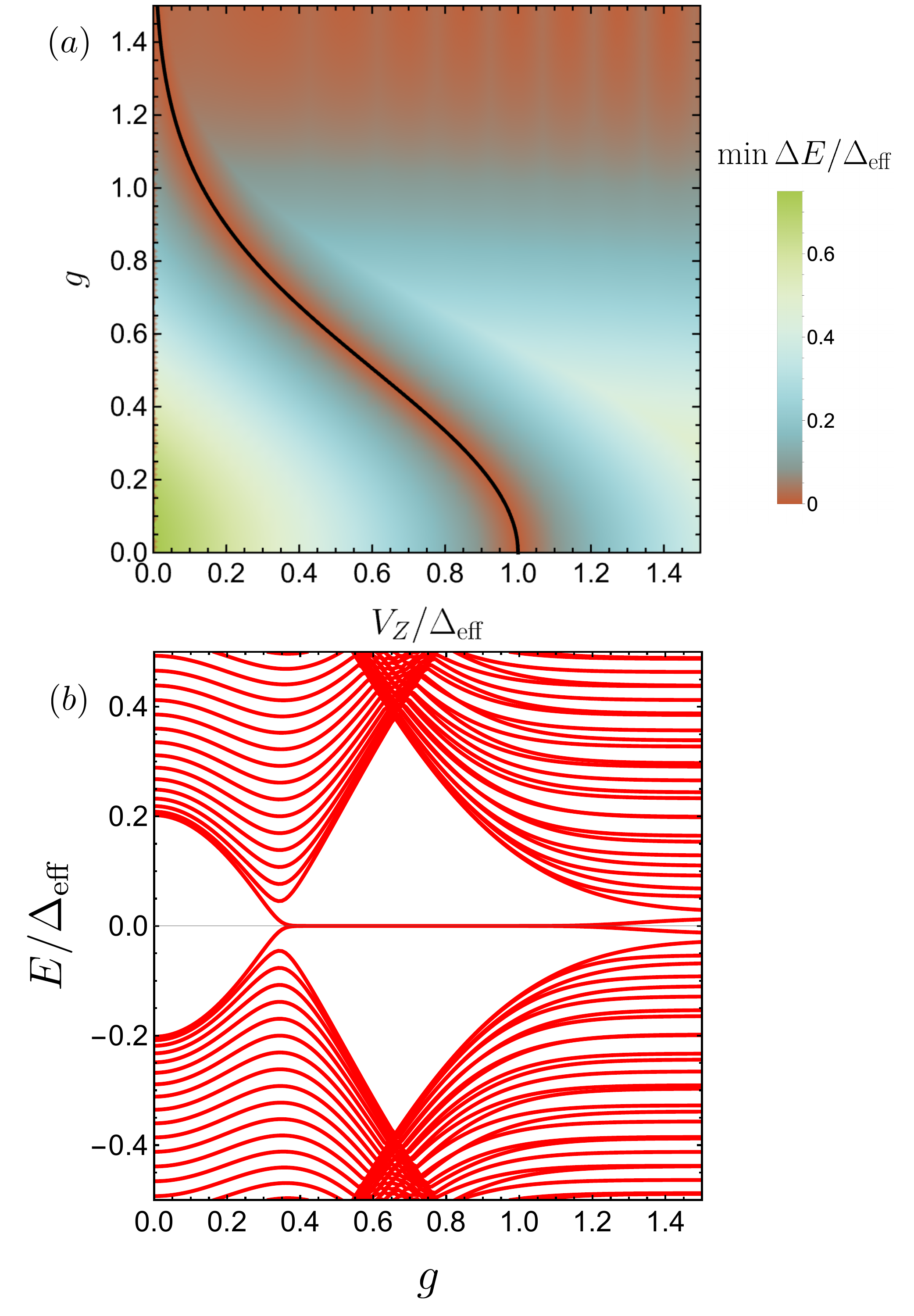}
    \caption{(a) Topological phase diagram and bulk gap of the proximitized wire embedded in a cavity as a function of Zeeman energy $V_Z/\Delta_\text{eff}$ and  light-matter coupling $g$. For $g=0$, we recover the phase boundary of the isolated nanowire, with the topological phase transition occurring at $V_{Z}=\Delta_\mathrm{eff}$.
  For $g$ smaller than the critical value $g_c$ the system is gapped except at the topological gap closing at $k=0$ (black solid line) given by Eq.~\eqref{eq:DarkCavitytransition}. The bulk gap increases when moving away from the phase boundaries. 
    For $g>g_{c}$  the superconducting pairing is suppressed and system becomes gapless.   
    (b) Energy spectrum of a finite-length proximitized wire embedded in a cavity as a function of $g$ for $V_{{Z}}/\Delta_{\mathrm{eff}} = 0.8$. The spectrum is obtained by diagonalizing Eq.~\eqref{eq:HeffCoulombDarkCavity} for a wire with open boundary conditions and $N_x=400$ sites. For $g<g_t$, the spectrum has a finite energy gap, consistent with the trivial phase below the solid black line in panel (a). As $g$ increases beyond $g_t$, the bulk gap closes and reopens, and zero-energy states emerge, signaling the transition into the topological phase. Upon further increasing the light-matter coupling to $g>g_{c}$, the gap closes and a continuum of states emerges, in agreement with the phase diagram shown in the upper region of panel (a). Other parameters  are chosen as $t/\Delta_\text{eff} = 10$, $\alpha/\Delta_\text{eff} = 2$, and $\mu = 0$. 
    }
    \label{fig:GapDarkCavity}
\end{figure}

To further illustrate our findings, we plot the bulk gap of the system as a function of the Zeeman energy and light-matter coupling in Fig.~\ref{fig:GapDarkCavity}~(a). For $g=0$, we recover the case of the isolated superconductor-semiconductor nanowire with zero bulk gap at $V_Z/\Delta_\text{eff}=1$ for $\mu=0$. As $g$ increases, the bulk gap closes at lower values of $V_Z$ that are exactly given  by Eq.~\eqref{eq:DarkCavitytransition}. Following the emergence of the gapless phase in the nanowires with strong supercurrents being driven through the bulk superconductor~\cite{romito2012manipulating,dmytruk2019majorana}, we examine numerically the evolution of the bulk gap for large values of the light-matter coupling strength. We find that the bulk gap closes at the critical value of $g\equiv g_c$ due to suppression of the superconducting pairing in the model.
We also note that the size of the bulk gap increases further away from the phase transition point making it more beneficial to work at smaller values of $g$. This gap closing takes place at finite momentum, therefore, to get more insights, we consider the strong SOI regime~\cite{klinovaja2012composite,schrade2017low} that allows us to find analytically the momentum for which the spectrum is gapless.

Linearizing the Hamiltonian $H_{\text{eff}}^{n_{\text{ph}}=0}$ given by Eq.~\eqref{eq:HeffCoulombDarkCavity} around the Fermi points $k_F = 0,\mp\frac{2\alpha m_w}{\hbar^2}$, we find that the bulk gap closes at momentum
\begin{equation}
(k/k_F) ^2 = \frac{(\Delta_ \text{eff}e^{-2g^2}\pm V_{Z})^2}{4\,\alpha^2 }.
\end{equation}
To provide more insight into the cavity-modified energy spectrum of the wire platform for MBSs, we consider a finite-length proximitized wire with open boundary conditions in the dark cavity regime. 
Using exact diagonalization of the effective Hamiltonian for a finite-size system (see Appendix~\ref{sec:finitelength} for the exact expression for the Hamiltonian), we compute the energy spectrum and demonstrate the emergence of zero-energy states in the topological phase (see Fig.~\ref{fig:GapDarkCavity}~(b)).
For $g<g_t$ the electronic spectrum shows an energy gap, without any in-gap states around zero energy, corresponding to the trivial phase with no MBSs. This corresponds to the region below the solid black line in Fig.~\ref{fig:GapDarkCavity}~(a). For $g_t<g<g_c$ the system enters the topological phase, signaled by the emergence of localized zero-energy modes inside the bulk gap. Upon further increasing $g>g_{c}$, the system transitions into a gapless phase where the bulk gap vanishes and a continuum of energy states develops, as illustrated in the top region of Fig.~\ref{fig:GapDarkCavity}~(a). Therefore, finite light-matter coupling can strongly influence the topological phase transition compared to the isolated nanowire.

\subsubsection{Semi-classical limit: $g\sqrt{n} = \text{const}$}\label{Sec:SemiClassical}
Next, we consider the semi-classical limit, corresponding to a large number of photons in the cavity. Following the approach of Ref.~\cite{sentef2020quantum,buonemani_poor_2026}, we send the number of photons to infinity while keeping the product $g\sqrt{n}$ constant. 
Starting again with the effective Hamiltonian Eq.~\eqref{eq:HeffCoulomb}, we project it
into the photonic Fock states $H_{n,m}=\braket{n|H|m}$ and consider the $n=m$ photon state. Then we look at the limit of $n$ going to infinity (see Appendix~\ref{sec:PhotonicOperators} for more details). Similarly to the dark cavity limit, the effective superconducting pairing is renormalized as 
$\Delta_{\text{eff}} \rightarrow \mathcal{J}_0(4g\sqrt{n})e^{-2g^2}\Delta_{\text{eff}}$, where $\mathcal{J}_0(x)$ is the zeroth order Bessel function of the first kind.

\begin{figure}[t!]
    \centering
    \includegraphics[width=\linewidth]{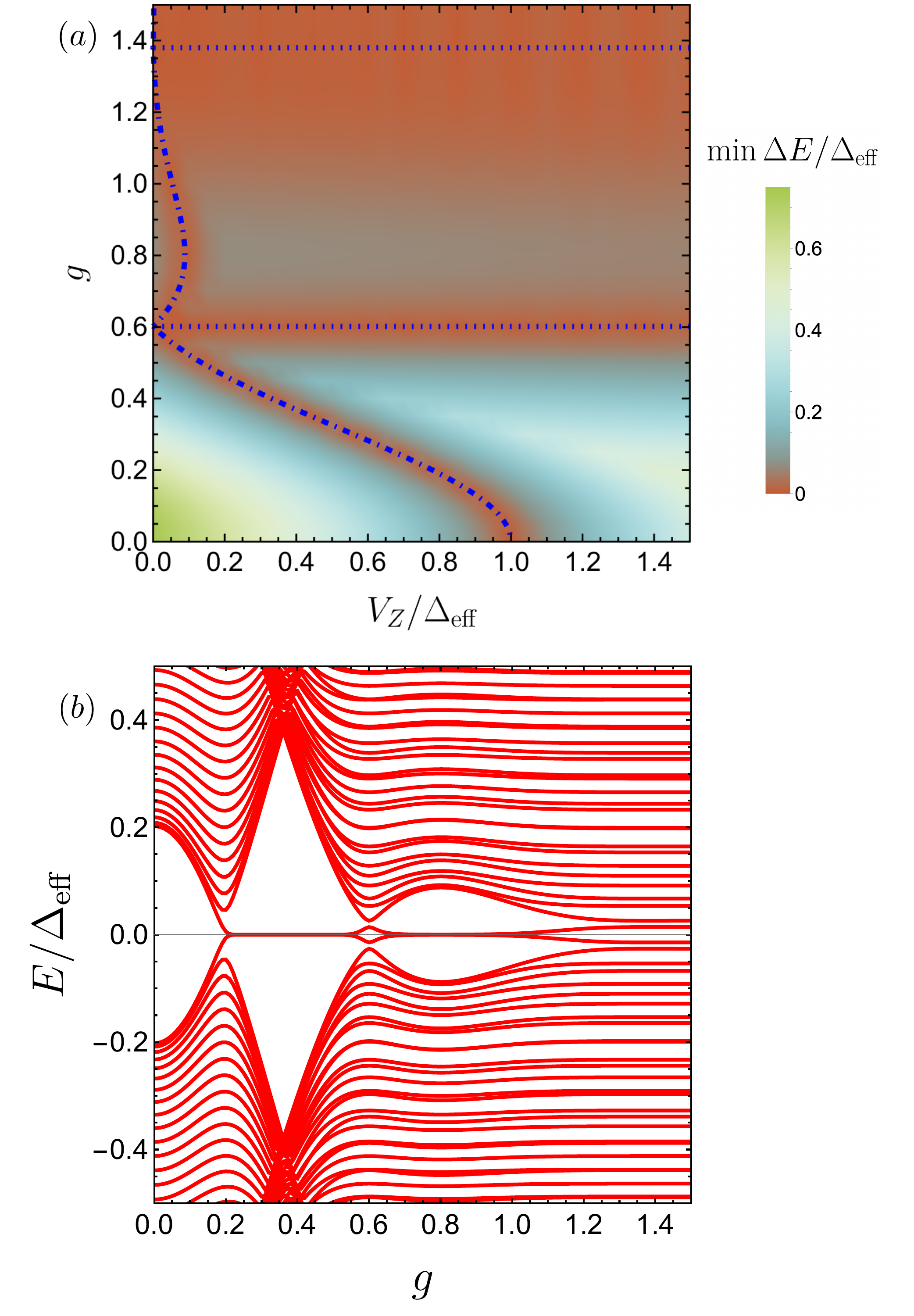}
   \caption{
   (a) Topological phase diagram and bulk gap of the proximitized wire coupled to photons as a function of $V_Z/\Delta_\text{eff}$ and $g$.   
   Gap closing at $k=0$ is given by Eq.~\eqref{eq:bulkgapsc} and corresponds to the phase boundary between topological and trivial phases (blue dotdashed line). 
   Blue dotted lines correspond  to the gap closing when $\Delta_\text{eff}\mathcal{J}_0(4g\sqrt{n}) = 0$ corresponding to the suppression of the superconducting pairing that is independent of $V_Z$. This gap closing at $g_b$ does not change the phase of the system, but reduces the bulk gap. 
 For  $g>g_{c}$ the system becomes gapless.
 (b) Energy spectrum as a function of $g$ for $V_Z/\Delta_\text{eff} =0.8$. The spectrum is obtained via numerical diagonalization of the effective Hamiltonian $H_\text{eff}$~\eqref{eq:HeffCoulomb} in the semiclassical regime for a system size of $N_x = 400$ sites with open boundary conditions. 
 The bulk gap closes at $g_t=0.25$, and zero-energy states emerge at $g>g_t$ marking the emergence of the topological phase. The gap closes again at $g_b=0.61$ that stems from the condition $\mathcal{J}_0(4g\sqrt{n}) = 0$. The zero-energy states persist until $g>g_{c}$, beyond which the system becomes gapless, though the bulk gap is smaller in the region $g_b<g<g_c$ than in the region $g_t<g<g_b$.
 The same feature is observable also in the topological phase diagram of panel (a). Other parameters are the same as in Fig.~\ref{fig:GapDarkCavity}.}
    \label{fig:GapSemiClass}
\end{figure}
By studying the bulk gap closing at $k=0$, we find that the critical value of the Zeeman energy for entering the topological phase becomes
\begin{equation}
V_Z^t(g) = \sqrt{\mu^2 +\left(\mathcal{J}_0(4g\sqrt{n})e^{-2g^2}\Delta_{\text{eff}}\right)^2}. 
\label{eq:bulkgapsc}
\end{equation}
To illustrate our findings, we plot the bulk gap for the wire coupled to a cavity as a function of $V_Z$ and $g$ (fixing $n=1$ in $g\sqrt{n}= \text{const}$) in Fig.~\ref{fig:GapSemiClass}~(a). The system remains gapped throughout most of the phase diagram, except along the blue lines, which correspond to two distinct types of bulk gap closing.
The horizontal blue dotted lines at $g\equiv g_b$ arise from the zeros of the Bessel function $\mathcal{J}_0(4g\sqrt{n})$, at which the effective induced superconducting pairing vanishes. In contrast, the dot-dashed blue lines correspond to gap closings at $k=0$, as described by Eq.~\eqref{eq:bulkgapsc}. Only the latter are associated with the topological phase transition.
Furthermore, the bulk gap decreases as the light-matter coupling $g$ increases and the system becomes gapless in the region with $g>g_c$. 
In the strong SOI regime~\cite{klinovaja2012composite,schrade2017low}, we can analytically determine the momentum at which the spectrum becomes gapless:
\begin{equation}
(k/k_\text{F})^2 = \frac{\left(\Delta_\text{eff}\,{J}_0(4g\sqrt{n})e^{-2g^2} \pm V_{Z}\right)^2}{4\,\alpha^2}.
\end{equation}

To clarify these features, we perform an exact diagonalization of a finite-sized chain (see Appendix~\ref{sec:finitelength} for more details on the Hamiltonian) as a function of the light-matter coupling $g$ for a fixed $V_{Z}/\Delta_\text{eff}=0.8$, as shown in Fig.~\ref{fig:GapSemiClass}~(b). 
We observe that the bulk gap closes and reopens twice as $g$ increases, with the spectrum becoming gapless at $g>g_c$ as also found in the topological phase diagram of Fig~\ref{fig:GapSemiClass}~(a). Zero-energy in-gap energy states appear after the gap closes for the first time at $g_t$ signaling the topological phase transition from the trivial to the topological phase. Next, the gap closes for the second time when  $\mathcal{J}_0(4g\sqrt{n}) = 0$ at $g_b$ and the superconducting pairing becomes zero. Interestingly, the zero-energy levels are still present after this second gap closing at $g_b$, though the size of the induced gap becomes smaller. Finally, at  values of $g>g_c$ the spectrum becomes gapless corresponding to zero bulk gap in Fig.~\ref{fig:GapSemiClass}~(a). We note that the effect of cavity coupling on the phase diagram is stronger in the semi-classical regime, shifting the phase boundary to even smaller values of the Zeeman energy compared to the dark cavity for the same value of $g$.

\begin{figure}[t]
    \centering
    \includegraphics[width=\linewidth]{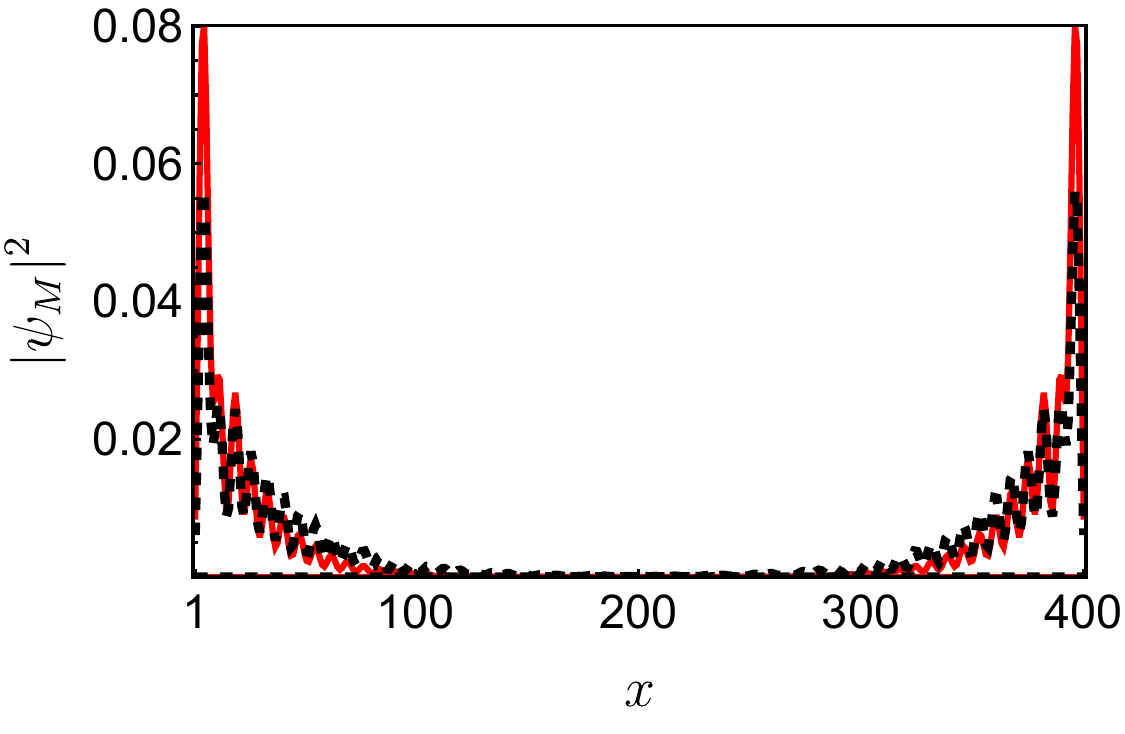}
    \caption{Probability density $|\psi_M|^2$ of the MBSs on the site $x$ for different values of the light-matter coupling strength $g$ and fixed value of Zeeman energy $V_{Z}/\Delta_\text{eff} = 0.8$.  For $g_t<g=0.5<g_b$ (red solid line), the system is in the topological phase, and the MBSs are localized at opposite ends of the wire. For $g_b<g=0.8<g_c$ (black dashed), the near zero-energy modes remain localized at the ends of the wire even after the bulk gap closes due to the suppression of the superconducting pairing when $\Delta_{\mathrm{eff}}\mathcal{J}_0(4g\sqrt{n})=0$. 
    The MBS amplitude is smaller for $g=0.8$ than for $g=0.5$ that reflects the suppression of the bulk gap with increasing light-matter coupling, as shown in Fig.~\ref{fig:GapSemiClass}. Other parameters are the same as in Fig.~\ref{fig:GapSemiClass}.  
    }
    \label{fig:mw_semiclassical}
\end{figure}

To gain further insight into the localization of the zero-energy states emerging in the regions with $g_t<g<g_b$ and $g_b<g<g_c$, we plot their probability density in 
Fig.~\ref{fig:mw_semiclassical}. In both cases, the zero-energy states are localized at the edges of the wire corresponding to the topological phase with the MBSs. Although the MBSs wavefunctions  remain well localized at the ends of the wire in both cases, their amplitude is  reduced after the gap closing at $g_b$ stemming from the Bessel function renormalization of the superconducting pairing. This behavior is consistent with the results shown in Fig.~\ref{fig:GapSemiClass}~(a,b): increasing $g$ suppresses the bulk gap, leading to a corresponding reduction in the Majorana wave-function amplitude.

\subsection{High-frequency expansion}\label{Sec:highfrequencyapproach}
In this section, we solve Eq.~\eqref{eq:HeffCoulomb} by constructing an effective Hamiltonian using the Brillouin-Wigner expansion~\cite{mikami2016brillouin,li2020manipulating,li2022effective,sueiro2025floquet,ritzzwilling2026topologicalmarkersonedimensionalfermionic}. Assuming that $\omega_c$ is the dominant energy scale, the high-frequency regime allows us to obtain the effective Hamiltonian from the Brillouin-Wigner expansion
\begin{equation}
	H_{\text{eff}}^{0} = \bra{0}H_{\text{eff}}\ket{0} +\sum_{l\neq 0} 
	\frac{\bra{l}H_{\text{eff}}\ket{0}\bra{0}H_{c}\ket{l}}{l \omega_c},
	\label{eq:BWapprox}
\end{equation}
where $\bra{0}H_{\text{eff}}\ket{0}$ is the Hamiltonian in the dark cavity regime ($n=0$).
To evaluate the second-order term in the expansion, it is helpful to introduce the following function
\begin{equation}
    D_{l,0}\left(g\right) = e^{-2g^2}(2ig)^l\sqrt{\frac{1}{l!}}L_0^l(4g^2), 
\end{equation}
where $L_n^\alpha(x)$ is the generalized Laguerre polynomials of degree $n$.
Therefore, $H_{\text{eff}}^{0}$ Eq.~\eqref{eq:BWapprox} becomes
\begin{align}
    H_{\text{eff}}^{0} &= \sum_{k,\sigma,\sigma'}b^\dagger_{k,\sigma}h_w(k)b_{k,\sigma'}-\Delta_\text{eff} e^{-2g^2}\left(b_{k,\uparrow}b_{-k,\downarrow}+\text{h.c.}\right)\notag\\&-\sum_{l,k} \frac{2\Delta_\text{eff}^2}{l\omega_c}|D_{l,0}(g)|^2n_{-k,\downarrow}n_{k,\uparrow},
    \label{eq:hfeHamiltonian}
\end{align}
with $n_{-k_x,\downarrow}=b_{-k_x,\downarrow}^\dagger b_{-k_x,\downarrow}$.
 Next, we are going to treat the electron-electron interaction term within the Hartree-Fock (HF) mean-field theory and replace:
\begin{align}
    n_{-k,\downarrow}n_{k,\uparrow} \approx \braket{ n_{-k,\downarrow}}n_{k,\uparrow}+\braket{n_{k,\uparrow}}n_{-k,\downarrow} - \braket{n_{-k,\downarrow}}\braket{n_{k,\uparrow}}.
\end{align}
Introducing $D\left(g\right)= \sum_{l>0}\frac{|D_{l,0}(g)|^2}{l\omega_c}$
and $n_{0\sigma}=\braket{b^\dagger_{k\sigma}b_{k\sigma}}$, the Hamiltonian Eq.~\eqref{eq:hfeHamiltonian} becomes 

\begin{align}
   \tilde{H}_{\text{eff}}^{0} &= \sum_{k,\sigma,\sigma'}b^\dagger_{k,\sigma}h_w(k)b_{k,\sigma'}-\Delta_\text{eff} e^{-2g^2}\left(b_{k,\uparrow}b_{-k,\downarrow}+\text{h.c.}\right)\notag\\&-\sum_{k} {2\Delta_\text{eff}^2}D(g)\left(\sum_\sigma n_{0\bar{\sigma}}b^\dagger_{k,\sigma}b_{k,\sigma} -n_{0\uparrow}n_{0\downarrow}\right).
   \label{eq:highfrequencyhf}
\end{align}

In the high-frequency limit, the cavity coupling results in a shift of the chemical potential and renormalization of the effective superconducting pairing.

Calculating the bulk spectrum of $\tilde{H}_{\text{eff}}^{0}$ Eq.~\eqref{eq:highfrequencyhf} and setting $k=0$, we find the gap closing condition associated with the topological phase transition

\begin{equation}
V_{{Z}}^\text{t}(g,\omega_c)=\sqrt{(\mu-\tilde{\mu})^2+(\Delta_\text{eff}e^{-2g^2})^2},
\label{eq:hfetransition}
\end{equation}
where $\tilde{\mu} =2\Delta_\text{eff}^2 D\left(g\right)n_{0\sigma}$.

We note that $\tilde{\mu}$ in Eq.~\eqref{eq:hfetransition} scales as $1/\omega_c$, and therefore constitutes only a small correction to the critical Zeeman energy obtained in the dark cavity regime, given by Eq.~\eqref{eq:DarkCavitytransition}.

\begin{figure}[t]
    \centering
    \includegraphics[width=\linewidth]{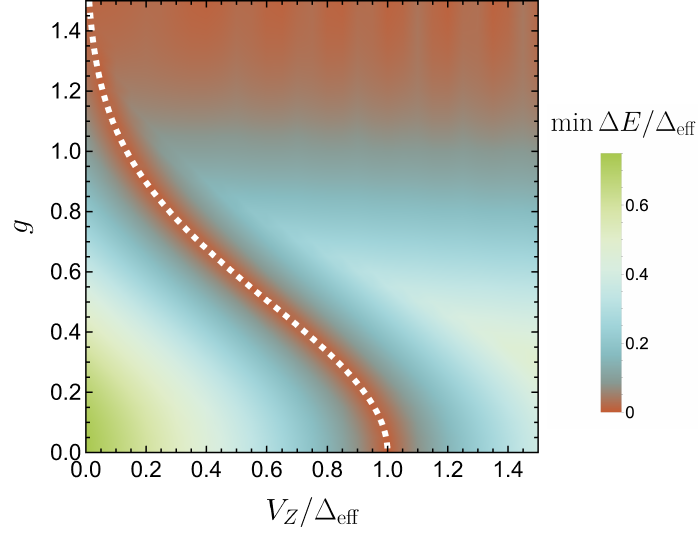}
    \caption{Topological phase diagram and bulk gap of the proximitized wire embedded in a cavity obtained from numerically solving Eq.~\eqref{eq:highfrequencyhf}  as a function of Zeeman energy $V_Z/\Delta$ and light-matter coupling $g$. The bulk gap closes for $g<g_c$ at the topological phase boundary given by Eq.~\eqref{eq:hfetransition} (dashed white line). The size of the bulk gap increases away from the phase boundary, but for large values of $g>g_{c}$ the gap vanishes for all values of $V_Z$, signaling a transition to a gapless phase.
   The photonic cutoff is set to be $l^\text{max}=10$ and other parameters are the same as in Fig.~\ref{fig:GapDarkCavity}. }
    \label{fig:hfphase_diagram}
\end{figure}
To illustrate our findings, we plot in Fig.~(\ref{fig:hfphase_diagram}) the bulk gap as a function of the Zeeman energy and the light-matter coupling. Similarly to the dark cavity regime, we find that at finite values of the light-matter coupling $g<g_c$ the system enters the topological phase at lower Zeeman energy than in the absence of the cavity. However, for sufficiently strong light-matter coupling, $g>g_{c}$, the bulk gap vanishes and the system enters a gapless phase. 
\subsection{Mean-field solution}\label{Sec:MeanField}
In this section, we focus only on infinite wire case and address the light-matter Hamiltonian~\eqref{eq:HeffCoulomb} using mean-field decoupling that neglects correlations between the electrons and photons~\cite{dmytruk2021gauge,dmytruk2023controlling,kozin2025cavity}. Therefore, a general state  $\ket{\Psi}$ of light-matter Hamiltonian $H_\text{eff}$ is
\begin{equation}
    \ket{\Psi} = \ket{\psi}\ket{n},
\end{equation}
where $\ket{\psi}$ is the electron state and $\ket{n}$ is a photonic Fock state, $a\ket{n}=\sqrt{n}\ket{n-1}$.
\begin{figure}[t]
    \centering
    \includegraphics[width=\linewidth]{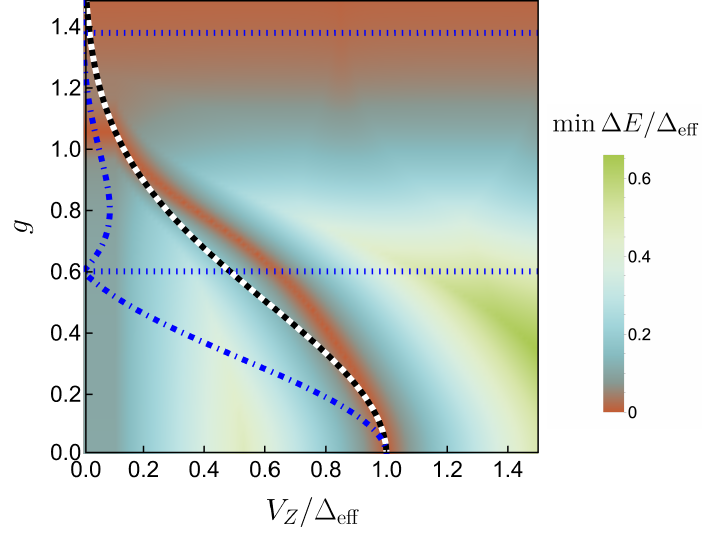}
    \caption{Topological phase diagram and bulk gap of the electronic mean-field Hamiltonian $H^\text{mf}_\text{el}$ as function of Zeeman energy $V_Z/\Delta$ and light-matter coupling $g$. 
     The solid black line indicates the bulk gap closing at $k=0$ in the dark cavity regime Eq.~\eqref{eq:DarkCavitytransition}, the blue dot-dashed line corresponds to the bulk gap closing in the semiclassical regime Eq~\eqref{eq:bulkgapsc}, and the white dotted line shows the bulk gap closing obtained via high-frequency expansion Eq.~\eqref{eq:hfetransition}. Within mean-field decoupling, $g<g_c$ shifts the bulk gap closing (brown line) towards lower values of Zeeman energy, reducing the value of $V_Z$ required to reach the topological phase. Compared with the other approaches, the mean-field decoupling shows a larger value of $g$ needed to reach the topological phase for a given Zeeman energy.  Upon further increasing $g$, the bulk gap is progressively suppressed and closes for $g>g_{c}$, as in other cases.
     Other parameters are the same of Fig.~\ref{fig:GapDarkCavity}.}
    \label{fig:meanfieldphase}
\end{figure}
As a result of the mean-field decoupling, we have to solve self-consistently a mean-field photonic Hamiltonian

\begin{equation}
    H^\text{mf}_\text{ph} = \omega_ca^\dagger a -B_{\text{el}}\,e^{2ig(a+a^\dagger)}+\text{h.c.},
\end{equation}

and a mean-field electronic Hamiltonian
\begin{equation}
    H^\text{mf}_\text{el}= \sum_{k,\sigma,\sigma'}h_w(k)b^\dagger_{k,\sigma}b_{k,\sigma'}- A_{\text{ph}}\Delta_\text{eff} b_{k,\uparrow}b_{-k,\downarrow}+\text{h.c},
    \label{eq:mfdelectronicHamiltonian}
\end{equation}

where we introduced $B_{\text{el}}=\sum_k\Delta_\text{eff}\braket{\psi|b_{k,\uparrow}b_{-k,\downarrow}|\psi}$ and $A_{\text{ph}}=\braket{n|e^{2ig(a+a^\dagger)}|n}$. For $k=0$ we can find the topological criterion for the electronic Hamiltonian
\begin{equation}
    V_{Z} ^\text{t} =\sqrt{\Delta_\text{eff}^2A_{\text{ph}}^2 + \mu^2},
\end{equation}
with $A_{\text{ph}}$ that depends on the photonic parameters to be determined self-consistently. 

Solving the electronic Hamiltonian $H^\text{mf}_\text{el}$~\eqref{eq:mfdelectronicHamiltonian} self-consistently, we compute the topological phase diagram presented in Fig.~\ref{fig:meanfieldphase}. 
We find numerically that cavity embedding  lowers the critical Zeeman energy required to enter into the topological phase. In the limit $g=0$, we recover the isolated-wire result.
Comparing the phase boundaries obtained from the different approaches, namely the dark cavity Eq.~\eqref{eq:DarkCavitytransition} (black line), the semiclassical approximation Eq~\eqref{eq:bulkgapsc} (dot-dashed blue line), and the high-frequency expansion Eq~\eqref{eq:hfetransition}  (dashed white line), we find that within the mean-field treatment a larger light-matter coupling is required for the system to enter the topological phase at a given Zeeman energy. Furthermore,  the bulk gap decreases vanishes for $g>g_{c}$, as in the previously considered cases.

\section{Conclusions}\label{Conclusions}

In this work, we study an effect of cavity embedding on a platform for MBSs based on a semiconducting wire proximity-coupled to a bulk superconductor. Integrating out the bulk superconductor yields an effective electron-photon Hamiltonian where the cavity coupling enters as a renormalization of the induced superconducting pairing. To find the modification of the topological phase diagram due to the coupling to a single mode photonic cavity, we analyze
the cavity-wire Hamiltonian in different regimes.

In the dark cavity regime, which corresponds to zero photons in the cavity, the superconducting pairing decreases with increasing the strength of the light-matter coupling strength, reducing the Zeeman energy required for reaching the topological phase with MBSs, though for very large light-matter coupling the system becomes gapless. Similar results are obtained in the semiclassical regime, with the additional bulk gap closing at specific values of the light-matter coupling strength stemming from the Bessel functions renormalization of the superconducting pairing, which vanishes at the zeros of the corresponding Bessel function. However, this additional gap closing does not change the topological phase of the system that is corroborated through exact diagonalization of a finite-size chain with open boundary conditions, where the emergence of zero-energy states is observed both before and after this gap closing. Furthermore, we consider a high frequency regime that allows us to derive an effective electronic Hamiltonian with cavity-induced electron interactions. Within HF mean-field theory we find that both chemical potential and superconducting pairing are modified  resulting in the reduction of the Zeeman energy for entering the topological phase compared to the bare electronic system. Finally, we consider the mean-field decoupling between photons and electrons and analyze the resulting topological phase diagram numerically. In this case we also find that cavity embedding drives the wire into the topological phase at lower values of the Zeeman energy. 

To summarize, we find that cavity embedding lowers the Zeeman energy at which the MBSs could emerge in a wire-superconductor platform even at small values of the light-matter coupling strength, though the system becomes gapless at strong couplings due to suppression for the superconducting pairing. Comparing different approaches to address the electron-photon problem, we find that for fixed Zeeman energy the light-matter coupling required to drive the system into the topological phase is smallest in the semiclassical regime. Moreover, dark cavity regime and high-frequency expansion give similar results. However, larger value of the light-matter coupling is needed to enter the topological phase within the MF approach. This work demonstrates that cavity embedding significantly influences the topological phase boundary in the  superconductor-semiconductor platform, offering a novel route for decreasing the values of the Zeeman energy required for entering the topological phase.

\section*{Acknowledgements}
This work is supported by ERC grant (Q-Light-Topo, Grant Agreement No.~101116525) (F.B. and O.D.).

\appendix

\section{Derivation of effective action}
\label{sec:Effective Electron-Photon Hamiltonian}
We are interested in integrating out the bulk superconductor to derive an effective Hamiltonian for a quantum wire with proximity-induced superconductivity embedded in a cavity. To do so, it is instructive to first perform a unitary transformation $H_{el-ph} = U^\dag H_C U$, with
$U=e^{-ig(a+a^\dag)\sum_{x=1}^{N}\sum_\sigma b_{x,\sigma}^\dag b_{x,\sigma}}$, 
to remove the photonic operators from the tunnelling term ~\cite{dmytruk2015cavity}. Applying $U$ to Eq.~\eqref{eq:HCoulombMain} and using that
\begin{align}
    U^\dag b_{x',\sigma'}U = e^{-i g (a+a^\dag)}b_{x',\sigma'},
    \label{eq:unitary_transformation_of_b},
\end{align}
we arrive at
\begin{align}
    &H_{el-ph}  = H_{ph} +H_w+H_{sc}+H_t+\notag\\&- i\omega_c g (a-a^\dag) \sum_{x=1}^{N_x}\sum_\sigma b_{x,\sigma}^\dag b_{x,\sigma}\notag\\
    &+ \omega_c g^2 \left(\sum_{x=1}^{N_x}\sum_\sigma b_{x,\sigma}^\dag b_{x,\sigma}\right)^2.
\label{eq:dipole_gauge_hamiltonian}
\end{align}

Next, we transform the electronic part of the Hamiltonian $H_{el} = H_w+H_{sc}+H_t$ in the momentum space, 
$\hat{o}_{\mathbf{R}} = \frac{1}{\sqrt{V}}\sum_{\mathbf{k}} e^{i \mathbf{k}\mathbf{R}} \hat{o}_{\mathbf{k}}$.
We obtain for the quantum wire Hamiltonian~\eqref{eq:Hqwire} 
\begin{align}
    H_w = \sum_{k_x,\sigma,\sigma'} b_{k_x,\sigma}^{\dagger} h_w(k_x) b_{k_x,\sigma'}.
\end{align}
with $h_w(k_x) = 2t_w - \mu - V_Z\sigma_x - 2t_w\cos{k_x} + 2\alpha \sin{k_x}\sigma_y$,
for the superconductor Hamiltonian~\eqref{eq:Hsc}
\begin{align}
   H_{sc} = \sum_{\mathbf{k}} \varepsilon_{k}c_{\mathbf{k}}^{\dagger}c_{\mathbf{k}} - \Delta(c_{\mathbf{k}\uparrow}c_{-\mathbf{k}\downarrow} + \text{h.c.}),
   \label{eq:HscMomentum}
\end{align}
where $\varepsilon_{k} = 6 t_{sc} - \mu_{sc} -2t_{sc}(\cos{k_x} + \cos{k_y} + \cos{k_z})$, and for the tunnelling Hamiltonian~\eqref{eq:Htun}

\begin{align}
    H_t = -\Gamma \sum_{\mathbf{k},\sigma} c_{\mathbf{k},\sigma}^{\dagger} b_{k_x,\sigma} + \text{h.c.}.
\end{align}

Next, we bring Eq.~\eqref{eq:HscMomentum} to a diagonal form $H_{sc} = \sum_{\mathbf{k}} E_k\left(\chi_{ \mathbf{k}1 }^\dag \chi_{ \mathbf{k}1 } + \chi_{ \mathbf{k}2}^\dag \chi_{ \mathbf{k}2 }\right)$ by performing the transformation~\cite{alicea2012new}
\begin{align}
    &c_{k,\uparrow} = -u_k\chi_{1,k}+v_{k}\chi_{2,-k}^\dagger,\\
    &c_{k,\downarrow} = v_k\chi_{1,-k}^\dagger+u_{k}\chi_{2,k},
\end{align}
with $u_k = \Delta/\sqrt{2 E_k(E_k - \epsilon_k)}$, $v_k = \Delta/\sqrt{2 E_k(E_k + \epsilon_k)}$, and $E_k = \sqrt{\epsilon_k^2 + \Delta^2}$.

The partition function of the electron-photon system reads
  \begin{align}
      Z = \int \mathcal{D}\left[\bar \eta,\eta,\bar \chi, \chi,\bar \Phi, \Phi\right] e^{ -S_{w}-S_{sc} - S_{t}},
      \label{eq:Z_nanosclight}
  \end{align}
  where $\eta,\bar\eta$, $\chi,\bar\chi$ and $\Phi,\bar\Phi$  correspond to the quantum wire, superconductor and photonic fields, respectively. Here, $S_{w}$ is the action for the quantum wire and
 \begin{align}
     &S_{sc}+S_t = \int\frac{d\omega}{2\pi}\sum_\mathbf{k} \bar\chi_{1k}(-i\omega + E_k) \chi_{1k} \notag\\&+ \bar\chi_{2k}(-i\omega + E_k) \chi_{2k}+\notag\\& 
+ \left(\bar J_{1k}\chi_{1k}+  
      \bar J_{2k}\chi_{2k} + \text{h.c.}\right),
 \end{align}
is the action for the superconductor and tunnelling, where
\begin{align}
        &\bar J_{1k} = \Gamma u_k \bar \eta_{k_x \uparrow} + \Gamma v_k\eta_{-k_x \downarrow},\\ 
        &\bar J_{2k} = \Gamma  v_k \bar \eta_{-k_x \uparrow} - \Gamma  u_k\eta_{k_x \downarrow}.
\end{align}

Next, we integrate out $\chi,\bar\chi$ corresponding to the bulk superconductor. We arrive at the effective action $S_{\text{eff}} = S_{w} + \delta S$, where
\begin{align}
&\delta S = -\Gamma^2 \int \dfrac{d\omega}{2\pi} \sum_{\mathbf{k}} \Big[\sum_\sigma \dfrac{ i\omega + \varepsilon_k}{\omega^2 +E_k^2}\bar\eta_{(k_x,\sigma)}(\omega)\eta_{(k_x,\sigma)}(\omega) \notag \\
&+  \dfrac{\Delta}{\omega^2 +E_k^2} \left( \eta_{(k_x,\uparrow)}(\omega)\eta_{(-k_x, \downarrow)}(\omega) + \text{h.c.}\right)\Big].  
\label{eq:EffectiveAction}
 \end{align}

Introducing $f(k_x,\omega) = \sum_{k_y,k_z} \Gamma^2/\left(\omega^2 +E_k^2\right)$ and $g(k_x,\omega) = \sum_{k_y,k_z} \varepsilon_k/\left(\omega^2+E_k^2\right)$, the effective action
$S_{\text{eff}}$ can be simplified as

\begin{align}
    &S_w+\delta S=\int\frac{d\omega}{2\pi}\sum_{k_x,\sigma,\sigma'}\bar\eta_{k_x,\sigma}(\omega)\Big[h_w(k_x)+\notag\\&-i\omega\left(1+f_{k_x}(\omega)\right)-g_{k_x}(\omega)\Big]\eta_{k_x,\sigma'}\notag\\&-\Gamma^2\Delta f_{k_x}(\omega)\left( \eta_{(k_x,\uparrow)}(\omega)\eta_{(-k_x, \downarrow)}(\omega) + \text{h.c.}\right)\Big].
\label{eq:EffectiveAction2}
\end{align}
Let us focus on $f(k_x,\omega)$ and $g(k_x,\omega)$.
The term proportional to $g(k_x,\omega)$ is  a shift of the chemical potential of the quantum wire Hamiltonian and can be absorbed into this term. Instead, $f(k_x,\omega)$ is proportional to the quasiparticle weight of the bulk superconductor~\cite{Reeg2017Transport}.
Upon the definition of the quasiparticle weight  $Z(\omega) =[ 1+f(k_x,\omega)]^{-1}$ we can express $S_{\text{eff}}$ as
\begin{align}
    &S_\text{eff}=\int\frac{d\omega}{2\pi}\sum_{k_x,\sigma,\sigma'}\bar\eta_{k_x,\sigma}\left(h_w(k_x)-i\frac{\omega}{Z(\omega)}\right)\eta_{k_x,\sigma'}\notag\\&-\Delta\left[ \frac{1}{Z(\omega)}-1\right]\left( \eta_{(k_x,\uparrow)}(\omega)\eta_{(-k_x, \downarrow)}(\omega) + \text{h.c.}\right)\Big].
\label{eq:EffectiveAction2}
\end{align}

Because we are interested in the low energy limit $E\ll~\Delta$ we rescale $\omega\to\omega/Z(\omega)$  and define $\Delta_\text{eff}~=~ \Delta\left[ 1/Z(\omega)-1\right]$~\cite{lutchyn_majorana_2010,stanescu_majorana_2011,stanescu_proximity_2010,Reeg2017Transport}.
Therefore, we are left with
\begin{align}
   &S_\text{eff}=\int\frac{d\omega}{2\pi}\sum_{k_x,\sigma,\sigma'}\bar\eta_{k_x,\sigma}(\omega)\left[h_w(k_x)-i\omega\right]\eta_{k_x,\sigma'}\notag\\&-\Delta_\text{eff}\left( \eta_{(k_x,\uparrow)}(\omega)\eta_{(-k_x, \downarrow)}(\omega) + \text{h.c.}\right).
\end{align}

From this action, in the static approximation $\omega\to 0$, we can read off the effective Hamiltonian for the wire $H_w$~\cite{lutchyn_majorana_2010,Oreg2010helical} 
\begin{align}
    H_w &= \sum_{k_x,\sigma,\sigma'}b^\dagger_{k_x,\sigma}(\omega)h_w(k_x)b_{k_x,\sigma'}\notag\\
    &-\Delta_{\text{eff}}\left( b_{k_x,\uparrow}b_{-k_x, \downarrow} + \text{h.c}\right).
\end{align}
Consequently, the effective Hamiltonian for the coupled electron-photon system is given by:
\begin{align}
H_\text{eff} &=  H_w + \omega_c a^\dagger a+\notag\\
&- i\omega_c g (a-a^\dag) \sum_{k_x,\sigma
} b_{k_x,\sigma}^\dag b_{k_x,\sigma}+\notag \\&+ \omega_c g^2 \left(\sum_{k_x,\sigma} b_{k_x,\sigma}^\dag b_{k_x,\sigma}\right)^2.
\end{align}
Applying then the inverse unitary transformation $U^{-1}$, defined by
 \begin{align}
    (U^{-1})^\dagger b_{k_x',\sigma'}U^{-1} = e^{i g (a+a^\dag)}b_{k_x',\sigma'},
    \label{eq:unitary_transformation_of_b}
\end{align}
to  $H_\text{eff}$ we arrive at the final form 
\begin{align}
    &H_{\text{eff}} = \sum_{k_x} \Big[\sum_{\sigma,\sigma'}  b_{k_x,\sigma}^{\dagger} h_w(k_x) b_{k_x,\sigma'} \notag\\
    &- \Delta_{\text{eff}} \left(e^{2ig (a + a^{\dagger})} b_{k_x, \uparrow} b_{-k_x, \downarrow} + \text{h.c.}\right)\Big]+\omega_c a^\dagger a.
    \label{eq:HeffCoulombapp}
\end{align}

\section{Evaluation of photonic expectation values}
\label{sec:PhotonicOperators}
\subsection{Dark cavity}
Before performing numerical studies of the electronic Hamiltonian, we want to get rid of the dependence on the photonic creation and annihilation operators. We start by considering a dark cavity with zero photons~\cite{sentef2020quantum}.
We project the Hamiltonian Eq.~\eqref{eq:HeffCoulomb} into the photonic Fock states $\ket{n},\,\ket{m}$ and then consider only the states with $n=m=0$. Then we replace the photonic operator with the expectation value 

\begin{align}
  &\bra{0}e^{2ig(a + a^{\dagger})} \ket{0} =\notag\\  &e^{-2{g^2}} \bra{0}\left(\sum_n  \dfrac{(2iga^{\dagger})^n}{n!}\right)\left(\sum_m \dfrac{(2iga)^m}{m!}\right)\ket{0}  =  e^{-{2g^2}}.
\end{align}
In the last step, we used the fact that only the $n=m = 0$ terms survive in the sum and vacuum normalization. Hence, we arrive at the final result
\begin{align}
            \bra{0} H_\text{eff} \ket{0} =  \sum_{k,\sigma,
            \sigma'} b_{k,\sigma}^{\dagger} (2t - \mu - V_Z\sigma_x - 2t\cos{k}+\notag \\ + 2\alpha \sin{k}\sigma_y) b_{k,\sigma'} -  (\Delta_\text{eff} e^{-2g^2} b_{k, \uparrow} b_{-k, \downarrow} + \text{h.c.}).
            \label{eq:effectiveham_momentum}
\end{align}
To lighten the notation, we dropped the subscript $x$ on the momentum since it is clear that electrons in the wire only propagate along this direction.\\
The net effect on the system is a renormalization of the superconducting pairing that get reduced while increasing the light-matter coupling strength. This new parameter changes the expression for the phase transition between trivial and non-trivial phase. We can derive the new phase transition point by looking at the value of the parameter for which the gap closes at $k=0$. The eigenvalues in the lower branch at $k =0$ are
\begin{align}
    E^2 &= \mu^2 + V_{Z}^2 + \Delta_\text{eff}^2e^{-2g^2} \notag\\&- 2 \sqrt{\mu^2 V_{Z}^2 + V_{Z}^2\Delta_\text{eff}^2e^{-2g^2}}.
 \end{align}
The condition for the phase transition is then easily determined by requiring the eigenvalues to vanish. The condition take a relatively simple form if $\mu = 0$, namely
\begin{equation}
    V_Z ={\Delta_\text{eff}} e^{-2g^2}.
    \label{eq:modified phase trans}
\end{equation}
\subsection{Semi-classical limit}
\begin{figure}[t]
    \centering
    \includegraphics[width=\linewidth]{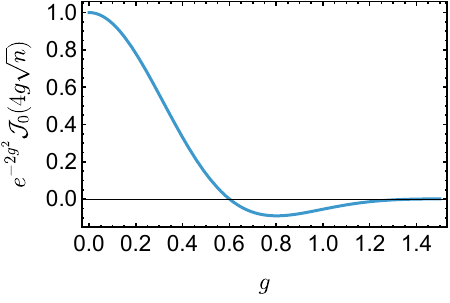}
    \caption{Plot of the semiclassical renormalization  $\mathcal{J}_0(4g\sqrt{n})$ Eq.~\eqref{eq:sc_renomalization_delta} of $\Delta_{\text{eff}}$ as function of the light-matter coupling $g$. We observe that the Bessel function $\mathcal{J}_0(4g\sqrt{n})$ vanishes for $g=0.6$ and $g=1.4$, resulting in a corresponding suppression of the superconducting pairing at $g=g_b$.}
    \label{fig:bessel_function}
\end{figure}

Another limit that we are interested in is the semi-classical limit. To perform this approximation, we follow the approach of \cite{sentef2020quantum,buonemani_poor_2026}. We send the number of photons to infinity by keeping the product $g\sqrt{n} = const$. Physically, we are considering a regime in which the light-matter coupling strength is weak, but the amount of photons is large. 
We project the Hamiltonian Eq.~\eqref{eq:HeffCoulomb} into the photonic Fock states $H_{n,m}=\braket{n|H|m}$ and consider the $n=m$ photon state. Therefore, the expectation value of the photonic operators becomes
\begin{align}
     &e^{{2g^2}} \bra{n}e^{2ig(a + a^{\dagger})} \ket{n} = \notag \\ &= \bra{n}\left(\sum_k  \dfrac{(2iga^{\dagger})^k}{k!}\right)\left(\sum_l \dfrac{(2iga)^l}{l!}\right)\ket{n} \notag \\& = \sum_{k} \frac{(-1)^k (2g)^{2k}}{(k!)^2}  \bra{n}a^{\dagger k} a^k \ket{n} = \sum_{k=0}^n \dfrac{(-1)^k (2g)^{2k} n!}{(k!)^2 (n-k)!} .
\end{align}
Here we used the orthogonality of the states $\ket{n}$, and in the last step the standard relationships $a\ket{n} = \sqrt{n}\ket{n-1}$, $a\ket{n} = \sqrt{n+1}\ket{n+1}$.
Taking $n\to\infty$ limit, the result becomes 
\begin{equation}
    \lim_{n \rightarrow \infty} \sum_{k=0}^{n} \dfrac{(-1)^k}{(k!)^2} \left( \dfrac{4g\sqrt{n}}{2}\right)^{2k} = \mathcal{J}_0(4g\sqrt{n}).
\end{equation}
We can finally put everything together to get 
\begin{equation}
    \lim_{n \rightarrow \infty} \bra{n}e^{2ig(a + a^{\dagger})} \ket{n} = 
    e^{-2g^2} \mathcal{J}_0(4g\sqrt{n}).
    \label{eq:sc_renomalization_delta}
\end{equation}
Plugging this into the Hamiltonian Eq.~\eqref{eq:HeffCoulomb}, we arrive at a Hamiltonian with a modified superconducting pairing
\begin{align}
    H_\text{sc}&=\sum_{k,\sigma,\sigma'} b_{k,\sigma}^{\dagger}\big(2t_w(1 - \cos k)\notag\\& - \mu - V_Z\sigma_x + 2 \alpha\sin k \sigma_y\big)b_{k,\sigma'} \notag\\&- (\Delta_\text{eff}\mathcal{J}_0({4g\sqrt{n}})e^{-2g^2}b_{k,\uparrow}b_{-k,\downarrow} + \text{h.c.}).
    \label{eq:effective ham semiclassical}
\end{align}
It is interesting to notice that, in contrast to the exponential suppression of the effective superconducting pairing, the Bessel function can go to zero at a finite value of $g\sqrt{n}$ as depicted in Fig. \ref{fig:bessel_function}.

From Fig.~\ref{fig:bessel_function} we can extrapolate that $\Delta_\text{eff}$ vanishes at a specific coupling strength of $g\approx0.6$ and $g\approx1.4$ for $n=1$. However, this gap closing does not correspond to a phase transition; as the coupling strength is further increased, the gap reopens while the system remains within the same topological phase.

\section{Effective Hamiltonian with open boundary conditions}
\label{sec:finitelength}
To complement the momentum-space analysis presented in the main text, we provide here the real-space formulation of the Hamiltonian Eq. (\ref{eq:HeffCoulomb}). Imposing the open boundary conditions will allow us to perform a numerical exact diagonalization of the system. By performing a Fourier transformation $b_{k,\sigma} = \frac{1}{\sqrt{N_x}}\sum_{x=1}^{N_x} e^{ikx}\,b_{x,\sigma}$,
we arrive at the Hamiltonian with open boundary conditions
\begin{align}
    H_\text{eff} &= \sum_{x=1}^{N_x} \sum_{\sigma\sigma'}\left[b_{x,\sigma}^{\dagger}(2t_w - \mu - V_Z \sigma_x)_{\sigma,\sigma'}\,b_{x,\sigma'}\right] \notag\\
    &-\sum_{x=1}^{N_x-1} \sum_{\sigma,\sigma'}\left[b_{x,\sigma}^{\dagger} (t_w + i\alpha \sigma_y)_{\sigma,\sigma'}\, b_{x+1,\sigma'} + \text{h.c.}\right]\notag\\&
    -\sum_{x=1}^{N_x}\Delta_\text{eff}e^{2ig(a^\dagger+a)}\left[b^\dagger_{x,\downarrow}b^\dagger_{x,\uparrow}+\text{h.c.}\right].
    \label{eq:finitesizechain}
\end{align}
Numerical exact diagonalization is then readily performed by rewriting Eq.~\eqref{eq:finitesizechain}
in the Bogoliubov-de Gennes~(BdG) basis using the spinor
$\Psi=\left(b_{1,\uparrow}b_{1,\downarrow}\dots b_{N_x,\uparrow}^\dagger b_{N_x,\downarrow}^\dagger\right)^T$.


\begin{thebibliography}{105}%
		\makeatletter
		\providecommand \@ifxundefined [1]{%
			\@ifx{#1\undefined}
		}%
		\providecommand \@ifnum [1]{%
			\ifnum #1\expandafter \@firstoftwo
			\else \expandafter \@secondoftwo
			\fi
		}%
		\providecommand \@ifx [1]{%
			\ifx #1\expandafter \@firstoftwo
			\else \expandafter \@secondoftwo
			\fi
		}%
		\providecommand \natexlab [1]{#1}%
		\providecommand \enquote  [1]{``#1''}%
		\providecommand \bibnamefont  [1]{#1}%
		\providecommand \bibfnamefont [1]{#1}%
		\providecommand \citenamefont [1]{#1}%
		\providecommand \href@noop [0]{\@secondoftwo}%
		\providecommand \href [0]{\begingroup \@sanitize@url \@href}%
		\providecommand \@href[1]{\@@startlink{#1}\@@href}%
		\providecommand \@@href[1]{\endgroup#1\@@endlink}%
		\providecommand \@sanitize@url [0]{\catcode `\\12\catcode `\$12\catcode
			`\&12\catcode `\#12\catcode `\^12\catcode `\_12\catcode `\%12\relax}%
		\providecommand \@@startlink[1]{}%
		\providecommand \@@endlink[0]{}%
		\providecommand \url  [0]{\begingroup\@sanitize@url \@url }%
		\providecommand \@url [1]{\endgroup\@href {#1}{\urlprefix }}%
		\providecommand \urlprefix  [0]{URL }%
		\providecommand \Eprint [0]{\href }%
		\providecommand \doibase [0]{https://doi.org/}%
		\providecommand \selectlanguage [0]{\@gobble}%
		\providecommand \bibinfo  [0]{\@secondoftwo}%
		\providecommand \bibfield  [0]{\@secondoftwo}%
		\providecommand \translation [1]{[#1]}%
		\providecommand \BibitemOpen [0]{}%
		\providecommand \bibitemStop [0]{}%
		\providecommand \bibitemNoStop [0]{.\EOS\space}%
		\providecommand \EOS [0]{\spacefactor3000\relax}%
		\providecommand \BibitemShut  [1]{\csname bibitem#1\endcsname}%
		\let\auto@bib@innerbib\@empty
		\bibitem [{\citenamefont {Alicea}(2012)}]{alicea2012new}%
		\BibitemOpen
		\bibfield  {author} {\bibinfo {author} {\bibfnamefont {J.}~\bibnamefont
				{Alicea}},\ }\bibfield  {title} {\bibinfo {title} {New directions in the
				pursuit of {M}ajorana fermions in solid state systems},\ }\href
		{https://doi.org/10.1088/0034-4885/75/7/076501} {\bibfield  {journal}
			{\bibinfo  {journal} {Reports on Progress in Physics}\ }\textbf {\bibinfo
				{volume} {75}},\ \bibinfo {pages} {076501} (\bibinfo {year}
			{2012})}\BibitemShut {NoStop}%
		\bibitem [{\citenamefont {Beenakker}(2013)}]{Beenakker_search_2013}%
		\BibitemOpen
		\bibfield  {author} {\bibinfo {author} {\bibfnamefont {C.}~\bibnamefont
				{Beenakker}},\ }\bibfield  {title} {\bibinfo {title} {Search for {M}ajorana
				{F}ermions in {S}uperconductors},\ }\href
		{https://doi.org/https://doi.org/10.1146/annurev-conmatphys-030212-184337}
		{\bibfield  {journal} {\bibinfo  {journal} {Annual Review of Condensed Matter
					Physics}\ }\textbf {\bibinfo {volume} {4}},\ \bibinfo {pages} {113} (\bibinfo
			{year} {2013})}\BibitemShut {NoStop}%
		\bibitem [{\citenamefont {Stanescu}\ and\ \citenamefont
			{Tewari}(2013)}]{Stanescu_2013}%
		\BibitemOpen
		\bibfield  {author} {\bibinfo {author} {\bibfnamefont {T.~D.}\ \bibnamefont
				{Stanescu}}\ and\ \bibinfo {author} {\bibfnamefont {S.}~\bibnamefont
				{Tewari}},\ }\bibfield  {title} {\bibinfo {title} {{M}ajorana fermions in
				semiconductor nanowires: fundamentals, modeling, and experiment},\ }\href
		{https://doi.org/10.1088/0953-8984/25/23/233201} {\bibfield  {journal}
			{\bibinfo  {journal} {Journal of Physics: Condensed Matter}\ }\textbf
			{\bibinfo {volume} {25}},\ \bibinfo {pages} {233201} (\bibinfo {year}
			{2013})}\BibitemShut {NoStop}%
		\bibitem [{\citenamefont {Aguado}(2017)}]{aguado_majorana_2017}%
		\BibitemOpen
		\bibfield  {author} {\bibinfo {author} {\bibfnamefont {R.}~\bibnamefont
				{Aguado}},\ }\bibfield  {title} {\bibinfo {title} {{M}ajorana quasiparticles
				in condensed matter},\ }\bibfield  {journal} {\bibinfo  {journal} {La Rivista
				del Nuovo Cimento}\ }\textbf {\bibinfo {volume} {40}},\ \href
		{https://doi.org/10.1393/ncr/i2017-10141-9} {10.1393/ncr/i2017-10141-9}
		(\bibinfo {year} {2017}),\ \bibinfo {note} {arXiv:1711.00011
			[cond-mat.supr-con]}\BibitemShut {NoStop}%
		\bibitem [{\citenamefont {Lutchyn}\ \emph {et~al.}(2018)\citenamefont
			{Lutchyn}, \citenamefont {Bakkers}, \citenamefont {Kouwenhoven},
			\citenamefont {Krogstrup}, \citenamefont {Marcus},\ and\ \citenamefont
			{Oreg}}]{lutchyn2018realizing}%
		\BibitemOpen
		\bibfield  {author} {\bibinfo {author} {\bibfnamefont {R.~M.}\ \bibnamefont
				{Lutchyn}}, \bibinfo {author} {\bibfnamefont {E.~P. A.~M.}\ \bibnamefont
				{Bakkers}}, \bibinfo {author} {\bibfnamefont {L.~P.}\ \bibnamefont
				{Kouwenhoven}}, \bibinfo {author} {\bibfnamefont {P.}~\bibnamefont
				{Krogstrup}}, \bibinfo {author} {\bibfnamefont {C.~M.}\ \bibnamefont
				{Marcus}},\ and\ \bibinfo {author} {\bibfnamefont {Y.}~\bibnamefont {Oreg}},\
		}\bibfield  {title} {\bibinfo {title} {Majorana zero modes in
				superconductor--semiconductor heterostructures},\ }\href
		{https://doi.org/10.1038/s41578-018-0003-1} {\bibfield  {journal} {\bibinfo
				{journal} {Nature Reviews Materials}\ }\textbf {\bibinfo {volume} {3}},\
			\bibinfo {pages} {52} (\bibinfo {year} {2018})}\BibitemShut {NoStop}%
		\bibitem [{\citenamefont {Prada}\ \emph {et~al.}(2020)\citenamefont {Prada},
			\citenamefont {San-Jose}, \citenamefont {de~Moor}, \citenamefont {Geresdi},
			\citenamefont {Lee}, \citenamefont {Klinovaja}, \citenamefont {Loss},
			\citenamefont {Nyg{\aa}rd}, \citenamefont {Aguado},\ and\ \citenamefont
			{Kouwenhoven}}]{prada2020from}%
		\BibitemOpen
		\bibfield  {author} {\bibinfo {author} {\bibfnamefont {E.}~\bibnamefont
				{Prada}}, \bibinfo {author} {\bibfnamefont {P.}~\bibnamefont {San-Jose}},
			\bibinfo {author} {\bibfnamefont {M.~W.~A.}\ \bibnamefont {de~Moor}},
			\bibinfo {author} {\bibfnamefont {A.}~\bibnamefont {Geresdi}}, \bibinfo
			{author} {\bibfnamefont {E.~J.~H.}\ \bibnamefont {Lee}}, \bibinfo {author}
			{\bibfnamefont {J.}~\bibnamefont {Klinovaja}}, \bibinfo {author}
			{\bibfnamefont {D.}~\bibnamefont {Loss}}, \bibinfo {author} {\bibfnamefont
				{J.}~\bibnamefont {Nyg{\aa}rd}}, \bibinfo {author} {\bibfnamefont
				{R.}~\bibnamefont {Aguado}},\ and\ \bibinfo {author} {\bibfnamefont {L.~P.}\
				\bibnamefont {Kouwenhoven}},\ }\bibfield  {title} {\bibinfo {title} {From
				andreev to majorana bound states in hybrid superconductor--semiconductor
				nanowires},\ }\href {https://doi.org/10.1038/s42254-020-0228-y} {\bibfield
			{journal} {\bibinfo  {journal} {Nature Reviews Physics}\ }\textbf {\bibinfo
				{volume} {2}},\ \bibinfo {pages} {575} (\bibinfo {year} {2020})}\BibitemShut
		{NoStop}%
		\bibitem [{\citenamefont {Laubscher}\ and\ \citenamefont
			{Klinovaja}(2021)}]{laubscher2021}%
		\BibitemOpen
		\bibfield  {author} {\bibinfo {author} {\bibfnamefont {K.}~\bibnamefont
				{Laubscher}}\ and\ \bibinfo {author} {\bibfnamefont {J.}~\bibnamefont
				{Klinovaja}},\ }\bibfield  {title} {\bibinfo {title} {Majorana bound states
				in semiconducting nanostructures},\ }\href
		{https://doi.org/10.1063/5.0055997} {\bibfield  {journal} {\bibinfo
				{journal} {Journal of Applied Physics}\ }\textbf {\bibinfo {volume} {130}},\
			\bibinfo {pages} {081101} (\bibinfo {year} {2021})}\BibitemShut {NoStop}%
		\bibitem [{\citenamefont {Flensberg}\ \emph {et~al.}(2021)\citenamefont
			{Flensberg}, \citenamefont {von Oppen},\ and\ \citenamefont
			{Stern}}]{Flensberg2021}%
		\BibitemOpen
		\bibfield  {author} {\bibinfo {author} {\bibfnamefont {K.}~\bibnamefont
				{Flensberg}}, \bibinfo {author} {\bibfnamefont {F.}~\bibnamefont {von
					Oppen}},\ and\ \bibinfo {author} {\bibfnamefont {A.}~\bibnamefont {Stern}},\
		}\bibfield  {title} {\bibinfo {title} {Engineered platforms for topological
				superconductivity and {M}ajorana zero modes},\ }\href
		{https://doi.org/10.1038/s41578-021-00336-6} {\bibfield  {journal} {\bibinfo
				{journal} {Nature Reviews Materials}\ }\textbf {\bibinfo {volume} {6}},\
			\bibinfo {pages} {944} (\bibinfo {year} {2021})}\BibitemShut {NoStop}%
		\bibitem [{\citenamefont {Kitaev}(2001)}]{kitaev2001unpaired}%
		\BibitemOpen
		\bibfield  {author} {\bibinfo {author} {\bibfnamefont {A.~Y.}\ \bibnamefont
				{Kitaev}},\ }\bibfield  {title} {\bibinfo {title} {Unpaired {M}ajorana
				fermions in quantum wires},\ }\href
		{https://doi.org/10.1070/1063-7869/44/10S/S29} {\bibfield  {journal}
			{\bibinfo  {journal} {Physics-Uspekhi}\ }\textbf {\bibinfo {volume} {44}},\
			\bibinfo {pages} {131} (\bibinfo {year} {2001})}\BibitemShut {NoStop}%
		\bibitem [{\citenamefont {Dvir}\ \emph {et~al.}(2023)\citenamefont {Dvir},
			\citenamefont {Wang}, \citenamefont {van Loo}, \citenamefont {Liu},
			\citenamefont {Mazur}, \citenamefont {Bordin}, \citenamefont {ten Haaf},
			\citenamefont {Wang}, \citenamefont {van Driel}, \citenamefont {Zatelli},
			\citenamefont {Li}, \citenamefont {Malinowski}, \citenamefont {Gazibegovic},
			\citenamefont {Badawy}, \citenamefont {Bakkers}, \citenamefont {Wimmer},\
			and\ \citenamefont {Kouwenhoven}}]{Dvir2023exp2site}%
		\BibitemOpen
		\bibfield  {author} {\bibinfo {author} {\bibfnamefont {T.}~\bibnamefont
				{Dvir}}, \bibinfo {author} {\bibfnamefont {G.}~\bibnamefont {Wang}}, \bibinfo
			{author} {\bibfnamefont {N.}~\bibnamefont {van Loo}}, \bibinfo {author}
			{\bibfnamefont {C.-X.}\ \bibnamefont {Liu}}, \bibinfo {author} {\bibfnamefont
				{G.~P.}\ \bibnamefont {Mazur}}, \bibinfo {author} {\bibfnamefont
				{A.}~\bibnamefont {Bordin}}, \bibinfo {author} {\bibfnamefont {S.~L.~D.}\
				\bibnamefont {ten Haaf}}, \bibinfo {author} {\bibfnamefont {J.-Y.}\
				\bibnamefont {Wang}}, \bibinfo {author} {\bibfnamefont {D.}~\bibnamefont {van
					Driel}}, \bibinfo {author} {\bibfnamefont {F.}~\bibnamefont {Zatelli}},
			\bibinfo {author} {\bibfnamefont {X.}~\bibnamefont {Li}}, \bibinfo {author}
			{\bibfnamefont {F.~K.}\ \bibnamefont {Malinowski}}, \bibinfo {author}
			{\bibfnamefont {S.}~\bibnamefont {Gazibegovic}}, \bibinfo {author}
			{\bibfnamefont {G.}~\bibnamefont {Badawy}}, \bibinfo {author} {\bibfnamefont
				{E.~P. A.~M.}\ \bibnamefont {Bakkers}}, \bibinfo {author} {\bibfnamefont
				{M.}~\bibnamefont {Wimmer}},\ and\ \bibinfo {author} {\bibfnamefont {L.~P.}\
				\bibnamefont {Kouwenhoven}},\ }\bibfield  {title} {\bibinfo {title}
			{Realization of a minimal {K}itaev chain in coupled quantum dots},\ }\href
		{https://doi.org/10.1038/s41586-022-05585-1} {\bibfield  {journal} {\bibinfo
				{journal} {Nature}\ }\textbf {\bibinfo {volume} {614}},\ \bibinfo {pages}
			{445} (\bibinfo {year} {2023})}\BibitemShut {NoStop}%
		\bibitem [{\citenamefont {ten Haaf}\ \emph {et~al.}(2024)\citenamefont {ten
				Haaf}, \citenamefont {Wang}, \citenamefont {Bozkurt}, \citenamefont {Liu},
			\citenamefont {Kulesh}, \citenamefont {Kim}, \citenamefont {Xiao},
			\citenamefont {Thomas}, \citenamefont {Manfra}, \citenamefont {Dvir},
			\citenamefont {Wimmer},\ and\ \citenamefont {Goswami}}]{tenHaaf2024exp2site}%
		\BibitemOpen
		\bibfield  {author} {\bibinfo {author} {\bibfnamefont {S.~L.~D.}\
				\bibnamefont {ten Haaf}}, \bibinfo {author} {\bibfnamefont {Q.}~\bibnamefont
				{Wang}}, \bibinfo {author} {\bibfnamefont {A.~M.}\ \bibnamefont {Bozkurt}},
			\bibinfo {author} {\bibfnamefont {C.-X.}\ \bibnamefont {Liu}}, \bibinfo
			{author} {\bibfnamefont {I.}~\bibnamefont {Kulesh}}, \bibinfo {author}
			{\bibfnamefont {P.}~\bibnamefont {Kim}}, \bibinfo {author} {\bibfnamefont
				{D.}~\bibnamefont {Xiao}}, \bibinfo {author} {\bibfnamefont {C.}~\bibnamefont
				{Thomas}}, \bibinfo {author} {\bibfnamefont {M.~J.}\ \bibnamefont {Manfra}},
			\bibinfo {author} {\bibfnamefont {T.}~\bibnamefont {Dvir}}, \bibinfo {author}
			{\bibfnamefont {M.}~\bibnamefont {Wimmer}},\ and\ \bibinfo {author}
			{\bibfnamefont {S.}~\bibnamefont {Goswami}},\ }\bibfield  {title} {\bibinfo
			{title} {A two-site {K}itaev chain in a two-dimensional electron gas},\
		}\href {https://doi.org/10.1038/s41586-024-07434-9} {\bibfield  {journal}
			{\bibinfo  {journal} {Nature}\ }\textbf {\bibinfo {volume} {630}},\ \bibinfo
			{pages} {329} (\bibinfo {year} {2024})}\BibitemShut {NoStop}%
		\bibitem [{\citenamefont {Zatelli}\ \emph {et~al.}(2024)\citenamefont
			{Zatelli}, \citenamefont {van Driel}, \citenamefont {Xu}, \citenamefont
			{Wang}, \citenamefont {Liu}, \citenamefont {Bordin}, \citenamefont {Roovers},
			\citenamefont {Mazur}, \citenamefont {van Loo}, \citenamefont {Wolff},
			\citenamefont {Bozkurt}, \citenamefont {Badawy}, \citenamefont {Gazibegovic},
			\citenamefont {Bakkers}, \citenamefont {Wimmer}, \citenamefont
			{Kouwenhoven},\ and\ \citenamefont {Dvir}}]{Zatelli2024exp2site}%
		\BibitemOpen
		\bibfield  {author} {\bibinfo {author} {\bibfnamefont {F.}~\bibnamefont
				{Zatelli}}, \bibinfo {author} {\bibfnamefont {D.}~\bibnamefont {van Driel}},
			\bibinfo {author} {\bibfnamefont {D.}~\bibnamefont {Xu}}, \bibinfo {author}
			{\bibfnamefont {G.}~\bibnamefont {Wang}}, \bibinfo {author} {\bibfnamefont
				{C.-X.}\ \bibnamefont {Liu}}, \bibinfo {author} {\bibfnamefont
				{A.}~\bibnamefont {Bordin}}, \bibinfo {author} {\bibfnamefont
				{B.}~\bibnamefont {Roovers}}, \bibinfo {author} {\bibfnamefont {G.~P.}\
				\bibnamefont {Mazur}}, \bibinfo {author} {\bibfnamefont {N.}~\bibnamefont
				{van Loo}}, \bibinfo {author} {\bibfnamefont {J.~C.}\ \bibnamefont {Wolff}},
			\bibinfo {author} {\bibfnamefont {A.~M.}\ \bibnamefont {Bozkurt}}, \bibinfo
			{author} {\bibfnamefont {G.}~\bibnamefont {Badawy}}, \bibinfo {author}
			{\bibfnamefont {S.}~\bibnamefont {Gazibegovic}}, \bibinfo {author}
			{\bibfnamefont {E.~P. A.~M.}\ \bibnamefont {Bakkers}}, \bibinfo {author}
			{\bibfnamefont {M.}~\bibnamefont {Wimmer}}, \bibinfo {author} {\bibfnamefont
				{L.~P.}\ \bibnamefont {Kouwenhoven}},\ and\ \bibinfo {author} {\bibfnamefont
				{T.}~\bibnamefont {Dvir}},\ }\bibfield  {title} {\bibinfo {title} {Robust
				poor man's {M}ajorana zero modes using {Y}u-{S}hiba-{R}usinov states},\
		}\href {https://doi.org/10.1038/s41467-024-52066-2} {\bibfield  {journal}
			{\bibinfo  {journal} {Nature Communications}\ }\textbf {\bibinfo {volume}
				{15}},\ \bibinfo {pages} {7933} (\bibinfo {year} {2024})}\BibitemShut
		{NoStop}%
		\bibitem [{\citenamefont {van Loo}\ \emph {et~al.}(2026)\citenamefont {van
				Loo}, \citenamefont {Zatelli}, \citenamefont {Steffensen}, \citenamefont
			{Roovers}, \citenamefont {Wang}, \citenamefont {Van~Caekenberghe},
			\citenamefont {Bordin}, \citenamefont {van Driel}, \citenamefont {Zhang},
			\citenamefont {Huisman}, \citenamefont {Badawy}, \citenamefont {Bakkers},
			\citenamefont {Mazur}, \citenamefont {Aguado},\ and\ \citenamefont
			{Kouwenhoven}}]{vanLoo2026single}%
		\BibitemOpen
		\bibfield  {author} {\bibinfo {author} {\bibfnamefont {N.}~\bibnamefont {van
					Loo}}, \bibinfo {author} {\bibfnamefont {F.}~\bibnamefont {Zatelli}},
			\bibinfo {author} {\bibfnamefont {G.~O.}\ \bibnamefont {Steffensen}},
			\bibinfo {author} {\bibfnamefont {B.}~\bibnamefont {Roovers}}, \bibinfo
			{author} {\bibfnamefont {G.}~\bibnamefont {Wang}}, \bibinfo {author}
			{\bibfnamefont {T.}~\bibnamefont {Van~Caekenberghe}}, \bibinfo {author}
			{\bibfnamefont {A.}~\bibnamefont {Bordin}}, \bibinfo {author} {\bibfnamefont
				{D.}~\bibnamefont {van Driel}}, \bibinfo {author} {\bibfnamefont
				{Y.}~\bibnamefont {Zhang}}, \bibinfo {author} {\bibfnamefont {W.~D.}\
				\bibnamefont {Huisman}}, \bibinfo {author} {\bibfnamefont {G.}~\bibnamefont
				{Badawy}}, \bibinfo {author} {\bibfnamefont {E.~P. A.~M.}\ \bibnamefont
				{Bakkers}}, \bibinfo {author} {\bibfnamefont {G.~P.}\ \bibnamefont {Mazur}},
			\bibinfo {author} {\bibfnamefont {R.}~\bibnamefont {Aguado}},\ and\ \bibinfo
			{author} {\bibfnamefont {L.~P.}\ \bibnamefont {Kouwenhoven}},\ }\bibfield
		{title} {\bibinfo {title} {Single-shot parity readout of a minimal {K}itaev
				chain},\ }\href {https://doi.org/10.1038/s41586-025-09927-7} {\bibfield
			{journal} {\bibinfo  {journal} {Nature}\ }\textbf {\bibinfo {volume} {650}},\
			\bibinfo {pages} {334} (\bibinfo {year} {2026})}\BibitemShut {NoStop}%
		\bibitem [{\citenamefont {Bordin}\ \emph {et~al.}(2025)\citenamefont {Bordin},
			\citenamefont {Liu}, \citenamefont {Dvir}, \citenamefont {Zatelli},
			\citenamefont {ten Haaf}, \citenamefont {van Driel}, \citenamefont {Wang},
			\citenamefont {van Loo}, \citenamefont {Zhang}, \citenamefont {Wolff},
			\citenamefont {Van~Caekenberghe}, \citenamefont {Badawy}, \citenamefont
			{Gazibegovic}, \citenamefont {Bakkers}, \citenamefont {Wimmer}, \citenamefont
			{Kouwenhoven},\ and\ \citenamefont {Mazur}}]{Bordin2025exp3site}%
		\BibitemOpen
		\bibfield  {author} {\bibinfo {author} {\bibfnamefont {A.}~\bibnamefont
				{Bordin}}, \bibinfo {author} {\bibfnamefont {C.-X.}\ \bibnamefont {Liu}},
			\bibinfo {author} {\bibfnamefont {T.}~\bibnamefont {Dvir}}, \bibinfo {author}
			{\bibfnamefont {F.}~\bibnamefont {Zatelli}}, \bibinfo {author} {\bibfnamefont
				{S.~L.~D.}\ \bibnamefont {ten Haaf}}, \bibinfo {author} {\bibfnamefont
				{D.}~\bibnamefont {van Driel}}, \bibinfo {author} {\bibfnamefont
				{G.}~\bibnamefont {Wang}}, \bibinfo {author} {\bibfnamefont {N.}~\bibnamefont
				{van Loo}}, \bibinfo {author} {\bibfnamefont {Y.}~\bibnamefont {Zhang}},
			\bibinfo {author} {\bibfnamefont {J.~C.}\ \bibnamefont {Wolff}}, \bibinfo
			{author} {\bibfnamefont {T.}~\bibnamefont {Van~Caekenberghe}}, \bibinfo
			{author} {\bibfnamefont {G.}~\bibnamefont {Badawy}}, \bibinfo {author}
			{\bibfnamefont {S.}~\bibnamefont {Gazibegovic}}, \bibinfo {author}
			{\bibfnamefont {E.~P. A.~M.}\ \bibnamefont {Bakkers}}, \bibinfo {author}
			{\bibfnamefont {M.}~\bibnamefont {Wimmer}}, \bibinfo {author} {\bibfnamefont
				{L.~P.}\ \bibnamefont {Kouwenhoven}},\ and\ \bibinfo {author} {\bibfnamefont
				{G.~P.}\ \bibnamefont {Mazur}},\ }\bibfield  {title} {\bibinfo {title}
			{Enhanced {M}ajorana stability in a three-site {K}itaev chain},\ }\bibfield
		{journal} {\bibinfo  {journal} {Nature Nanotechnology}\ }\href
		{https://doi.org/10.1038/s41565-025-01894-4} {10.1038/s41565-025-01894-4}
		(\bibinfo {year} {2025})\BibitemShut {NoStop}%
		\bibitem [{\citenamefont {Ten~Haaf}\ \emph {et~al.}(2025)\citenamefont
			{Ten~Haaf}, \citenamefont {Zhang}, \citenamefont {Wang}, \citenamefont
			{Bordin}, \citenamefont {Liu}, \citenamefont {Kulesh}, \citenamefont
			{Sietses}, \citenamefont {Prosko}, \citenamefont {Xiao}, \citenamefont
			{Thomas} \emph {et~al.}}]{ten2025exp3site}%
		\BibitemOpen
		\bibfield  {author} {\bibinfo {author} {\bibfnamefont {S.~L.}\ \bibnamefont
				{Ten~Haaf}}, \bibinfo {author} {\bibfnamefont {Y.}~\bibnamefont {Zhang}},
			\bibinfo {author} {\bibfnamefont {Q.}~\bibnamefont {Wang}}, \bibinfo {author}
			{\bibfnamefont {A.}~\bibnamefont {Bordin}}, \bibinfo {author} {\bibfnamefont
				{C.-X.}\ \bibnamefont {Liu}}, \bibinfo {author} {\bibfnamefont
				{I.}~\bibnamefont {Kulesh}}, \bibinfo {author} {\bibfnamefont {V.~P.}\
				\bibnamefont {Sietses}}, \bibinfo {author} {\bibfnamefont {C.~G.}\
				\bibnamefont {Prosko}}, \bibinfo {author} {\bibfnamefont {D.}~\bibnamefont
				{Xiao}}, \bibinfo {author} {\bibfnamefont {C.}~\bibnamefont {Thomas}}, \emph
			{et~al.},\ }\bibfield  {title} {\bibinfo {title} {Observation of edge and
				bulk states in a three-site {K}itaev chain},\ }\href
		{https://doi.org/10.1038/s41586-025-08892-5} {\bibfield  {journal} {\bibinfo
				{journal} {Nature}\ ,\ \bibinfo {pages} {1}} (\bibinfo {year}
			{2025})}\BibitemShut {NoStop}%
		\bibitem [{\citenamefont {Leijnse}\ and\ \citenamefont
			{Flensberg}(2012)}]{Leijnse2012Parityqubits}%
		\BibitemOpen
		\bibfield  {author} {\bibinfo {author} {\bibfnamefont {M.}~\bibnamefont
				{Leijnse}}\ and\ \bibinfo {author} {\bibfnamefont {K.}~\bibnamefont
				{Flensberg}},\ }\bibfield  {title} {\bibinfo {title} {Parity qubits and poor
				man's {M}ajorana bound states in double quantum dots},\ }\href
		{https://doi.org/10.1103/PhysRevB.86.134528} {\bibfield  {journal} {\bibinfo
				{journal} {Phys. Rev. B}\ }\textbf {\bibinfo {volume} {86}},\ \bibinfo
			{pages} {134528} (\bibinfo {year} {2012})}\BibitemShut {NoStop}%
		\bibitem [{\citenamefont {Seoane~Souto}\ and\ \citenamefont
			{Aguado}(2024)}]{SeoaneSouto2024}%
		\BibitemOpen
		\bibfield  {author} {\bibinfo {author} {\bibfnamefont {R.}~\bibnamefont
				{Seoane~Souto}}\ and\ \bibinfo {author} {\bibfnamefont {R.}~\bibnamefont
				{Aguado}},\ }\bibinfo {title} {Subgap {S}tates in
			{S}emiconductor-{S}uperconductor {D}evices for {Q}uantum {T}echnologies:
			{A}ndreev {Q}ubits and {M}inimal {M}ajorana {C}hains},\ in\ \href
		{https://doi.org/10.1007/978-3-031-55657-9_3} {\emph {\bibinfo {booktitle}
				{New Trends and Platforms for Quantum Technologies}}},\ \bibinfo {editor}
		{edited by\ \bibinfo {editor} {\bibfnamefont {R.}~\bibnamefont {Aguado}},
			\bibinfo {editor} {\bibfnamefont {R.}~\bibnamefont {Citro}}, \bibinfo
			{editor} {\bibfnamefont {M.}~\bibnamefont {Lewenstein}},\ and\ \bibinfo
			{editor} {\bibfnamefont {M.}~\bibnamefont {Stern}}}\ (\bibinfo  {publisher}
		{Springer Nature Switzerland},\ \bibinfo {address} {Cham},\ \bibinfo {year}
		{2024})\ pp.\ \bibinfo {pages} {133--223}\BibitemShut {NoStop}%
		\bibitem [{\citenamefont {Luethi}\ \emph {et~al.}(2024)\citenamefont {Luethi},
			\citenamefont {Legg}, \citenamefont {Loss},\ and\ \citenamefont
			{Klinovaja}}]{luethi2024from}%
		\BibitemOpen
		\bibfield  {author} {\bibinfo {author} {\bibfnamefont {M.}~\bibnamefont
				{Luethi}}, \bibinfo {author} {\bibfnamefont {H.~F.}\ \bibnamefont {Legg}},
			\bibinfo {author} {\bibfnamefont {D.}~\bibnamefont {Loss}},\ and\ \bibinfo
			{author} {\bibfnamefont {J.}~\bibnamefont {Klinovaja}},\ }\bibfield  {title}
		{\bibinfo {title} {From perfect to imperfect poor man's {M}ajoranas in
				minimal {K}itaev chains},\ }\href
		{https://doi.org/10.1103/PhysRevB.110.245412} {\bibfield  {journal} {\bibinfo
				{journal} {Phys. Rev. B}\ }\textbf {\bibinfo {volume} {110}},\ \bibinfo
			{pages} {245412} (\bibinfo {year} {2024})}\BibitemShut {NoStop}%
		\bibitem [{\citenamefont {Luethi}\ \emph {et~al.}(2025)\citenamefont {Luethi},
			\citenamefont {Legg}, \citenamefont {Loss},\ and\ \citenamefont
			{Klinovaja}}]{luethi2025fate}%
		\BibitemOpen
		\bibfield  {author} {\bibinfo {author} {\bibfnamefont {M.}~\bibnamefont
				{Luethi}}, \bibinfo {author} {\bibfnamefont {H.~F.}\ \bibnamefont {Legg}},
			\bibinfo {author} {\bibfnamefont {D.}~\bibnamefont {Loss}},\ and\ \bibinfo
			{author} {\bibfnamefont {J.}~\bibnamefont {Klinovaja}},\ }\bibfield  {title}
		{\bibinfo {title} {Fate of poor man's {M}ajoranas in the long {K}itaev chain
				limit},\ }\href {https://doi.org/10.1103/PhysRevB.111.115419} {\bibfield
			{journal} {\bibinfo  {journal} {Phys. Rev. B}\ }\textbf {\bibinfo {volume}
				{111}},\ \bibinfo {pages} {115419} (\bibinfo {year} {2025})}\BibitemShut
		{NoStop}%
		\bibitem [{\citenamefont {Dutreix}\ \emph {et~al.}(2014)\citenamefont
			{Dutreix}, \citenamefont {Guigou}, \citenamefont {Chevallier},\ and\
			\citenamefont {Bena}}]{graphene2014}%
		\BibitemOpen
		\bibfield  {author} {\bibinfo {author} {\bibfnamefont {C.}~\bibnamefont
				{Dutreix}}, \bibinfo {author} {\bibfnamefont {M.}~\bibnamefont {Guigou}},
			\bibinfo {author} {\bibfnamefont {D.}~\bibnamefont {Chevallier}},\ and\
			\bibinfo {author} {\bibfnamefont {C.}~\bibnamefont {Bena}},\ }\bibfield
		{title} {\bibinfo {title} {Majorana fermions in honeycomb lattices},\ }\href
		{https://doi.org/10.1140/epjb/e2014-50243-9} {\bibfield  {journal} {\bibinfo
				{journal} {The European Physical Journal B}\ }\textbf {\bibinfo {volume}
				{87}},\ \bibinfo {pages} {296} (\bibinfo {year} {2014})}\BibitemShut
		{NoStop}%
		\bibitem [{\citenamefont {San-Jose}\ \emph {et~al.}(2015)\citenamefont
			{San-Jose}, \citenamefont {Lado}, \citenamefont {Aguado}, \citenamefont
			{Guinea},\ and\ \citenamefont {Fern\'andez-Rossier}}]{graphene2015}%
		\BibitemOpen
		\bibfield  {author} {\bibinfo {author} {\bibfnamefont {P.}~\bibnamefont
				{San-Jose}}, \bibinfo {author} {\bibfnamefont {J.~L.}\ \bibnamefont {Lado}},
			\bibinfo {author} {\bibfnamefont {R.}~\bibnamefont {Aguado}}, \bibinfo
			{author} {\bibfnamefont {F.}~\bibnamefont {Guinea}},\ and\ \bibinfo {author}
			{\bibfnamefont {J.}~\bibnamefont {Fern\'andez-Rossier}},\ }\bibfield  {title}
		{\bibinfo {title} {Majorana {Zero} {Modes} in {G}raphene},\ }\href
		{https://doi.org/10.1103/PhysRevX.5.041042} {\bibfield  {journal} {\bibinfo
				{journal} {Phys. Rev. X}\ }\textbf {\bibinfo {volume} {5}},\ \bibinfo {pages}
			{041042} (\bibinfo {year} {2015})}\BibitemShut {NoStop}%
		\bibitem [{\citenamefont {Nadj-Perge}\ \emph {et~al.}(2013)\citenamefont
			{Nadj-Perge}, \citenamefont {Drozdov}, \citenamefont {Bernevig},\ and\
			\citenamefont {Yazdani}}]{magneticatom1}%
		\BibitemOpen
		\bibfield  {author} {\bibinfo {author} {\bibfnamefont {S.}~\bibnamefont
				{Nadj-Perge}}, \bibinfo {author} {\bibfnamefont {I.~K.}\ \bibnamefont
				{Drozdov}}, \bibinfo {author} {\bibfnamefont {B.~A.}\ \bibnamefont
				{Bernevig}},\ and\ \bibinfo {author} {\bibfnamefont {A.}~\bibnamefont
				{Yazdani}},\ }\bibfield  {title} {\bibinfo {title} {Proposal for realizing
				{M}ajorana fermions in chains of magnetic atoms on a superconductor},\ }\href
		{https://doi.org/10.1103/PhysRevB.88.020407} {\bibfield  {journal} {\bibinfo
				{journal} {Phys. Rev. B}\ }\textbf {\bibinfo {volume} {88}},\ \bibinfo
			{pages} {020407(R)} (\bibinfo {year} {2013})}\BibitemShut {NoStop}%
		\bibitem [{\citenamefont {Klinovaja}\ \emph {et~al.}(2013)\citenamefont
			{Klinovaja}, \citenamefont {Stano}, \citenamefont {Yazdani},\ and\
			\citenamefont {Loss}}]{magneticatom2}%
		\BibitemOpen
		\bibfield  {author} {\bibinfo {author} {\bibfnamefont {J.}~\bibnamefont
				{Klinovaja}}, \bibinfo {author} {\bibfnamefont {P.}~\bibnamefont {Stano}},
			\bibinfo {author} {\bibfnamefont {A.}~\bibnamefont {Yazdani}},\ and\ \bibinfo
			{author} {\bibfnamefont {D.}~\bibnamefont {Loss}},\ }\bibfield  {title}
		{\bibinfo {title} {Topological {Superconductivity} and {Majorana} {Fermions}
				in {R}{K}{K}{Y} {S}ystems},\ }\href
		{https://doi.org/10.1103/PhysRevLett.111.186805} {\bibfield  {journal}
			{\bibinfo  {journal} {Phys. Rev. Lett.}\ }\textbf {\bibinfo {volume} {111}},\
			\bibinfo {pages} {186805} (\bibinfo {year} {2013})}\BibitemShut {NoStop}%
		\bibitem [{\citenamefont {Braunecker}\ and\ \citenamefont
			{Simon}(2013)}]{magneticatom3}%
		\BibitemOpen
		\bibfield  {author} {\bibinfo {author} {\bibfnamefont {B.}~\bibnamefont
				{Braunecker}}\ and\ \bibinfo {author} {\bibfnamefont {P.}~\bibnamefont
				{Simon}},\ }\bibfield  {title} {\bibinfo {title} {Interplay between
				{C}lassical {M}agnetic {M}oments and {Superconductivity} in {Quantum}
				{One}-{Dimensional} {Conductors}: {Toward} a {Self}-{Sustained} {Topological}
				{Majorana} {Phase}},\ }\href {https://doi.org/10.1103/PhysRevLett.111.147202}
		{\bibfield  {journal} {\bibinfo  {journal} {Phys. Rev. Lett.}\ }\textbf
			{\bibinfo {volume} {111}},\ \bibinfo {pages} {147202} (\bibinfo {year}
			{2013})}\BibitemShut {NoStop}%
		\bibitem [{\citenamefont {Lutchyn}\ \emph {et~al.}(2010)\citenamefont
			{Lutchyn}, \citenamefont {Sau},\ and\ \citenamefont
			{Das~Sarma}}]{lutchyn_majorana_2010}%
		\BibitemOpen
		\bibfield  {author} {\bibinfo {author} {\bibfnamefont {R.~M.}\ \bibnamefont
				{Lutchyn}}, \bibinfo {author} {\bibfnamefont {J.~D.}\ \bibnamefont {Sau}},\
			and\ \bibinfo {author} {\bibfnamefont {S.}~\bibnamefont {Das~Sarma}},\
		}\bibfield  {title} {\bibinfo {title} {{M}ajorana {Fermions} and a
				{Topological} {Phase} {Transition} in {Semiconductor}-{Superconductor}
				{Heterostructures}},\ }\href {https://doi.org/10.1103/PhysRevLett.105.077001}
		{\bibfield  {journal} {\bibinfo  {journal} {Physical Review Letters}\
			}\textbf {\bibinfo {volume} {105}},\ \bibinfo {pages} {077001} (\bibinfo
			{year} {2010})}\BibitemShut {NoStop}%
		\bibitem [{\citenamefont {Oreg}\ \emph {et~al.}(2010)\citenamefont {Oreg},
			\citenamefont {Refael},\ and\ \citenamefont {von Oppen}}]{Oreg2010helical}%
		\BibitemOpen
		\bibfield  {author} {\bibinfo {author} {\bibfnamefont {Y.}~\bibnamefont
				{Oreg}}, \bibinfo {author} {\bibfnamefont {G.}~\bibnamefont {Refael}},\ and\
			\bibinfo {author} {\bibfnamefont {F.}~\bibnamefont {von Oppen}},\ }\bibfield
		{title} {\bibinfo {title} {Helical {L}iquids and {M}ajorana {B}ound {S}tates
				in {Q}uantum {W}ires},\ }\href
		{https://doi.org/10.1103/PhysRevLett.105.177002} {\bibfield  {journal}
			{\bibinfo  {journal} {Phys. Rev. Lett.}\ }\textbf {\bibinfo {volume} {105}},\
			\bibinfo {pages} {177002} (\bibinfo {year} {2010})}\BibitemShut {NoStop}%
		\bibitem [{\citenamefont {Mourik}\ \emph {et~al.}(2012)\citenamefont {Mourik},
			\citenamefont {Zuo}, \citenamefont {Frolov}, \citenamefont {Plissard},
			\citenamefont {Bakkers},\ and\ \citenamefont {Kouwenhoven}}]{Mourik_2012}%
		\BibitemOpen
		\bibfield  {author} {\bibinfo {author} {\bibfnamefont {V.}~\bibnamefont
				{Mourik}}, \bibinfo {author} {\bibfnamefont {K.}~\bibnamefont {Zuo}},
			\bibinfo {author} {\bibfnamefont {S.~M.}\ \bibnamefont {Frolov}}, \bibinfo
			{author} {\bibfnamefont {S.~R.}\ \bibnamefont {Plissard}}, \bibinfo {author}
			{\bibfnamefont {E.~P. A.~M.}\ \bibnamefont {Bakkers}},\ and\ \bibinfo
			{author} {\bibfnamefont {L.~P.}\ \bibnamefont {Kouwenhoven}},\ }\bibfield
		{title} {\bibinfo {title} {Signatures of {M}ajorana {Fermions} in {Hybrid}
				{Superconductor}-{Semiconductor} {Nanowire} {Devices}},\ }\href
		{https://doi.org/10.1126/science.1222360} {\bibfield  {journal} {\bibinfo
				{journal} {Science}\ }\textbf {\bibinfo {volume} {336}},\ \bibinfo {pages}
			{1003–1007} (\bibinfo {year} {2012})}\BibitemShut {NoStop}%
		\bibitem [{\citenamefont {Das}\ \emph {et~al.}(2012)\citenamefont {Das},
			\citenamefont {Ronen}, \citenamefont {Most}, \citenamefont {Oreg},
			\citenamefont {Heiblum},\ and\ \citenamefont {Shtrikman}}]{Das_2012}%
		\BibitemOpen
		\bibfield  {author} {\bibinfo {author} {\bibfnamefont {A.}~\bibnamefont
				{Das}}, \bibinfo {author} {\bibfnamefont {Y.}~\bibnamefont {Ronen}}, \bibinfo
			{author} {\bibfnamefont {Y.}~\bibnamefont {Most}}, \bibinfo {author}
			{\bibfnamefont {Y.}~\bibnamefont {Oreg}}, \bibinfo {author} {\bibfnamefont
				{M.}~\bibnamefont {Heiblum}},\ and\ \bibinfo {author} {\bibfnamefont
				{H.}~\bibnamefont {Shtrikman}},\ }\bibfield  {title} {\bibinfo {title}
			{Zero-bias peaks and splitting in an {A}l–{I}n{A}s nanowire topological
				superconductor as a signature of {M}ajorana fermions},\ }\href
		{https://doi.org/10.1038/nphys2479} {\bibfield  {journal} {\bibinfo
				{journal} {Nature Physics}\ }\textbf {\bibinfo {volume} {8}},\ \bibinfo
			{pages} {887–895} (\bibinfo {year} {2012})}\BibitemShut {NoStop}%
		\bibitem [{\citenamefont {Deng}\ \emph {et~al.}(2012)\citenamefont {Deng},
			\citenamefont {Yu}, \citenamefont {Huang}, \citenamefont {Larsson},
			\citenamefont {Caroff},\ and\ \citenamefont {Xu}}]{Deng_2012}%
		\BibitemOpen
		\bibfield  {author} {\bibinfo {author} {\bibfnamefont {M.~T.}\ \bibnamefont
				{Deng}}, \bibinfo {author} {\bibfnamefont {C.~L.}\ \bibnamefont {Yu}},
			\bibinfo {author} {\bibfnamefont {G.~Y.}\ \bibnamefont {Huang}}, \bibinfo
			{author} {\bibfnamefont {M.}~\bibnamefont {Larsson}}, \bibinfo {author}
			{\bibfnamefont {P.}~\bibnamefont {Caroff}},\ and\ \bibinfo {author}
			{\bibfnamefont {H.~Q.}\ \bibnamefont {Xu}},\ }\bibfield  {title} {\bibinfo
			{title} {Anomalous {Zero}-{Bias} {Conductance} {Peak} in a {N}b–{I}n{S}b
				{Nanowire}–{Nb} {Hybrid} {Device}},\ }\href
		{https://doi.org/10.1021/nl303758w} {\bibfield  {journal} {\bibinfo
				{journal} {Nano Letters}\ }\textbf {\bibinfo {volume} {12}},\ \bibinfo
			{pages} {6414–6419} (\bibinfo {year} {2012})}\BibitemShut {NoStop}%
		\bibitem [{\citenamefont {Churchill}\ \emph {et~al.}(2013)\citenamefont
			{Churchill}, \citenamefont {Fatemi}, \citenamefont {Grove-Rasmussen},
			\citenamefont {Deng}, \citenamefont {Caroff}, \citenamefont {Xu},\ and\
			\citenamefont {Marcus}}]{Churchill_2013}%
		\BibitemOpen
		\bibfield  {author} {\bibinfo {author} {\bibfnamefont {H.~O.~H.}\
				\bibnamefont {Churchill}}, \bibinfo {author} {\bibfnamefont {V.}~\bibnamefont
				{Fatemi}}, \bibinfo {author} {\bibfnamefont {K.}~\bibnamefont
				{Grove-Rasmussen}}, \bibinfo {author} {\bibfnamefont {M.~T.}\ \bibnamefont
				{Deng}}, \bibinfo {author} {\bibfnamefont {P.}~\bibnamefont {Caroff}},
			\bibinfo {author} {\bibfnamefont {H.~Q.}\ \bibnamefont {Xu}},\ and\ \bibinfo
			{author} {\bibfnamefont {C.~M.}\ \bibnamefont {Marcus}},\ }\bibfield  {title}
		{\bibinfo {title} {Superconductor-nanowire devices from tunneling to the
				multichannel regime: {Z}ero-bias oscillations and magnetoconductance
				crossover},\ }\bibfield  {journal} {\bibinfo  {journal} {Physical Review B}\
		}\textbf {\bibinfo {volume} {87}},\ \href
		{https://doi.org/10.1103/physrevb.87.241401} {10.1103/physrevb.87.241401}
		(\bibinfo {year} {2013})\BibitemShut {NoStop}%
		\bibitem [{\citenamefont {Finck}\ \emph {et~al.}(2013)\citenamefont {Finck},
			\citenamefont {Van~Harlingen}, \citenamefont {Mohseni}, \citenamefont
			{Jung},\ and\ \citenamefont {Li}}]{Finck_2013}%
		\BibitemOpen
		\bibfield  {author} {\bibinfo {author} {\bibfnamefont {A.~D.~K.}\
				\bibnamefont {Finck}}, \bibinfo {author} {\bibfnamefont {D.~J.}\ \bibnamefont
				{Van~Harlingen}}, \bibinfo {author} {\bibfnamefont {P.~K.}\ \bibnamefont
				{Mohseni}}, \bibinfo {author} {\bibfnamefont {K.}~\bibnamefont {Jung}},\ and\
			\bibinfo {author} {\bibfnamefont {X.}~\bibnamefont {Li}},\ }\bibfield
		{title} {\bibinfo {title} {Anomalous {Modulation} of a {Z}ero-{Bias} {Peak}
				in a {H}ybrid {N}anowire-{S}uperconductor device},\ }\bibfield  {journal}
		{\bibinfo  {journal} {Physical Review Letters}\ }\textbf {\bibinfo {volume}
			{110}},\ \href {https://doi.org/10.1103/physrevlett.110.126406}
		{10.1103/physrevlett.110.126406} (\bibinfo {year} {2013})\BibitemShut
		{NoStop}%
		\bibitem [{\citenamefont {Deng}\ \emph {et~al.}(2016)\citenamefont {Deng},
			\citenamefont {Vaitiekėnas}, \citenamefont {Hansen}, \citenamefont {Danon},
			\citenamefont {Leijnse}, \citenamefont {Flensberg}, \citenamefont {Nygård},
			\citenamefont {Krogstrup},\ and\ \citenamefont {Marcus}}]{deng2016}%
		\BibitemOpen
		\bibfield  {author} {\bibinfo {author} {\bibfnamefont {M.~T.}\ \bibnamefont
				{Deng}}, \bibinfo {author} {\bibfnamefont {S.}~\bibnamefont {Vaitiekėnas}},
			\bibinfo {author} {\bibfnamefont {E.~B.}\ \bibnamefont {Hansen}}, \bibinfo
			{author} {\bibfnamefont {J.}~\bibnamefont {Danon}}, \bibinfo {author}
			{\bibfnamefont {M.}~\bibnamefont {Leijnse}}, \bibinfo {author} {\bibfnamefont
				{K.}~\bibnamefont {Flensberg}}, \bibinfo {author} {\bibfnamefont
				{J.}~\bibnamefont {Nygård}}, \bibinfo {author} {\bibfnamefont
				{P.}~\bibnamefont {Krogstrup}},\ and\ \bibinfo {author} {\bibfnamefont
				{C.~M.}\ \bibnamefont {Marcus}},\ }\bibfield  {title} {\bibinfo {title}
			{Majorana bound state in a coupled quantum-dot hybrid-nanowire system},\
		}\href {https://doi.org/10.1126/science.aaf3961} {\bibfield  {journal}
			{\bibinfo  {journal} {Science}\ }\textbf {\bibinfo {volume} {354}},\ \bibinfo
			{pages} {1557} (\bibinfo {year} {2016})}\BibitemShut {NoStop}%
		\bibitem [{\citenamefont {de~Moor}\ \emph {et~al.}(2018)\citenamefont
			{de~Moor}, \citenamefont {Bommer}, \citenamefont {Xu}, \citenamefont
			{Winkler}, \citenamefont {Antipov}, \citenamefont {Bargerbos}, \citenamefont
			{Wang}, \citenamefont {Loo}, \citenamefont {Op~het Veld}, \citenamefont
			{Gazibegovic}, \citenamefont {Car}, \citenamefont {Logan}, \citenamefont
			{Pendharkar}, \citenamefont {Lee}, \citenamefont {M~Bakkers}, \citenamefont
			{Palmstrøm}, \citenamefont {Lutchyn}, \citenamefont {Kouwenhoven},\ and\
			\citenamefont {Zhang}}]{deMoor2018}%
		\BibitemOpen
		\bibfield  {author} {\bibinfo {author} {\bibfnamefont {M.~W.~A.}\
				\bibnamefont {de~Moor}}, \bibinfo {author} {\bibfnamefont {J.~D.~S.}\
				\bibnamefont {Bommer}}, \bibinfo {author} {\bibfnamefont {D.}~\bibnamefont
				{Xu}}, \bibinfo {author} {\bibfnamefont {G.~W.}\ \bibnamefont {Winkler}},
			\bibinfo {author} {\bibfnamefont {A.~E.}\ \bibnamefont {Antipov}}, \bibinfo
			{author} {\bibfnamefont {A.}~\bibnamefont {Bargerbos}}, \bibinfo {author}
			{\bibfnamefont {G.}~\bibnamefont {Wang}}, \bibinfo {author} {\bibfnamefont
				{N.~v.}\ \bibnamefont {Loo}}, \bibinfo {author} {\bibfnamefont {R.~L.~M.}\
				\bibnamefont {Op~het Veld}}, \bibinfo {author} {\bibfnamefont
				{S.}~\bibnamefont {Gazibegovic}}, \bibinfo {author} {\bibfnamefont
				{D.}~\bibnamefont {Car}}, \bibinfo {author} {\bibfnamefont {J.~A.}\
				\bibnamefont {Logan}}, \bibinfo {author} {\bibfnamefont {M.}~\bibnamefont
				{Pendharkar}}, \bibinfo {author} {\bibfnamefont {J.~S.}\ \bibnamefont {Lee}},
			\bibinfo {author} {\bibfnamefont {E.~P.~A.}\ \bibnamefont {M~Bakkers}},
			\bibinfo {author} {\bibfnamefont {C.~J.}\ \bibnamefont {Palmstrøm}},
			\bibinfo {author} {\bibfnamefont {R.~M.}\ \bibnamefont {Lutchyn}}, \bibinfo
			{author} {\bibfnamefont {L.~P.}\ \bibnamefont {Kouwenhoven}},\ and\ \bibinfo
			{author} {\bibfnamefont {H.}~\bibnamefont {Zhang}},\ }\bibfield  {title}
		{\bibinfo {title} {Electric field tunable superconductor-semiconductor
				coupling in {M}xajorana nanowires},\ }\href
		{https://doi.org/10.1088/1367-2630/aae61d} {\bibfield  {journal} {\bibinfo
				{journal} {New Journal of Physics}\ }\textbf {\bibinfo {volume} {20}},\
			\bibinfo {pages} {103049} (\bibinfo {year} {2018})}\BibitemShut {NoStop}%
		\bibitem [{\citenamefont {Aghaee}\ \emph {et~al.}(2023)\citenamefont {Aghaee},
			\citenamefont {Akkala}, \citenamefont {Alam}, \citenamefont {Ali},
			\citenamefont {Ramirez}, \citenamefont {Andrzejczuk}, \citenamefont
			{Antipov}, \citenamefont {Astafev}, \citenamefont {Bauer}, \citenamefont
			{Becker} \emph {et~al.}}]{MicrosoftQ2023}%
		\BibitemOpen
		\bibfield  {author} {\bibinfo {author} {\bibfnamefont {M.}~\bibnamefont
				{Aghaee}}, \bibinfo {author} {\bibfnamefont {A.}~\bibnamefont {Akkala}},
			\bibinfo {author} {\bibfnamefont {Z.}~\bibnamefont {Alam}}, \bibinfo {author}
			{\bibfnamefont {R.}~\bibnamefont {Ali}}, \bibinfo {author} {\bibfnamefont
				{A.}~\bibnamefont {Ramirez}}, \bibinfo {author} {\bibfnamefont
				{M.}~\bibnamefont {Andrzejczuk}}, \bibinfo {author} {\bibfnamefont
				{A.}~\bibnamefont {Antipov}}, \bibinfo {author} {\bibfnamefont
				{M.}~\bibnamefont {Astafev}}, \bibinfo {author} {\bibfnamefont
				{B.}~\bibnamefont {Bauer}}, \bibinfo {author} {\bibfnamefont
				{J.}~\bibnamefont {Becker}}, \emph {et~al.} (\bibinfo {collaboration}
			{Microsoft Quantum}),\ }\bibfield  {title} {\bibinfo {title} {{I}n{A}s-{A}l
				hybrid devices passing the topological gap protocol},\ }\href
		{https://doi.org/10.1103/PhysRevB.107.245423} {\bibfield  {journal} {\bibinfo
				{journal} {Phys. Rev. B}\ }\textbf {\bibinfo {volume} {107}},\ \bibinfo
			{pages} {245423} (\bibinfo {year} {2023})}\BibitemShut {NoStop}%
		\bibitem [{\citenamefont {Aghaee}\ \emph
			{et~al.}(2025{\natexlab{a}})\citenamefont {Aghaee}, \citenamefont {Akkala},
			\citenamefont {Alam}, \citenamefont {Ali}, \citenamefont {Ramirez},
			\citenamefont {Andrzejczuk}, \citenamefont {Antipov}, \citenamefont
			{Astafev}, \citenamefont {Bauer}, \citenamefont {Becker} \emph
			{et~al.}}]{MicrosoftQ2025}%
		\BibitemOpen
		\bibfield  {author} {\bibinfo {author} {\bibfnamefont {M.}~\bibnamefont
				{Aghaee}}, \bibinfo {author} {\bibfnamefont {A.}~\bibnamefont {Akkala}},
			\bibinfo {author} {\bibfnamefont {Z.}~\bibnamefont {Alam}}, \bibinfo {author}
			{\bibfnamefont {R.}~\bibnamefont {Ali}}, \bibinfo {author} {\bibfnamefont
				{A.}~\bibnamefont {Ramirez}}, \bibinfo {author} {\bibfnamefont
				{M.}~\bibnamefont {Andrzejczuk}}, \bibinfo {author} {\bibfnamefont
				{A.}~\bibnamefont {Antipov}}, \bibinfo {author} {\bibfnamefont
				{M.}~\bibnamefont {Astafev}}, \bibinfo {author} {\bibfnamefont
				{B.}~\bibnamefont {Bauer}}, \bibinfo {author} {\bibfnamefont
				{J.}~\bibnamefont {Becker}}, \emph {et~al.} (\bibinfo {collaboration}
			{Microsoft Azure Quantum}),\ }\bibfield  {title} {\bibinfo {title}
			{Interferometric single-shot parity measurement in {I}n{A}s--{A}l hybrid
				devices},\ }\href {https://doi.org/10.1038/s41586-024-08445-2} {\bibfield
			{journal} {\bibinfo  {journal} {Nature}\ }\textbf {\bibinfo {volume} {638}},\
			\bibinfo {pages} {651} (\bibinfo {year} {2025}{\natexlab{a}})}\BibitemShut
		{NoStop}%
		\bibitem [{\citenamefont {Aghaee}\ \emph
			{et~al.}(2025{\natexlab{b}})\citenamefont {Aghaee}, \citenamefont {Alam},
			\citenamefont {Andersen}, \citenamefont {Andrzejczuk}, \citenamefont
			{Antipov}, \citenamefont {Astafev}, \citenamefont {Avilovas}, \citenamefont
			{Azizimanesh}, \citenamefont {Banek}, \citenamefont {Bauer} \emph
			{et~al.}}]{aghaee2025distinctlifetimesxz}%
		\BibitemOpen
		\bibfield  {author} {\bibinfo {author} {\bibfnamefont {M.}~\bibnamefont
				{Aghaee}}, \bibinfo {author} {\bibfnamefont {Z.}~\bibnamefont {Alam}},
			\bibinfo {author} {\bibfnamefont {R.}~\bibnamefont {Andersen}}, \bibinfo
			{author} {\bibfnamefont {M.}~\bibnamefont {Andrzejczuk}}, \bibinfo {author}
			{\bibfnamefont {A.}~\bibnamefont {Antipov}}, \bibinfo {author} {\bibfnamefont
				{M.}~\bibnamefont {Astafev}}, \bibinfo {author} {\bibfnamefont
				{L.}~\bibnamefont {Avilovas}}, \bibinfo {author} {\bibfnamefont
				{A.}~\bibnamefont {Azizimanesh}}, \bibinfo {author} {\bibfnamefont
				{E.}~\bibnamefont {Banek}}, \bibinfo {author} {\bibfnamefont
				{B.}~\bibnamefont {Bauer}}, \emph {et~al.} (\bibinfo {collaboration}
			{Microsoft Quantum}),\ }\href {https://arxiv.org/abs/2507.08795} {\bibinfo
			{title} {Distinct {L}ifetimes for $x$ and $z$ {L}oop {M}easurements in a
				{M}ajorana {T}etron {D}evice}} (\bibinfo {year} {2025}{\natexlab{b}}),\
		\Eprint {https://arxiv.org/abs/2507.08795} {arXiv:2507.08795
			[cond-mat.mes-hall]} \BibitemShut {NoStop}%
		\bibitem [{\citenamefont {Schrade}\ \emph {et~al.}(2017)\citenamefont
			{Schrade}, \citenamefont {Thakurathi}, \citenamefont {Reeg}, \citenamefont
			{Hoffman}, \citenamefont {Klinovaja},\ and\ \citenamefont
			{Loss}}]{schrade2017low}%
		\BibitemOpen
		\bibfield  {author} {\bibinfo {author} {\bibfnamefont {C.}~\bibnamefont
				{Schrade}}, \bibinfo {author} {\bibfnamefont {M.}~\bibnamefont {Thakurathi}},
			\bibinfo {author} {\bibfnamefont {C.}~\bibnamefont {Reeg}}, \bibinfo {author}
			{\bibfnamefont {S.}~\bibnamefont {Hoffman}}, \bibinfo {author} {\bibfnamefont
				{J.}~\bibnamefont {Klinovaja}},\ and\ \bibinfo {author} {\bibfnamefont
				{D.}~\bibnamefont {Loss}},\ }\bibfield  {title} {\bibinfo {title} {Low-field
				topological threshold in {M}ajorana double nanowires},\ }\href
		{https://doi.org/10.1103/PhysRevB.96.035306} {\bibfield  {journal} {\bibinfo
				{journal} {Phys. Rev. B}\ }\textbf {\bibinfo {volume} {96}},\ \bibinfo
			{pages} {035306} (\bibinfo {year} {2017})}\BibitemShut {NoStop}%
		\bibitem [{\citenamefont {Romito}\ \emph {et~al.}(2012)\citenamefont {Romito},
			\citenamefont {Alicea}, \citenamefont {Refael},\ and\ \citenamefont {von
				Oppen}}]{romito2012manipulating}%
		\BibitemOpen
		\bibfield  {author} {\bibinfo {author} {\bibfnamefont {A.}~\bibnamefont
				{Romito}}, \bibinfo {author} {\bibfnamefont {J.}~\bibnamefont {Alicea}},
			\bibinfo {author} {\bibfnamefont {G.}~\bibnamefont {Refael}},\ and\ \bibinfo
			{author} {\bibfnamefont {F.}~\bibnamefont {von Oppen}},\ }\bibfield  {title}
		{\bibinfo {title} {Manipulating {M}ajorana fermions using supercurrents},\
		}\href {https://doi.org/10.1103/PhysRevB.85.020502} {\bibfield  {journal}
			{\bibinfo  {journal} {Phys. Rev. B}\ }\textbf {\bibinfo {volume} {85}},\
			\bibinfo {pages} {020502} (\bibinfo {year} {2012})}\BibitemShut {NoStop}%
		\bibitem [{\citenamefont {Dmytruk}\ \emph {et~al.}(2019)\citenamefont
			{Dmytruk}, \citenamefont {Thakurathi}, \citenamefont {Loss},\ and\
			\citenamefont {Klinovaja}}]{dmytruk2019majorana}%
		\BibitemOpen
		\bibfield  {author} {\bibinfo {author} {\bibfnamefont {O.}~\bibnamefont
				{Dmytruk}}, \bibinfo {author} {\bibfnamefont {M.}~\bibnamefont {Thakurathi}},
			\bibinfo {author} {\bibfnamefont {D.}~\bibnamefont {Loss}},\ and\ \bibinfo
			{author} {\bibfnamefont {J.}~\bibnamefont {Klinovaja}},\ }\bibfield  {title}
		{\bibinfo {title} {{M}ajorana bound states in double nanowires with reduced
				{Z}eeman thresholds due to supercurrents},\ }\href
		{https://doi.org/10.1103/PhysRevB.99.245416} {\bibfield  {journal} {\bibinfo
				{journal} {Phys. Rev. B}\ }\textbf {\bibinfo {volume} {99}},\ \bibinfo
			{pages} {245416} (\bibinfo {year} {2019})}\BibitemShut {NoStop}%
		\bibitem [{\citenamefont {Kells}\ \emph {et~al.}(2012)\citenamefont {Kells},
			\citenamefont {Meidan},\ and\ \citenamefont {Brouwer}}]{kells2012}%
		\BibitemOpen
		\bibfield  {author} {\bibinfo {author} {\bibfnamefont {G.}~\bibnamefont
				{Kells}}, \bibinfo {author} {\bibfnamefont {D.}~\bibnamefont {Meidan}},\ and\
			\bibinfo {author} {\bibfnamefont {P.~W.}\ \bibnamefont {Brouwer}},\
		}\bibfield  {title} {\bibinfo {title} {Near-zero-energy end states in
				topologically trivial spin-orbit coupled superconducting nanowires with a
				smooth confinement},\ }\href {https://doi.org/10.1103/PhysRevB.86.100503}
		{\bibfield  {journal} {\bibinfo  {journal} {Phys. Rev. B}\ }\textbf {\bibinfo
				{volume} {86}},\ \bibinfo {pages} {100503(R)} (\bibinfo {year}
			{2012})}\BibitemShut {NoStop}%
		\bibitem [{\citenamefont {Prada}\ \emph {et~al.}(2012)\citenamefont {Prada},
			\citenamefont {San-Jose},\ and\ \citenamefont {Aguado}}]{prada2012transport}%
		\BibitemOpen
		\bibfield  {author} {\bibinfo {author} {\bibfnamefont {E.}~\bibnamefont
				{Prada}}, \bibinfo {author} {\bibfnamefont {P.}~\bibnamefont {San-Jose}},\
			and\ \bibinfo {author} {\bibfnamefont {R.}~\bibnamefont {Aguado}},\
		}\bibfield  {title} {\bibinfo {title} {Transport spectroscopy of $ns$
				nanowire junctions with majorana fermions},\ }\href
		{https://doi.org/10.1103/PhysRevB.86.180503} {\bibfield  {journal} {\bibinfo
				{journal} {Phys. Rev. B}\ }\textbf {\bibinfo {volume} {86}},\ \bibinfo
			{pages} {180503(R)} (\bibinfo {year} {2012})}\BibitemShut {NoStop}%
		\bibitem [{\citenamefont {Pe\~naranda}\ \emph {et~al.}(2018)\citenamefont
			{Pe\~naranda}, \citenamefont {Aguado}, \citenamefont {San-Jose},\ and\
			\citenamefont {Prada}}]{penaranda2018}%
		\BibitemOpen
		\bibfield  {author} {\bibinfo {author} {\bibfnamefont {F.}~\bibnamefont
				{Pe\~naranda}}, \bibinfo {author} {\bibfnamefont {R.}~\bibnamefont {Aguado}},
			\bibinfo {author} {\bibfnamefont {P.}~\bibnamefont {San-Jose}},\ and\
			\bibinfo {author} {\bibfnamefont {E.}~\bibnamefont {Prada}},\ }\bibfield
		{title} {\bibinfo {title} {Quantifying wave-function overlaps in
				inhomogeneous majorana nanowires},\ }\href
		{https://doi.org/10.1103/PhysRevB.98.235406} {\bibfield  {journal} {\bibinfo
				{journal} {Phys. Rev. B}\ }\textbf {\bibinfo {volume} {98}},\ \bibinfo
			{pages} {235406} (\bibinfo {year} {2018})}\BibitemShut {NoStop}%
		\bibitem [{\citenamefont {Vuik}\ \emph {et~al.}(2019)\citenamefont {Vuik},
			\citenamefont {Nijholt}, \citenamefont {Akhmerov},\ and\ \citenamefont
			{Wimmer}}]{vuik2019}%
		\BibitemOpen
		\bibfield  {author} {\bibinfo {author} {\bibfnamefont {A.}~\bibnamefont
				{Vuik}}, \bibinfo {author} {\bibfnamefont {B.}~\bibnamefont {Nijholt}},
			\bibinfo {author} {\bibfnamefont {A.~R.}\ \bibnamefont {Akhmerov}},\ and\
			\bibinfo {author} {\bibfnamefont {M.}~\bibnamefont {Wimmer}},\ }\bibfield
		{title} {\bibinfo {title} {Reproducing topological properties with
				quasi-majorana states},\ }\href
		{https://doi.org/10.21468/SciPostPhys.7.5.061} {\bibfield  {journal}
			{\bibinfo  {journal} {SciPost Phys.}\ }\textbf {\bibinfo {volume} {7}},\
			\bibinfo {pages} {061} (\bibinfo {year} {2019})}\BibitemShut {NoStop}%
		\bibitem [{\citenamefont {Liu}\ \emph {et~al.}(2012)\citenamefont {Liu},
			\citenamefont {Potter}, \citenamefont {Law},\ and\ \citenamefont
			{Lee}}]{liu2012zero}%
		\BibitemOpen
		\bibfield  {author} {\bibinfo {author} {\bibfnamefont {J.}~\bibnamefont
				{Liu}}, \bibinfo {author} {\bibfnamefont {A.~C.}\ \bibnamefont {Potter}},
			\bibinfo {author} {\bibfnamefont {K.~T.}\ \bibnamefont {Law}},\ and\ \bibinfo
			{author} {\bibfnamefont {P.~A.}\ \bibnamefont {Lee}},\ }\bibfield  {title}
		{\bibinfo {title} {Zero-bias peaks in the tunneling conductance of
				spin-orbit-coupled superconducting wires with and without majorana
				end-states},\ }\href {https://doi.org/10.1103/PhysRevLett.109.267002}
		{\bibfield  {journal} {\bibinfo  {journal} {Phys. Rev. Lett.}\ }\textbf
			{\bibinfo {volume} {109}},\ \bibinfo {pages} {267002} (\bibinfo {year}
			{2012})}\BibitemShut {NoStop}%
		\bibitem [{\citenamefont {Liu}\ \emph {et~al.}(2017)\citenamefont {Liu},
			\citenamefont {Sau}, \citenamefont {Stanescu},\ and\ \citenamefont
			{Das~Sarma}}]{liu2017andreev}%
		\BibitemOpen
		\bibfield  {author} {\bibinfo {author} {\bibfnamefont {C.-X.}\ \bibnamefont
				{Liu}}, \bibinfo {author} {\bibfnamefont {J.~D.}\ \bibnamefont {Sau}},
			\bibinfo {author} {\bibfnamefont {T.~D.}\ \bibnamefont {Stanescu}},\ and\
			\bibinfo {author} {\bibfnamefont {S.}~\bibnamefont {Das~Sarma}},\ }\bibfield
		{title} {\bibinfo {title} {Andreev bound states versus majorana bound states
				in quantum dot-nanowire-superconductor hybrid structures: Trivial versus
				topological zero-bias conductance peaks},\ }\href
		{https://doi.org/10.1103/PhysRevB.96.075161} {\bibfield  {journal} {\bibinfo
				{journal} {Phys. Rev. B}\ }\textbf {\bibinfo {volume} {96}},\ \bibinfo
			{pages} {075161} (\bibinfo {year} {2017})}\BibitemShut {NoStop}%
		\bibitem [{\citenamefont {Reeg}\ \emph {et~al.}(2018)\citenamefont {Reeg},
			\citenamefont {Dmytruk}, \citenamefont {Chevallier}, \citenamefont {Loss},\
			and\ \citenamefont {Klinovaja}}]{reeg2018zero}%
		\BibitemOpen
		\bibfield  {author} {\bibinfo {author} {\bibfnamefont {C.}~\bibnamefont
				{Reeg}}, \bibinfo {author} {\bibfnamefont {O.}~\bibnamefont {Dmytruk}},
			\bibinfo {author} {\bibfnamefont {D.}~\bibnamefont {Chevallier}}, \bibinfo
			{author} {\bibfnamefont {D.}~\bibnamefont {Loss}},\ and\ \bibinfo {author}
			{\bibfnamefont {J.}~\bibnamefont {Klinovaja}},\ }\bibfield  {title} {\bibinfo
			{title} {Zero-energy {A}ndreev bound states from quantum dots in proximitized
				{R}ashba nanowires},\ }\href {https://doi.org/10.1103/PhysRevB.98.245407}
		{\bibfield  {journal} {\bibinfo  {journal} {Phys. Rev. B}\ }\textbf {\bibinfo
				{volume} {98}},\ \bibinfo {pages} {245407} (\bibinfo {year}
			{2018})}\BibitemShut {NoStop}%
		\bibitem [{\citenamefont {Hess}\ \emph {et~al.}(2021)\citenamefont {Hess},
			\citenamefont {Legg}, \citenamefont {Loss},\ and\ \citenamefont
			{Klinovaja}}]{hess2021local}%
		\BibitemOpen
		\bibfield  {author} {\bibinfo {author} {\bibfnamefont {R.}~\bibnamefont
				{Hess}}, \bibinfo {author} {\bibfnamefont {H.~F.}\ \bibnamefont {Legg}},
			\bibinfo {author} {\bibfnamefont {D.}~\bibnamefont {Loss}},\ and\ \bibinfo
			{author} {\bibfnamefont {J.}~\bibnamefont {Klinovaja}},\ }\bibfield  {title}
		{\bibinfo {title} {Local and nonlocal quantum transport due to andreev bound
				states in finite rashba nanowires with superconducting and normal sections},\
		}\href {https://doi.org/10.1103/PhysRevB.104.075405} {\bibfield  {journal}
			{\bibinfo  {journal} {Phys. Rev. B}\ }\textbf {\bibinfo {volume} {104}},\
			\bibinfo {pages} {075405} (\bibinfo {year} {2021})}\BibitemShut {NoStop}%
		\bibitem [{\citenamefont {Hess}\ \emph {et~al.}(2023)\citenamefont {Hess},
			\citenamefont {Legg}, \citenamefont {Loss},\ and\ \citenamefont
			{Klinovaja}}]{hess2023trivial}%
		\BibitemOpen
		\bibfield  {author} {\bibinfo {author} {\bibfnamefont {R.}~\bibnamefont
				{Hess}}, \bibinfo {author} {\bibfnamefont {H.~F.}\ \bibnamefont {Legg}},
			\bibinfo {author} {\bibfnamefont {D.}~\bibnamefont {Loss}},\ and\ \bibinfo
			{author} {\bibfnamefont {J.}~\bibnamefont {Klinovaja}},\ }\bibfield  {title}
		{\bibinfo {title} {Trivial andreev band mimicking topological bulk gap
				reopening in the nonlocal conductance of long rashba nanowires},\ }\href
		{https://doi.org/10.1103/PhysRevLett.130.207001} {\bibfield  {journal}
			{\bibinfo  {journal} {Phys. Rev. Lett.}\ }\textbf {\bibinfo {volume} {130}},\
			\bibinfo {pages} {207001} (\bibinfo {year} {2023})}\BibitemShut {NoStop}%
		\bibitem [{\citenamefont {Sahu}\ \emph {et~al.}(2023)\citenamefont {Sahu},
			\citenamefont {Khade},\ and\ \citenamefont {Gangadharaiah}}]{sahu2023effect}%
		\BibitemOpen
		\bibfield  {author} {\bibinfo {author} {\bibfnamefont {D.}~\bibnamefont
				{Sahu}}, \bibinfo {author} {\bibfnamefont {V.}~\bibnamefont {Khade}},\ and\
			\bibinfo {author} {\bibfnamefont {S.}~\bibnamefont {Gangadharaiah}},\
		}\bibfield  {title} {\bibinfo {title} {Effect of topological length on bound
				state signatures in a topological nanowire},\ }\href
		{https://doi.org/10.1103/PhysRevB.108.205426} {\bibfield  {journal} {\bibinfo
				{journal} {Phys. Rev. B}\ }\textbf {\bibinfo {volume} {108}},\ \bibinfo
			{pages} {205426} (\bibinfo {year} {2023})}\BibitemShut {NoStop}%
		\bibitem [{\citenamefont {Cottet}\ \emph {et~al.}(2013)\citenamefont {Cottet},
			\citenamefont {Kontos},\ and\ \citenamefont {Dou\ifmmode~\mbox{\c{c}}\else
				\c{c}\fi{}ot}}]{Cottet2013Squeezing}%
		\BibitemOpen
		\bibfield  {author} {\bibinfo {author} {\bibfnamefont {A.}~\bibnamefont
				{Cottet}}, \bibinfo {author} {\bibfnamefont {T.}~\bibnamefont {Kontos}},\
			and\ \bibinfo {author} {\bibfnamefont {B.}~\bibnamefont
				{Dou\ifmmode~\mbox{\c{c}}\else \c{c}\fi{}ot}},\ }\bibfield  {title} {\bibinfo
			{title} {Squeezing light with {M}ajorana fermions},\ }\href
		{https://doi.org/10.1103/PhysRevB.88.195415} {\bibfield  {journal} {\bibinfo
				{journal} {Phys. Rev. B}\ }\textbf {\bibinfo {volume} {88}},\ \bibinfo
			{pages} {195415} (\bibinfo {year} {2013})}\BibitemShut {NoStop}%
		\bibitem [{\citenamefont {Dmytruk}\ \emph {et~al.}(2015)\citenamefont
			{Dmytruk}, \citenamefont {Trif},\ and\ \citenamefont
			{Simon}}]{dmytruk2015cavity}%
		\BibitemOpen
		\bibfield  {author} {\bibinfo {author} {\bibfnamefont {O.}~\bibnamefont
				{Dmytruk}}, \bibinfo {author} {\bibfnamefont {M.}~\bibnamefont {Trif}},\ and\
			\bibinfo {author} {\bibfnamefont {P.}~\bibnamefont {Simon}},\ }\bibfield
		{title} {\bibinfo {title} {Cavity quantum electrodynamics with mesoscopic
				topological superconductors},\ }\href
		{https://doi.org/10.1103/PhysRevB.92.245432} {\bibfield  {journal} {\bibinfo
				{journal} {Phys. Rev. B}\ }\textbf {\bibinfo {volume} {92}},\ \bibinfo
			{pages} {245432} (\bibinfo {year} {2015})}\BibitemShut {NoStop}%
		\bibitem [{\citenamefont {Dartiailh}\ \emph {et~al.}(2017)\citenamefont
			{Dartiailh}, \citenamefont {Kontos}, \citenamefont
			{Dou\ifmmode~\mbox{\c{c}}\else \c{c}\fi{}ot},\ and\ \citenamefont
			{Cottet}}]{Dartiailh2017Cavity}%
		\BibitemOpen
		\bibfield  {author} {\bibinfo {author} {\bibfnamefont {M.~C.}\ \bibnamefont
				{Dartiailh}}, \bibinfo {author} {\bibfnamefont {T.}~\bibnamefont {Kontos}},
			\bibinfo {author} {\bibfnamefont {B.}~\bibnamefont
				{Dou\ifmmode~\mbox{\c{c}}\else \c{c}\fi{}ot}},\ and\ \bibinfo {author}
			{\bibfnamefont {A.}~\bibnamefont {Cottet}},\ }\bibfield  {title} {\bibinfo
			{title} {Direct {C}avity {D}etection of {M}ajorana {P}airs},\ }\href
		{https://doi.org/10.1103/PhysRevLett.118.126803} {\bibfield  {journal}
			{\bibinfo  {journal} {Phys. Rev. Lett.}\ }\textbf {\bibinfo {volume} {118}},\
			\bibinfo {pages} {126803} (\bibinfo {year} {2017})}\BibitemShut {NoStop}%
		\bibitem [{\citenamefont {Cottet}\ \emph {et~al.}(2017)\citenamefont {Cottet},
			\citenamefont {Dartiailh}, \citenamefont {Desjardins}, \citenamefont
			{Cubaynes}, \citenamefont {Contamin}, \citenamefont {Delbecq}, \citenamefont
			{Viennot}, \citenamefont {Bruhat}, \citenamefont {Douçot},\ and\
			\citenamefont {Kontos}}]{Cottet2017cavity}%
		\BibitemOpen
		\bibfield  {author} {\bibinfo {author} {\bibfnamefont {A.}~\bibnamefont
				{Cottet}}, \bibinfo {author} {\bibfnamefont {M.~C.}\ \bibnamefont
				{Dartiailh}}, \bibinfo {author} {\bibfnamefont {M.~M.}\ \bibnamefont
				{Desjardins}}, \bibinfo {author} {\bibfnamefont {T.}~\bibnamefont
				{Cubaynes}}, \bibinfo {author} {\bibfnamefont {L.~C.}\ \bibnamefont
				{Contamin}}, \bibinfo {author} {\bibfnamefont {M.}~\bibnamefont {Delbecq}},
			\bibinfo {author} {\bibfnamefont {J.~J.}\ \bibnamefont {Viennot}}, \bibinfo
			{author} {\bibfnamefont {L.~E.}\ \bibnamefont {Bruhat}}, \bibinfo {author}
			{\bibfnamefont {B.}~\bibnamefont {Douçot}},\ and\ \bibinfo {author}
			{\bibfnamefont {T.}~\bibnamefont {Kontos}},\ }\bibfield  {title} {\bibinfo
			{title} {Cavity qed with hybrid nanocircuits: from atomic-like physics to
				condensed matter phenomena},\ }\href
		{https://doi.org/10.1088/1361-648X/aa7b4d} {\bibfield  {journal} {\bibinfo
				{journal} {Journal of Physics: Condensed Matter}\ }\textbf {\bibinfo {volume}
				{29}},\ \bibinfo {pages} {433002} (\bibinfo {year} {2017})}\BibitemShut
		{NoStop}%
		\bibitem [{\citenamefont {Dmytruk}\ and\ \citenamefont
			{Trif}(2023)}]{Dmytruk2023microwave}%
		\BibitemOpen
		\bibfield  {author} {\bibinfo {author} {\bibfnamefont {O.}~\bibnamefont
				{Dmytruk}}\ and\ \bibinfo {author} {\bibfnamefont {M.}~\bibnamefont {Trif}},\
		}\bibfield  {title} {\bibinfo {title} {Microwave detection of gliding
				{M}ajorana zero modes in nanowires},\ }\href
		{https://doi.org/10.1103/PhysRevB.107.115418} {\bibfield  {journal} {\bibinfo
				{journal} {Phys. Rev. B}\ }\textbf {\bibinfo {volume} {107}},\ \bibinfo
			{pages} {115418} (\bibinfo {year} {2023})}\BibitemShut {NoStop}%
		\bibitem [{\citenamefont {Prem}\ \emph {et~al.}(2026)\citenamefont {Prem},
			\citenamefont {Dmytruk},\ and\ \citenamefont
			{Trif}}]{Prem2026Distinguishing}%
		\BibitemOpen
		\bibfield  {author} {\bibinfo {author} {\bibfnamefont {S.}~\bibnamefont
				{Prem}}, \bibinfo {author} {\bibfnamefont {O.}~\bibnamefont {Dmytruk}},\ and\
			\bibinfo {author} {\bibfnamefont {M.}~\bibnamefont {Trif}},\ }\bibfield
		{title} {\bibinfo {title} {Distinguishing {M}ajorana bound states from
				accidental zero-energy modes with a microwave cavity},\ }\href
		{https://doi.org/10.1103/q7l6-ytrr} {\bibfield  {journal} {\bibinfo
				{journal} {Phys. Rev. B}\ }\textbf {\bibinfo {volume} {113}},\ \bibinfo
			{pages} {085420} (\bibinfo {year} {2026})}\BibitemShut {NoStop}%
		\bibitem [{\citenamefont {Trif}\ and\ \citenamefont
			{Tserkovnyak}(2012)}]{Trif2012Resonantly}%
		\BibitemOpen
		\bibfield  {author} {\bibinfo {author} {\bibfnamefont {M.}~\bibnamefont
				{Trif}}\ and\ \bibinfo {author} {\bibfnamefont {Y.}~\bibnamefont
				{Tserkovnyak}},\ }\bibfield  {title} {\bibinfo {title} {Resonantly {T}unable
				{M}ajorana {P}olariton in a {M}icrowave {C}avity},\ }\href
		{https://doi.org/10.1103/PhysRevLett.109.257002} {\bibfield  {journal}
			{\bibinfo  {journal} {Phys. Rev. Lett.}\ }\textbf {\bibinfo {volume} {109}},\
			\bibinfo {pages} {257002} (\bibinfo {year} {2012})}\BibitemShut {NoStop}%
		\bibitem [{\citenamefont {Bacciconi}\ \emph {et~al.}(2024)\citenamefont
			{Bacciconi}, \citenamefont {Andolina},\ and\ \citenamefont
			{Mora}}]{bacciconi2024}%
		\BibitemOpen
		\bibfield  {author} {\bibinfo {author} {\bibfnamefont {Z.}~\bibnamefont
				{Bacciconi}}, \bibinfo {author} {\bibfnamefont {G.~M.}\ \bibnamefont
				{Andolina}},\ and\ \bibinfo {author} {\bibfnamefont {C.}~\bibnamefont
				{Mora}},\ }\bibfield  {title} {\bibinfo {title} {Topological protection of
				{M}ajorana polaritons in a cavity},\ }\href
		{https://doi.org/10.1103/PhysRevB.109.165434} {\bibfield  {journal} {\bibinfo
				{journal} {Phys. Rev. B}\ }\textbf {\bibinfo {volume} {109}},\ \bibinfo
			{pages} {165434} (\bibinfo {year} {2024})}\BibitemShut {NoStop}%
		\bibitem [{\citenamefont {Dmytruk}\ and\ \citenamefont
			{Schir\`o}(2024)}]{dmytruk2024hybrid}%
		\BibitemOpen
		\bibfield  {author} {\bibinfo {author} {\bibfnamefont {O.}~\bibnamefont
				{Dmytruk}}\ and\ \bibinfo {author} {\bibfnamefont {M.}~\bibnamefont
				{Schir\`o}},\ }\bibfield  {title} {\bibinfo {title} {Hybrid light-matter
				states in topological superconductors coupled to cavity photons},\ }\href
		{https://doi.org/10.1103/PhysRevB.110.075416} {\bibfield  {journal} {\bibinfo
				{journal} {Phys. Rev. B}\ }\textbf {\bibinfo {volume} {110}},\ \bibinfo
			{pages} {075416} (\bibinfo {year} {2024})}\BibitemShut {NoStop}%
		\bibitem [{\citenamefont {Garcia-Vidal}\ \emph {et~al.}(2021)\citenamefont
			{Garcia-Vidal}, \citenamefont {Ciuti},\ and\ \citenamefont
			{Ebbesen}}]{Francisco2021manipulating}%
		\BibitemOpen
		\bibfield  {author} {\bibinfo {author} {\bibfnamefont {F.~J.}\ \bibnamefont
				{Garcia-Vidal}}, \bibinfo {author} {\bibfnamefont {C.}~\bibnamefont
				{Ciuti}},\ and\ \bibinfo {author} {\bibfnamefont {T.~W.}\ \bibnamefont
				{Ebbesen}},\ }\bibfield  {title} {\bibinfo {title} {Manipulating matter by
				strong coupling to vacuum fields},\ }\href
		{https://doi.org/10.1126/science.abd0336} {\bibfield  {journal} {\bibinfo
				{journal} {Science}\ }\textbf {\bibinfo {volume} {373}},\ \bibinfo {pages}
			{eabd0336} (\bibinfo {year} {2021})}\BibitemShut {NoStop}%
		\bibitem [{\citenamefont {Schlawin}\ \emph {et~al.}(2022)\citenamefont
			{Schlawin}, \citenamefont {Kennes},\ and\ \citenamefont
			{Sentef}}]{Schlawin2022quantummaterials}%
		\BibitemOpen
		\bibfield  {author} {\bibinfo {author} {\bibfnamefont {F.}~\bibnamefont
				{Schlawin}}, \bibinfo {author} {\bibfnamefont {D.~M.}\ \bibnamefont
				{Kennes}},\ and\ \bibinfo {author} {\bibfnamefont {M.~A.}\ \bibnamefont
				{Sentef}},\ }\bibfield  {title} {\bibinfo {title} {Cavity quantum
				materials},\ }\href {https://doi.org/10.1063/5.0083825} {\bibfield  {journal}
			{\bibinfo  {journal} {Applied Physics Reviews}\ }\textbf {\bibinfo {volume}
				{9}},\ \bibinfo {pages} {011312} (\bibinfo {year} {2022})}\BibitemShut
		{NoStop}%
		\bibitem [{\citenamefont {Lu}\ \emph {et~al.}(2025)\citenamefont {Lu},
			\citenamefont {Shin}, \citenamefont {Svendsen}, \citenamefont {Latini},
			\citenamefont {H\"{u}bener}, \citenamefont {Ruggenthaler},\ and\
			\citenamefont {Rubio}}]{Lu2025cavity}%
		\BibitemOpen
		\bibfield  {author} {\bibinfo {author} {\bibfnamefont {I.-T.}\ \bibnamefont
				{Lu}}, \bibinfo {author} {\bibfnamefont {D.}~\bibnamefont {Shin}}, \bibinfo
			{author} {\bibfnamefont {M.~K.}\ \bibnamefont {Svendsen}}, \bibinfo {author}
			{\bibfnamefont {S.}~\bibnamefont {Latini}}, \bibinfo {author} {\bibfnamefont
				{H.}~\bibnamefont {H\"{u}bener}}, \bibinfo {author} {\bibfnamefont
				{M.}~\bibnamefont {Ruggenthaler}},\ and\ \bibinfo {author} {\bibfnamefont
				{A.}~\bibnamefont {Rubio}},\ }\bibfield  {title} {\bibinfo {title} {Cavity
				engineering of solid-state materials without external driving},\ }\href
		{https://doi.org/10.1364/AOP.544138} {\bibfield  {journal} {\bibinfo
				{journal} {Adv. Opt. Photon.}\ }\textbf {\bibinfo {volume} {17}},\ \bibinfo
			{pages} {441} (\bibinfo {year} {2025})}\BibitemShut {NoStop}%
		\bibitem [{\citenamefont {Bretscher}\ \emph {et~al.}(2026)\citenamefont
			{Bretscher}, \citenamefont {Graziotto}, \citenamefont {Michael},
			\citenamefont {Montanaro}, \citenamefont {Lu}, \citenamefont {Grankin},
			\citenamefont {McIver}, \citenamefont {Faist}, \citenamefont {Fausti},
			\citenamefont {Eckstein}, \citenamefont {Ruggenthaler}, \citenamefont
			{Rubio}, \citenamefont {Basov}, \citenamefont {Hafezi}, \citenamefont
			{Claassen}, \citenamefont {Kennes},\ and\ \citenamefont
			{Sentef}}]{bretscher2026fluctuationengineeringcavityquantum}%
		\BibitemOpen
		\bibfield  {author} {\bibinfo {author} {\bibfnamefont {H.~M.}\ \bibnamefont
				{Bretscher}}, \bibinfo {author} {\bibfnamefont {L.}~\bibnamefont
				{Graziotto}}, \bibinfo {author} {\bibfnamefont {M.~H.}\ \bibnamefont
				{Michael}}, \bibinfo {author} {\bibfnamefont {A.}~\bibnamefont {Montanaro}},
			\bibinfo {author} {\bibfnamefont {I.-T.}\ \bibnamefont {Lu}}, \bibinfo
			{author} {\bibfnamefont {A.}~\bibnamefont {Grankin}}, \bibinfo {author}
			{\bibfnamefont {J.~W.}\ \bibnamefont {McIver}}, \bibinfo {author}
			{\bibfnamefont {J.}~\bibnamefont {Faist}}, \bibinfo {author} {\bibfnamefont
				{D.}~\bibnamefont {Fausti}}, \bibinfo {author} {\bibfnamefont
				{M.}~\bibnamefont {Eckstein}}, \bibinfo {author} {\bibfnamefont
				{M.}~\bibnamefont {Ruggenthaler}}, \bibinfo {author} {\bibfnamefont
				{A.}~\bibnamefont {Rubio}}, \bibinfo {author} {\bibfnamefont
				{D.}~\bibnamefont {Basov}}, \bibinfo {author} {\bibfnamefont
				{M.}~\bibnamefont {Hafezi}}, \bibinfo {author} {\bibfnamefont
				{M.}~\bibnamefont {Claassen}}, \bibinfo {author} {\bibfnamefont {D.~M.}\
				\bibnamefont {Kennes}},\ and\ \bibinfo {author} {\bibfnamefont {M.~A.}\
				\bibnamefont {Sentef}},\ }\href {https://arxiv.org/abs/2604.08666} {\bibinfo
			{title} {Fluctuation engineering in cavity quantum materials}} (\bibinfo
		{year} {2026}),\ \Eprint {https://arxiv.org/abs/2604.08666} {arXiv:2604.08666
			[cond-mat.mes-hall]} \BibitemShut {NoStop}%
		\bibitem [{\citenamefont {Keren}\ \emph {et~al.}(2026)\citenamefont {Keren},
			\citenamefont {Webb}, \citenamefont {Zhang}, \citenamefont {Xu},
			\citenamefont {Sun}, \citenamefont {Kim}, \citenamefont {Shin}, \citenamefont
			{Zhang}, \citenamefont {Zhang}, \citenamefont {Pereira}, \citenamefont {Yao},
			\citenamefont {Okugawa}, \citenamefont {Michael}, \citenamefont
			{Vi{\~{n}}as~Bostr{\"o}m}, \citenamefont {Edgar}, \citenamefont {Wolf},
			\citenamefont {Julian}, \citenamefont {Prasankumar}, \citenamefont
			{Miyagawa}, \citenamefont {Kanoda}, \citenamefont {Gu}, \citenamefont
			{Cothrine}, \citenamefont {Mandrus}, \citenamefont {Buzzi}, \citenamefont
			{Cavalleri}, \citenamefont {Dean}, \citenamefont {Kennes}, \citenamefont
			{Millis}, \citenamefont {Li}, \citenamefont {Sentef}, \citenamefont {Rubio},
			\citenamefont {Pasupathy},\ and\ \citenamefont
			{Basov}}]{Keren2026Cavitysuperconductivity}%
		\BibitemOpen
		\bibfield  {author} {\bibinfo {author} {\bibfnamefont {I.}~\bibnamefont
				{Keren}}, \bibinfo {author} {\bibfnamefont {T.~A.}\ \bibnamefont {Webb}},
			\bibinfo {author} {\bibfnamefont {S.}~\bibnamefont {Zhang}}, \bibinfo
			{author} {\bibfnamefont {J.}~\bibnamefont {Xu}}, \bibinfo {author}
			{\bibfnamefont {D.}~\bibnamefont {Sun}}, \bibinfo {author} {\bibfnamefont
				{B.~S.~Y.}\ \bibnamefont {Kim}}, \bibinfo {author} {\bibfnamefont
				{D.}~\bibnamefont {Shin}}, \bibinfo {author} {\bibfnamefont {S.~S.}\
				\bibnamefont {Zhang}}, \bibinfo {author} {\bibfnamefont {J.}~\bibnamefont
				{Zhang}}, \bibinfo {author} {\bibfnamefont {G.}~\bibnamefont {Pereira}},
			\bibinfo {author} {\bibfnamefont {J.}~\bibnamefont {Yao}}, \bibinfo {author}
			{\bibfnamefont {T.}~\bibnamefont {Okugawa}}, \bibinfo {author} {\bibfnamefont
				{M.~H.}\ \bibnamefont {Michael}}, \bibinfo {author} {\bibfnamefont
				{E.}~\bibnamefont {Vi{\~{n}}as~Bostr{\"o}m}}, \bibinfo {author}
			{\bibfnamefont {J.~H.}\ \bibnamefont {Edgar}}, \bibinfo {author}
			{\bibfnamefont {S.}~\bibnamefont {Wolf}}, \bibinfo {author} {\bibfnamefont
				{M.}~\bibnamefont {Julian}}, \bibinfo {author} {\bibfnamefont {R.~P.}\
				\bibnamefont {Prasankumar}}, \bibinfo {author} {\bibfnamefont
				{K.}~\bibnamefont {Miyagawa}}, \bibinfo {author} {\bibfnamefont
				{K.}~\bibnamefont {Kanoda}}, \bibinfo {author} {\bibfnamefont
				{G.}~\bibnamefont {Gu}}, \bibinfo {author} {\bibfnamefont {M.}~\bibnamefont
				{Cothrine}}, \bibinfo {author} {\bibfnamefont {D.}~\bibnamefont {Mandrus}},
			\bibinfo {author} {\bibfnamefont {M.}~\bibnamefont {Buzzi}}, \bibinfo
			{author} {\bibfnamefont {A.}~\bibnamefont {Cavalleri}}, \bibinfo {author}
			{\bibfnamefont {C.~R.}\ \bibnamefont {Dean}}, \bibinfo {author}
			{\bibfnamefont {D.~M.}\ \bibnamefont {Kennes}}, \bibinfo {author}
			{\bibfnamefont {A.~J.}\ \bibnamefont {Millis}}, \bibinfo {author}
			{\bibfnamefont {Q.}~\bibnamefont {Li}}, \bibinfo {author} {\bibfnamefont
				{M.~A.}\ \bibnamefont {Sentef}}, \bibinfo {author} {\bibfnamefont
				{A.}~\bibnamefont {Rubio}}, \bibinfo {author} {\bibfnamefont {A.~N.}\
				\bibnamefont {Pasupathy}},\ and\ \bibinfo {author} {\bibfnamefont {D.~N.}\
				\bibnamefont {Basov}},\ }\bibfield  {title} {\bibinfo {title} {Cavity-altered
				superconductivity},\ }\href {https://doi.org/10.1038/s41586-025-10062-6}
		{\bibfield  {journal} {\bibinfo  {journal} {Nature}\ }\textbf {\bibinfo
				{volume} {650}},\ \bibinfo {pages} {864} (\bibinfo {year}
			{2026})}\BibitemShut {NoStop}%
		\bibitem [{\citenamefont {Montanaro}\ \emph {et~al.}(2026)\citenamefont
			{Montanaro}, \citenamefont {Plastovets}, \citenamefont {Khatiwada},
			\citenamefont {Fiore}, \citenamefont {Jarc}, \citenamefont {Alabbadi},
			\citenamefont {Mastropasqua}, \citenamefont {Rigoni}, \citenamefont
			{Mathengattil}, \citenamefont {Zilio}, \citenamefont {Olsen}, \citenamefont
			{Novelli}, \citenamefont {Winnerl}, \citenamefont {Sentef}, \citenamefont
			{Kennes}, \citenamefont {Millis}, \citenamefont {Piazza},\ and\ \citenamefont
			{Fausti}}]{montanaro2026cavity}%
		\BibitemOpen
		\bibfield  {author} {\bibinfo {author} {\bibfnamefont {A.}~\bibnamefont
				{Montanaro}}, \bibinfo {author} {\bibfnamefont {V.}~\bibnamefont
				{Plastovets}}, \bibinfo {author} {\bibfnamefont {N.}~\bibnamefont
				{Khatiwada}}, \bibinfo {author} {\bibfnamefont {J.}~\bibnamefont {Fiore}},
			\bibinfo {author} {\bibfnamefont {G.}~\bibnamefont {Jarc}}, \bibinfo {author}
			{\bibfnamefont {A.}~\bibnamefont {Alabbadi}}, \bibinfo {author}
			{\bibfnamefont {A.}~\bibnamefont {Mastropasqua}}, \bibinfo {author}
			{\bibfnamefont {E.~M.}\ \bibnamefont {Rigoni}}, \bibinfo {author}
			{\bibfnamefont {S.~Y.}\ \bibnamefont {Mathengattil}}, \bibinfo {author}
			{\bibfnamefont {S.~D.}\ \bibnamefont {Zilio}}, \bibinfo {author}
			{\bibfnamefont {F.~F.}\ \bibnamefont {Olsen}}, \bibinfo {author}
			{\bibfnamefont {F.}~\bibnamefont {Novelli}}, \bibinfo {author} {\bibfnamefont
				{S.}~\bibnamefont {Winnerl}}, \bibinfo {author} {\bibfnamefont {M.~A.}\
				\bibnamefont {Sentef}}, \bibinfo {author} {\bibfnamefont {D.~M.}\
				\bibnamefont {Kennes}}, \bibinfo {author} {\bibfnamefont {A.~J.}\
				\bibnamefont {Millis}}, \bibinfo {author} {\bibfnamefont {F.}~\bibnamefont
				{Piazza}},\ and\ \bibinfo {author} {\bibfnamefont {D.}~\bibnamefont
				{Fausti}},\ }\href {https://arxiv.org/abs/2606.18084} {\bibinfo {title}
			{Cavity-enhanced superconducting response in an underdoped cuprate}}
		(\bibinfo {year} {2026}),\ \Eprint {https://arxiv.org/abs/2606.18084}
		{arXiv:2606.18084 [cond-mat.supr-con]} \BibitemShut {NoStop}%
		\bibitem [{\citenamefont {Zhang}\ \emph {et~al.}(2026)\citenamefont {Zhang},
			\citenamefont {Feng}, \citenamefont {Lu}, \citenamefont {Li}, \citenamefont
			{Guo}, \citenamefont {Shang}, \citenamefont {Tan}, \citenamefont {Lyu},
			\citenamefont {Shen}, \citenamefont {Luan}, \citenamefont {Panmai},
			\citenamefont {Watanabe}, \citenamefont {Taniguchi}, \citenamefont {Singh},
			\citenamefont {Rubio},\ and\ \citenamefont {Gao}}]{zhang2026cavity}%
		\BibitemOpen
		\bibfield  {author} {\bibinfo {author} {\bibfnamefont {H.}~\bibnamefont
				{Zhang}}, \bibinfo {author} {\bibfnamefont {Z.}~\bibnamefont {Feng}},
			\bibinfo {author} {\bibfnamefont {I.-T.}\ \bibnamefont {Lu}}, \bibinfo
			{author} {\bibfnamefont {Z.}~\bibnamefont {Li}}, \bibinfo {author}
			{\bibfnamefont {S.}~\bibnamefont {Guo}}, \bibinfo {author} {\bibfnamefont
				{Q.}~\bibnamefont {Shang}}, \bibinfo {author} {\bibfnamefont
				{T.}~\bibnamefont {Tan}}, \bibinfo {author} {\bibfnamefont {X.}~\bibnamefont
				{Lyu}}, \bibinfo {author} {\bibfnamefont {X.}~\bibnamefont {Shen}}, \bibinfo
			{author} {\bibfnamefont {D.}~\bibnamefont {Luan}}, \bibinfo {author}
			{\bibfnamefont {M.}~\bibnamefont {Panmai}}, \bibinfo {author} {\bibfnamefont
				{K.}~\bibnamefont {Watanabe}}, \bibinfo {author} {\bibfnamefont
				{T.}~\bibnamefont {Taniguchi}}, \bibinfo {author} {\bibfnamefont
				{R.}~\bibnamefont {Singh}}, \bibinfo {author} {\bibfnamefont
				{A.}~\bibnamefont {Rubio}},\ and\ \bibinfo {author} {\bibfnamefont
				{W.}~\bibnamefont {Gao}},\ }\href {https://arxiv.org/abs/2606.19171}
		{\bibinfo {title} {Cavity-enhanced superconductivity in the two-dimensional
				limit of nbse2}} (\bibinfo {year} {2026}),\ \Eprint
		{https://arxiv.org/abs/2606.19171} {arXiv:2606.19171 [cond-mat.supr-con]}
		\BibitemShut {NoStop}%
		\bibitem [{\citenamefont {Wang}\ \emph {et~al.}(2026)\citenamefont {Wang},
			\citenamefont {Cardoso}, \citenamefont {Yang}, \citenamefont {Gong},
			\citenamefont {Zhang}, \citenamefont {Zhu}, \citenamefont {Zhang},
			\citenamefont {Pan}, \citenamefont {Cai}, \citenamefont {Chen}, \citenamefont
			{Jiang}, \citenamefont {Cheng}, \citenamefont {Wilczek},\ and\ \citenamefont
			{Zeng}}]{wang2026vacuum}%
		\BibitemOpen
		\bibfield  {author} {\bibinfo {author} {\bibfnamefont {Z.}~\bibnamefont
				{Wang}}, \bibinfo {author} {\bibfnamefont {G.}~\bibnamefont {Cardoso}},
			\bibinfo {author} {\bibfnamefont {L.}~\bibnamefont {Yang}}, \bibinfo {author}
			{\bibfnamefont {X.}~\bibnamefont {Gong}}, \bibinfo {author} {\bibfnamefont
				{C.}~\bibnamefont {Zhang}}, \bibinfo {author} {\bibfnamefont
				{Y.}~\bibnamefont {Zhu}}, \bibinfo {author} {\bibfnamefont {D.}~\bibnamefont
				{Zhang}}, \bibinfo {author} {\bibfnamefont {N.}~\bibnamefont {Pan}}, \bibinfo
			{author} {\bibfnamefont {H.}~\bibnamefont {Cai}}, \bibinfo {author}
			{\bibfnamefont {Y.~P.}\ \bibnamefont {Chen}}, \bibinfo {author}
			{\bibfnamefont {Q.-D.}\ \bibnamefont {Jiang}}, \bibinfo {author}
			{\bibfnamefont {G.}~\bibnamefont {Cheng}}, \bibinfo {author} {\bibfnamefont
				{F.}~\bibnamefont {Wilczek}},\ and\ \bibinfo {author} {\bibfnamefont
				{C.}~\bibnamefont {Zeng}},\ }\bibfield  {title} {\bibinfo {title} {Evidence
				for vacuum-enhanced superconductivity in {N}b{S}e2},\ }\bibfield  {journal}
		{\bibinfo  {journal} {Nature}\ }\href
		{https://doi.org/10.1038/s41586-026-11037-x} {10.1038/s41586-026-11037-x}
		(\bibinfo {year} {2026})\BibitemShut {NoStop}%
		\bibitem [{\citenamefont {Jarc}\ \emph {et~al.}(2023)\citenamefont {Jarc},
			\citenamefont {Mathengattil}, \citenamefont {Montanaro}, \citenamefont
			{Giusti}, \citenamefont {Rigoni}, \citenamefont {Sergo}, \citenamefont
			{Fassioli}, \citenamefont {Winnerl}, \citenamefont {Dal~Zilio}, \citenamefont
			{Mihailovic}, \citenamefont {Prelov{\v{s}}ek}, \citenamefont {Eckstein},\
			and\ \citenamefont {Fausti}}]{Jarc2023metalinsulator}%
		\BibitemOpen
		\bibfield  {author} {\bibinfo {author} {\bibfnamefont {G.}~\bibnamefont
				{Jarc}}, \bibinfo {author} {\bibfnamefont {S.~Y.}\ \bibnamefont
				{Mathengattil}}, \bibinfo {author} {\bibfnamefont {A.}~\bibnamefont
				{Montanaro}}, \bibinfo {author} {\bibfnamefont {F.}~\bibnamefont {Giusti}},
			\bibinfo {author} {\bibfnamefont {E.~M.}\ \bibnamefont {Rigoni}}, \bibinfo
			{author} {\bibfnamefont {R.}~\bibnamefont {Sergo}}, \bibinfo {author}
			{\bibfnamefont {F.}~\bibnamefont {Fassioli}}, \bibinfo {author}
			{\bibfnamefont {S.}~\bibnamefont {Winnerl}}, \bibinfo {author} {\bibfnamefont
				{S.}~\bibnamefont {Dal~Zilio}}, \bibinfo {author} {\bibfnamefont
				{D.}~\bibnamefont {Mihailovic}}, \bibinfo {author} {\bibfnamefont
				{P.}~\bibnamefont {Prelov{\v{s}}ek}}, \bibinfo {author} {\bibfnamefont
				{M.}~\bibnamefont {Eckstein}},\ and\ \bibinfo {author} {\bibfnamefont
				{D.}~\bibnamefont {Fausti}},\ }\bibfield  {title} {\bibinfo {title}
			{Cavity-mediated thermal control of metal-to-insulator transition in
				1{T}-{T}a{S}2},\ }\href {https://doi.org/10.1038/s41586-023-06596-2}
		{\bibfield  {journal} {\bibinfo  {journal} {Nature}\ }\textbf {\bibinfo
				{volume} {622}},\ \bibinfo {pages} {487} (\bibinfo {year}
			{2023})}\BibitemShut {NoStop}%
		\bibitem [{\citenamefont {Appugliese}\ \emph {et~al.}(2022)\citenamefont
			{Appugliese}, \citenamefont {Enkner}, \citenamefont {Paravicini-Bagliani},
			\citenamefont {Beck}, \citenamefont {Reichl}, \citenamefont {Wegscheider},
			\citenamefont {Scalari}, \citenamefont {Ciuti},\ and\ \citenamefont
			{Faist}}]{Appugliese2022Breakdown}%
		\BibitemOpen
		\bibfield  {author} {\bibinfo {author} {\bibfnamefont {F.}~\bibnamefont
				{Appugliese}}, \bibinfo {author} {\bibfnamefont {J.}~\bibnamefont {Enkner}},
			\bibinfo {author} {\bibfnamefont {G.~L.}\ \bibnamefont
				{Paravicini-Bagliani}}, \bibinfo {author} {\bibfnamefont {M.}~\bibnamefont
				{Beck}}, \bibinfo {author} {\bibfnamefont {C.}~\bibnamefont {Reichl}},
			\bibinfo {author} {\bibfnamefont {W.}~\bibnamefont {Wegscheider}}, \bibinfo
			{author} {\bibfnamefont {G.}~\bibnamefont {Scalari}}, \bibinfo {author}
			{\bibfnamefont {C.}~\bibnamefont {Ciuti}},\ and\ \bibinfo {author}
			{\bibfnamefont {J.}~\bibnamefont {Faist}},\ }\bibfield  {title} {\bibinfo
			{title} {Breakdown of topological protection by cavity vacuum fields in the
				integer quantum hall effect},\ }\href
		{https://doi.org/10.1126/science.abl5818} {\bibfield  {journal} {\bibinfo
				{journal} {Science}\ }\textbf {\bibinfo {volume} {375}},\ \bibinfo {pages}
			{1030} (\bibinfo {year} {2022})}\BibitemShut {NoStop}%
		\bibitem [{\citenamefont {Enkner}\ \emph {et~al.}(2025)\citenamefont {Enkner},
			\citenamefont {Graziotto}, \citenamefont {Bori{\c{c}}i}, \citenamefont
			{Appugliese}, \citenamefont {Reichl}, \citenamefont {Scalari}, \citenamefont
			{Regnault}, \citenamefont {Wegscheider}, \citenamefont {Ciuti},\ and\
			\citenamefont {Faist}}]{enkner2025tunable}%
		\BibitemOpen
		\bibfield  {author} {\bibinfo {author} {\bibfnamefont {J.}~\bibnamefont
				{Enkner}}, \bibinfo {author} {\bibfnamefont {L.}~\bibnamefont {Graziotto}},
			\bibinfo {author} {\bibfnamefont {D.}~\bibnamefont {Bori{\c{c}}i}}, \bibinfo
			{author} {\bibfnamefont {F.}~\bibnamefont {Appugliese}}, \bibinfo {author}
			{\bibfnamefont {C.}~\bibnamefont {Reichl}}, \bibinfo {author} {\bibfnamefont
				{G.}~\bibnamefont {Scalari}}, \bibinfo {author} {\bibfnamefont
				{N.}~\bibnamefont {Regnault}}, \bibinfo {author} {\bibfnamefont
				{W.}~\bibnamefont {Wegscheider}}, \bibinfo {author} {\bibfnamefont
				{C.}~\bibnamefont {Ciuti}},\ and\ \bibinfo {author} {\bibfnamefont
				{J.}~\bibnamefont {Faist}},\ }\bibfield  {title} {\bibinfo {title} {Tunable
				vacuum-field control of fractional and integer quantum {H}all phases},\
		}\href {https://doi.org/10.1038/s41586-025-08894-3} {\bibfield  {journal}
			{\bibinfo  {journal} {Nature}\ }\textbf {\bibinfo {volume} {641}},\ \bibinfo
			{pages} {884} (\bibinfo {year} {2025})}\BibitemShut {NoStop}%
		\bibitem [{\citenamefont {Sentef}\ \emph {et~al.}()\citenamefont {Sentef},
			\citenamefont {Ruggenthaler},\ and\ \citenamefont {Rubio}}]{sentef2018}%
		\BibitemOpen
		\bibfield  {author} {\bibinfo {author} {\bibfnamefont {M.~A.}\ \bibnamefont
				{Sentef}}, \bibinfo {author} {\bibfnamefont {M.}~\bibnamefont
				{Ruggenthaler}},\ and\ \bibinfo {author} {\bibfnamefont {A.}~\bibnamefont
				{Rubio}},\ }\bibfield  {title} {\bibinfo {title} {Cavity
				quantum-electrodynamical polaritonically enhanced electron-phonon coupling
				and its influence on superconductivity},\ }\href
		{https://doi.org/10.1126/sciadv.aau6969} {\bibfield  {journal} {\bibinfo
				{journal} {Science Advances}\ }\textbf {\bibinfo {volume} {4}},\ \bibinfo
			{pages} {eaau6969}}\BibitemShut {NoStop}%
		\bibitem [{\citenamefont {Schlawin}\ \emph {et~al.}(2019)\citenamefont
			{Schlawin}, \citenamefont {Cavalleri},\ and\ \citenamefont
			{Jaksch}}]{schlawin2019cavity}%
		\BibitemOpen
		\bibfield  {author} {\bibinfo {author} {\bibfnamefont {F.}~\bibnamefont
				{Schlawin}}, \bibinfo {author} {\bibfnamefont {A.}~\bibnamefont
				{Cavalleri}},\ and\ \bibinfo {author} {\bibfnamefont {D.}~\bibnamefont
				{Jaksch}},\ }\bibfield  {title} {\bibinfo {title} {Cavity-mediated
				electron-photon superconductivity},\ }\href
		{https://doi.org/10.1103/PhysRevLett.122.133602} {\bibfield  {journal}
			{\bibinfo  {journal} {Phys. Rev. Lett.}\ }\textbf {\bibinfo {volume} {122}},\
			\bibinfo {pages} {133602} (\bibinfo {year} {2019})}\BibitemShut {NoStop}%
		\bibitem [{\citenamefont {Curtis}\ \emph {et~al.}(2019)\citenamefont {Curtis},
			\citenamefont {Raines}, \citenamefont {Allocca}, \citenamefont {Hafezi},\
			and\ \citenamefont {Galitski}}]{Curtis2019Eliashberg}%
		\BibitemOpen
		\bibfield  {author} {\bibinfo {author} {\bibfnamefont {J.~B.}\ \bibnamefont
				{Curtis}}, \bibinfo {author} {\bibfnamefont {Z.~M.}\ \bibnamefont {Raines}},
			\bibinfo {author} {\bibfnamefont {A.~A.}\ \bibnamefont {Allocca}}, \bibinfo
			{author} {\bibfnamefont {M.}~\bibnamefont {Hafezi}},\ and\ \bibinfo {author}
			{\bibfnamefont {V.~M.}\ \bibnamefont {Galitski}},\ }\bibfield  {title}
		{\bibinfo {title} {Cavity {Q}uantum {E}liashberg {E}nhancement of
				{S}uperconductivity},\ }\href
		{https://doi.org/10.1103/PhysRevLett.122.167002} {\bibfield  {journal}
			{\bibinfo  {journal} {Phys. Rev. Lett.}\ }\textbf {\bibinfo {volume} {122}},\
			\bibinfo {pages} {167002} (\bibinfo {year} {2019})}\BibitemShut {NoStop}%
		\bibitem [{\citenamefont {Allocca}\ \emph {et~al.}(2019)\citenamefont
			{Allocca}, \citenamefont {Raines}, \citenamefont {Curtis},\ and\
			\citenamefont {Galitski}}]{alloca2019}%
		\BibitemOpen
		\bibfield  {author} {\bibinfo {author} {\bibfnamefont {A.~A.}\ \bibnamefont
				{Allocca}}, \bibinfo {author} {\bibfnamefont {Z.~M.}\ \bibnamefont {Raines}},
			\bibinfo {author} {\bibfnamefont {J.~B.}\ \bibnamefont {Curtis}},\ and\
			\bibinfo {author} {\bibfnamefont {V.~M.}\ \bibnamefont {Galitski}},\
		}\bibfield  {title} {\bibinfo {title} {Cavity superconductor-polaritons},\
		}\href {https://doi.org/10.1103/PhysRevB.99.020504} {\bibfield  {journal}
			{\bibinfo  {journal} {Phys. Rev. B}\ }\textbf {\bibinfo {volume} {99}},\
			\bibinfo {pages} {020504} (\bibinfo {year} {2019})}\BibitemShut {NoStop}%
		\bibitem [{\citenamefont {Kozin}\ \emph {et~al.}(2025)\citenamefont {Kozin},
			\citenamefont {Thingstad}, \citenamefont {Loss},\ and\ \citenamefont
			{Klinovaja}}]{kozin2025cavity}%
		\BibitemOpen
		\bibfield  {author} {\bibinfo {author} {\bibfnamefont {V.~K.}\ \bibnamefont
				{Kozin}}, \bibinfo {author} {\bibfnamefont {E.}~\bibnamefont {Thingstad}},
			\bibinfo {author} {\bibfnamefont {D.}~\bibnamefont {Loss}},\ and\ \bibinfo
			{author} {\bibfnamefont {J.}~\bibnamefont {Klinovaja}},\ }\bibfield  {title}
		{\bibinfo {title} {Cavity-enhanced superconductivity via band engineering},\
		}\href {https://doi.org/10.1103/PhysRevB.111.035410} {\bibfield  {journal}
			{\bibinfo  {journal} {Phys. Rev. B}\ }\textbf {\bibinfo {volume} {111}},\
			\bibinfo {pages} {035410} (\bibinfo {year} {2025})}\BibitemShut {NoStop}%
		\bibitem [{\citenamefont {Passetti}\ \emph {et~al.}(2023)\citenamefont
			{Passetti}, \citenamefont {Eckhardt}, \citenamefont {Sentef},\ and\
			\citenamefont {Kennes}}]{passetti2023cavity}%
		\BibitemOpen
		\bibfield  {author} {\bibinfo {author} {\bibfnamefont {G.}~\bibnamefont
				{Passetti}}, \bibinfo {author} {\bibfnamefont {C.~J.}\ \bibnamefont
				{Eckhardt}}, \bibinfo {author} {\bibfnamefont {M.~A.}\ \bibnamefont
				{Sentef}},\ and\ \bibinfo {author} {\bibfnamefont {D.~M.}\ \bibnamefont
				{Kennes}},\ }\bibfield  {title} {\bibinfo {title} {Cavity {L}ight-{M}atter
				{E}ntanglement through {Q}uantum {F}luctuations},\ }\href
		{https://doi.org/10.1103/PhysRevLett.131.023601} {\bibfield  {journal}
			{\bibinfo  {journal} {Phys. Rev. Lett.}\ }\textbf {\bibinfo {volume} {131}},\
			\bibinfo {pages} {023601} (\bibinfo {year} {2023})}\BibitemShut {NoStop}%
		\bibitem [{\citenamefont {Fadler}\ \emph {et~al.}(2024)\citenamefont {Fadler},
			\citenamefont {Schmidt}, \citenamefont {Li},\ and\ \citenamefont
			{Eckstein}}]{Fadler2024}%
		\BibitemOpen
		\bibfield  {author} {\bibinfo {author} {\bibfnamefont {P.}~\bibnamefont
				{Fadler}}, \bibinfo {author} {\bibfnamefont {K.~P.}\ \bibnamefont {Schmidt}},
			\bibinfo {author} {\bibfnamefont {J.}~\bibnamefont {Li}},\ and\ \bibinfo
			{author} {\bibfnamefont {M.}~\bibnamefont {Eckstein}},\ }\bibfield  {title}
		{\bibinfo {title} {Engineering photon-mediated long-range spin interactions
				in {M}ott insulators},\ }\href {https://doi.org/10.1103/PhysRevB.109.085149}
		{\bibfield  {journal} {\bibinfo  {journal} {Phys. Rev. B}\ }\textbf {\bibinfo
				{volume} {109}},\ \bibinfo {pages} {085149} (\bibinfo {year}
			{2024})}\BibitemShut {NoStop}%
		\bibitem [{\citenamefont {Bori\ifmmode~\mbox{\c{c}}\else \c{c}\fi{}i}\ \emph
			{et~al.}(2025)\citenamefont {Bori\ifmmode~\mbox{\c{c}}\else \c{c}\fi{}i},
			\citenamefont {Arwas},\ and\ \citenamefont {Ciuti}}]{borici2025Cavity}%
		\BibitemOpen
		\bibfield  {author} {\bibinfo {author} {\bibfnamefont {D.}~\bibnamefont
				{Bori\ifmmode~\mbox{\c{c}}\else \c{c}\fi{}i}}, \bibinfo {author}
			{\bibfnamefont {G.}~\bibnamefont {Arwas}},\ and\ \bibinfo {author}
			{\bibfnamefont {C.}~\bibnamefont {Ciuti}},\ }\bibfield  {title} {\bibinfo
			{title} {Cavity-modified quantum electron transport in multiterminal devices
				and interferometers},\ }\href {https://doi.org/10.1103/4l6d-gqkw} {\bibfield
			{journal} {\bibinfo  {journal} {Phys. Rev. B}\ }\textbf {\bibinfo {volume}
				{112}},\ \bibinfo {pages} {045301} (\bibinfo {year} {2025})}\BibitemShut
		{NoStop}%
		\bibitem [{\citenamefont {Ciuti}(2021)}]{ciuti2021Cavity}%
		\BibitemOpen
		\bibfield  {author} {\bibinfo {author} {\bibfnamefont {C.}~\bibnamefont
				{Ciuti}},\ }\bibfield  {title} {\bibinfo {title} {Cavity-mediated electron
				hopping in disordered quantum hall systems},\ }\href
		{https://doi.org/10.1103/PhysRevB.104.155307} {\bibfield  {journal} {\bibinfo
				{journal} {Phys. Rev. B}\ }\textbf {\bibinfo {volume} {104}},\ \bibinfo
			{pages} {155307} (\bibinfo {year} {2021})}\BibitemShut {NoStop}%
		\bibitem [{\citenamefont {Winter}\ and\ \citenamefont
			{Zilberberg}(2025)}]{Winter2025Fractional}%
		\BibitemOpen
		\bibfield  {author} {\bibinfo {author} {\bibfnamefont {L.}~\bibnamefont
				{Winter}}\ and\ \bibinfo {author} {\bibfnamefont {O.}~\bibnamefont
				{Zilberberg}},\ }\bibfield  {title} {\bibinfo {title} {Fractional quantum
				{H}all edge polaritons},\ }\href {https://doi.org/10.1103/qc6z-87c6}
		{\bibfield  {journal} {\bibinfo  {journal} {Phys. Rev. B}\ }\textbf {\bibinfo
				{volume} {112}},\ \bibinfo {pages} {L241105} (\bibinfo {year}
			{2025})}\BibitemShut {NoStop}%
		\bibitem [{\citenamefont {M\'endez-C\'ordoba}\ \emph
			{et~al.}(2020)\citenamefont {M\'endez-C\'ordoba}, \citenamefont
			{Mendoza-Arenas}, \citenamefont {G\'omez-Ruiz}, \citenamefont
			{Rodr\'{\i}guez}, \citenamefont {Tejedor},\ and\ \citenamefont
			{Quiroga}}]{cordoba2020entropy}%
		\BibitemOpen
		\bibfield  {author} {\bibinfo {author} {\bibfnamefont {F.~P.~M.}\
				\bibnamefont {M\'endez-C\'ordoba}}, \bibinfo {author} {\bibfnamefont {J.~J.}\
				\bibnamefont {Mendoza-Arenas}}, \bibinfo {author} {\bibfnamefont {F.~J.}\
				\bibnamefont {G\'omez-Ruiz}}, \bibinfo {author} {\bibfnamefont {F.~J.}\
				\bibnamefont {Rodr\'{\i}guez}}, \bibinfo {author} {\bibfnamefont
				{C.}~\bibnamefont {Tejedor}},\ and\ \bibinfo {author} {\bibfnamefont
				{L.}~\bibnamefont {Quiroga}},\ }\bibfield  {title} {\bibinfo {title} {R\'enyi
				entropy singularities as signatures of topological criticality in coupled
				photon-fermion systems},\ }\href
		{https://doi.org/10.1103/PhysRevResearch.2.043264} {\bibfield  {journal}
			{\bibinfo  {journal} {Phys. Rev. Res.}\ }\textbf {\bibinfo {volume} {2}},\
			\bibinfo {pages} {043264} (\bibinfo {year} {2020})}\BibitemShut {NoStop}%
		\bibitem [{\citenamefont {Vlasiuk}\ \emph {et~al.}(2023)\citenamefont
			{Vlasiuk}, \citenamefont {Kozin}, \citenamefont {Klinovaja}, \citenamefont
			{Loss}, \citenamefont {Iorsh},\ and\ \citenamefont
			{Tokatly}}]{Vlasiuk2023cavity}%
		\BibitemOpen
		\bibfield  {author} {\bibinfo {author} {\bibfnamefont {E.}~\bibnamefont
				{Vlasiuk}}, \bibinfo {author} {\bibfnamefont {V.~K.}\ \bibnamefont {Kozin}},
			\bibinfo {author} {\bibfnamefont {J.}~\bibnamefont {Klinovaja}}, \bibinfo
			{author} {\bibfnamefont {D.}~\bibnamefont {Loss}}, \bibinfo {author}
			{\bibfnamefont {I.~V.}\ \bibnamefont {Iorsh}},\ and\ \bibinfo {author}
			{\bibfnamefont {I.~V.}\ \bibnamefont {Tokatly}},\ }\bibfield  {title}
		{\bibinfo {title} {Cavity-induced charge transfer in periodic systems:
				{L}ength-gauge formalism},\ }\href
		{https://doi.org/10.1103/PhysRevB.108.085410} {\bibfield  {journal} {\bibinfo
				{journal} {Phys. Rev. B}\ }\textbf {\bibinfo {volume} {108}},\ \bibinfo
			{pages} {085410} (\bibinfo {year} {2023})}\BibitemShut {NoStop}%
		\bibitem [{\citenamefont {Fernandez~Becerra}\ and\ \citenamefont
			{Dmytruk}(2026)}]{becerra2026fermion}%
		\BibitemOpen
		\bibfield  {author} {\bibinfo {author} {\bibfnamefont {V.}~\bibnamefont
				{Fernandez~Becerra}}\ and\ \bibinfo {author} {\bibfnamefont {O.}~\bibnamefont
				{Dmytruk}},\ }\bibfield  {title} {\bibinfo {title} {Fermion parity switching
				in a short {K}itaev chain coupled to a photonic cavity},\ }\href
		{https://doi.org/10.1103/hszk-ymf3} {\bibfield  {journal} {\bibinfo
				{journal} {Phys. Rev. B}\ }\textbf {\bibinfo {volume} {113}},\ \bibinfo
			{pages} {165410} (\bibinfo {year} {2026})}\BibitemShut {NoStop}%
		\bibitem [{\citenamefont {Yang}\ \emph {et~al.}(2026)\citenamefont {Yang},
			\citenamefont {Zhang}, \citenamefont {Ding},\ and\ \citenamefont
			{Li}}]{yang2026emergence}%
		\BibitemOpen
		\bibfield  {author} {\bibinfo {author} {\bibfnamefont {X.-X.}\ \bibnamefont
				{Yang}}, \bibinfo {author} {\bibfnamefont {S.}~\bibnamefont {Zhang}},
			\bibinfo {author} {\bibfnamefont {K.}~\bibnamefont {Ding}},\ and\ \bibinfo
			{author} {\bibfnamefont {X.}~\bibnamefont {Li}},\ }\href
		{https://arxiv.org/abs/2605.24439} {\bibinfo {title} {Emergence of {T}riplet
				{S}uperconductivity from {C}avity {V}acuum {F}luctuations}} (\bibinfo {year}
		{2026}),\ \Eprint {https://arxiv.org/abs/2605.24439} {arXiv:2605.24439
			[cond-mat.supr-con]} \BibitemShut {NoStop}%
		\bibitem [{\citenamefont {Dmytruk}\ and\ \citenamefont
			{Schir{\`o}}(2022)}]{dmytruk2023controlling}%
		\BibitemOpen
		\bibfield  {author} {\bibinfo {author} {\bibfnamefont {O.}~\bibnamefont
				{Dmytruk}}\ and\ \bibinfo {author} {\bibfnamefont {M.}~\bibnamefont
				{Schir{\`o}}},\ }\bibfield  {title} {\bibinfo {title} {Controlling
				topological phases of matter with quantum light},\ }\href
		{https://doi.org/10.1038/s42005-022-01049-0} {\bibfield  {journal} {\bibinfo
				{journal} {Communications Physics}\ }\textbf {\bibinfo {volume} {5}},\
			\bibinfo {pages} {271} (\bibinfo {year} {2022})}\BibitemShut {NoStop}%
		\bibitem [{\citenamefont {P{\'{e}}rez-Gonz{\'{a}}lez}\ \emph
			{et~al.}(2025)\citenamefont {P{\'{e}}rez-Gonz{\'{a}}lez}, \citenamefont
			{Platero},\ and\ \citenamefont
			{Gomez-Le{\'{o}}n}}]{PerezGonzalez2025lightmatter}%
		\BibitemOpen
		\bibfield  {author} {\bibinfo {author} {\bibfnamefont {B.}~\bibnamefont
				{P{\'{e}}rez-Gonz{\'{a}}lez}}, \bibinfo {author} {\bibfnamefont
				{G.}~\bibnamefont {Platero}},\ and\ \bibinfo {author} {\bibfnamefont
				{{\'{A}}.}~\bibnamefont {Gomez-Le{\'{o}}n}},\ }\bibfield  {title} {\bibinfo
			{title} {Light-matter correlations in {Q}uantum {F}loquet engineering of
				cavity quantum materials},\ }\href
		{https://doi.org/10.22331/q-2025-02-17-1633} {\bibfield  {journal} {\bibinfo
				{journal} {{Quantum}}\ }\textbf {\bibinfo {volume} {9}},\ \bibinfo {pages}
			{1633} (\bibinfo {year} {2025})}\BibitemShut {NoStop}%
		\bibitem [{\citenamefont {Nguyen}\ \emph {et~al.}(2024)\citenamefont {Nguyen},
			\citenamefont {Arwas},\ and\ \citenamefont {Ciuti}}]{Nguyen2024Electron}%
		\BibitemOpen
		\bibfield  {author} {\bibinfo {author} {\bibfnamefont {D.-P.}\ \bibnamefont
				{Nguyen}}, \bibinfo {author} {\bibfnamefont {G.}~\bibnamefont {Arwas}},\ and\
			\bibinfo {author} {\bibfnamefont {C.}~\bibnamefont {Ciuti}},\ }\bibfield
		{title} {\bibinfo {title} {Electron conductance and many-body marker of a
				cavity-embedded topological one-dimensional chain},\ }\href
		{https://doi.org/10.1103/PhysRevB.110.195416} {\bibfield  {journal} {\bibinfo
				{journal} {Phys. Rev. B}\ }\textbf {\bibinfo {volume} {110}},\ \bibinfo
			{pages} {195416} (\bibinfo {year} {2024})}\BibitemShut {NoStop}%
		\bibitem [{\citenamefont {Shaffer}\ \emph {et~al.}(2024)\citenamefont
			{Shaffer}, \citenamefont {Claassen}, \citenamefont {Srivastava},\ and\
			\citenamefont {Santos}}]{Shaffer2024Entanglement}%
		\BibitemOpen
		\bibfield  {author} {\bibinfo {author} {\bibfnamefont {D.}~\bibnamefont
				{Shaffer}}, \bibinfo {author} {\bibfnamefont {M.}~\bibnamefont {Claassen}},
			\bibinfo {author} {\bibfnamefont {A.}~\bibnamefont {Srivastava}},\ and\
			\bibinfo {author} {\bibfnamefont {L.~H.}\ \bibnamefont {Santos}},\ }\bibfield
		{title} {\bibinfo {title} {Entanglement and topology in
				{S}u-{S}chrieffer-{H}eeger cavity quantum electrodynamics},\ }\href
		{https://doi.org/10.1103/PhysRevB.109.155160} {\bibfield  {journal} {\bibinfo
				{journal} {Phys. Rev. B}\ }\textbf {\bibinfo {volume} {109}},\ \bibinfo
			{pages} {155160} (\bibinfo {year} {2024})}\BibitemShut {NoStop}%
		\bibitem [{\citenamefont {Sueiro}\ \emph {et~al.}(2025)\citenamefont {Sueiro},
			\citenamefont {Andolina},\ and\ \citenamefont {Schirò}}]{sueiro2025floquet}%
		\BibitemOpen
		\bibfield  {author} {\bibinfo {author} {\bibfnamefont {J.}~\bibnamefont
				{Sueiro}}, \bibinfo {author} {\bibfnamefont {G.~M.}\ \bibnamefont
				{Andolina}},\ and\ \bibinfo {author} {\bibfnamefont {M.}~\bibnamefont
				{Schirò}},\ }\href {https://arxiv.org/abs/2507.22715} {\bibinfo {title}
			{Floquet theory of lattice electrons coupled to an off-resonant cavity}}
		(\bibinfo {year} {2025}),\ \Eprint {https://arxiv.org/abs/2507.22715}
		{arXiv:2507.22715 [cond-mat.str-el]} \BibitemShut {NoStop}%
		\bibitem [{\citenamefont {Ritz-Zwilling}\ and\ \citenamefont
			{Dmytruk}(2026)}]{ritzzwilling2026topologicalmarkersonedimensionalfermionic}%
		\BibitemOpen
		\bibfield  {author} {\bibinfo {author} {\bibfnamefont {A.}~\bibnamefont
				{Ritz-Zwilling}}\ and\ \bibinfo {author} {\bibfnamefont {O.}~\bibnamefont
				{Dmytruk}},\ }\href {https://arxiv.org/abs/2604.13936} {\bibinfo {title}
			{Topological markers for a one-dimensional fermionic chain coupled to a
				single-mode cavity}} (\bibinfo {year} {2026}),\ \Eprint
		{https://arxiv.org/abs/2604.13936} {arXiv:2604.13936 [cond-mat.mes-hall]}
		\BibitemShut {NoStop}%
		\bibitem [{\citenamefont {Kobiałka}\ \emph {et~al.}(2026)\citenamefont
			{Kobiałka}, \citenamefont {Ghosh}, \citenamefont {Arouca},\ and\
			\citenamefont {Black-Schaffer}}]{kobialka2026topology}%
		\BibitemOpen
		\bibfield  {author} {\bibinfo {author} {\bibfnamefont {A.}~\bibnamefont
				{Kobiałka}}, \bibinfo {author} {\bibfnamefont {A.~K.}\ \bibnamefont
				{Ghosh}}, \bibinfo {author} {\bibfnamefont {R.}~\bibnamefont {Arouca}},\ and\
			\bibinfo {author} {\bibfnamefont {A.~M.}\ \bibnamefont {Black-Schaffer}},\
		}\href {https://arxiv.org/abs/2602.03553} {\bibinfo {title} {Topology and
				energy dependence of {M}ajorana bound states in a photonic cavity}} (\bibinfo
		{year} {2026}),\ \Eprint {https://arxiv.org/abs/2602.03553} {arXiv:2602.03553
			[cond-mat.mes-hall]} \BibitemShut {NoStop}%
		\bibitem [{\citenamefont {Stanescu}\ \emph
			{et~al.}(2011{\natexlab{a}})\citenamefont {Stanescu}, \citenamefont
			{Lutchyn},\ and\ \citenamefont {Das~Sarma}}]{stanescu2011majorana}%
		\BibitemOpen
		\bibfield  {author} {\bibinfo {author} {\bibfnamefont {T.~D.}\ \bibnamefont
				{Stanescu}}, \bibinfo {author} {\bibfnamefont {R.~M.}\ \bibnamefont
				{Lutchyn}},\ and\ \bibinfo {author} {\bibfnamefont {S.}~\bibnamefont
				{Das~Sarma}},\ }\bibfield  {title} {\bibinfo {title} {{M}ajorana fermions in
				semiconductor nanowires},\ }\href
		{https://doi.org/10.1103/PhysRevB.84.144522} {\bibfield  {journal} {\bibinfo
				{journal} {Phys. Rev. B}\ }\textbf {\bibinfo {volume} {84}},\ \bibinfo
			{pages} {144522} (\bibinfo {year} {2011}{\natexlab{a}})}\BibitemShut
		{NoStop}%
		\bibitem [{\citenamefont {Zyuzin}\ \emph {et~al.}(2013)\citenamefont {Zyuzin},
			\citenamefont {Rainis}, \citenamefont {Klinovaja},\ and\ \citenamefont
			{Loss}}]{zyuzin2013correlations}%
		\BibitemOpen
		\bibfield  {author} {\bibinfo {author} {\bibfnamefont {A.~A.}\ \bibnamefont
				{Zyuzin}}, \bibinfo {author} {\bibfnamefont {D.}~\bibnamefont {Rainis}},
			\bibinfo {author} {\bibfnamefont {J.}~\bibnamefont {Klinovaja}},\ and\
			\bibinfo {author} {\bibfnamefont {D.}~\bibnamefont {Loss}},\ }\bibfield
		{title} {\bibinfo {title} {Correlations between {M}ajorana {Fermions}
				{Through} a {Superconductor}},\ }\href
		{https://doi.org/10.1103/PhysRevLett.111.056802} {\bibfield  {journal}
			{\bibinfo  {journal} {Phys. Rev. Lett.}\ }\textbf {\bibinfo {volume} {111}},\
			\bibinfo {pages} {056802} (\bibinfo {year} {2013})}\BibitemShut {NoStop}%
		\bibitem [{\citenamefont {Chevallier}\ \emph {et~al.}(2013)\citenamefont
			{Chevallier}, \citenamefont {Simon},\ and\ \citenamefont
			{Bena}}]{chevallier2013from}%
		\BibitemOpen
		\bibfield  {author} {\bibinfo {author} {\bibfnamefont {D.}~\bibnamefont
				{Chevallier}}, \bibinfo {author} {\bibfnamefont {P.}~\bibnamefont {Simon}},\
			and\ \bibinfo {author} {\bibfnamefont {C.}~\bibnamefont {Bena}},\ }\bibfield
		{title} {\bibinfo {title} {From {A}ndreev bound states to {M}ajorana fermions
				in topological wires on superconducting substrates: {A} story of mutation},\
		}\href {https://doi.org/10.1103/PhysRevB.88.165401} {\bibfield  {journal}
			{\bibinfo  {journal} {Phys. Rev. B}\ }\textbf {\bibinfo {volume} {88}},\
			\bibinfo {pages} {165401} (\bibinfo {year} {2013})}\BibitemShut {NoStop}%
		\bibitem [{\citenamefont {Cole}\ \emph {et~al.}(2016)\citenamefont {Cole},
			\citenamefont {Sau},\ and\ \citenamefont {Das~Sarma}}]{cole2016proximity}%
		\BibitemOpen
		\bibfield  {author} {\bibinfo {author} {\bibfnamefont {W.~S.}\ \bibnamefont
				{Cole}}, \bibinfo {author} {\bibfnamefont {J.~D.}\ \bibnamefont {Sau}},\ and\
			\bibinfo {author} {\bibfnamefont {S.}~\bibnamefont {Das~Sarma}},\ }\bibfield
		{title} {\bibinfo {title} {Proximity effect and {M}ajorana bound states in
				clean semiconductor nanowires coupled to disordered superconductors},\ }\href
		{https://doi.org/10.1103/PhysRevB.94.140505} {\bibfield  {journal} {\bibinfo
				{journal} {Phys. Rev. B}\ }\textbf {\bibinfo {volume} {94}},\ \bibinfo
			{pages} {140505(R)} (\bibinfo {year} {2016})}\BibitemShut {NoStop}%
		\bibitem [{\citenamefont {Nakosai}\ \emph {et~al.}(2013)\citenamefont
			{Nakosai}, \citenamefont {Budich}, \citenamefont {Tanaka}, \citenamefont
			{Trauzettel},\ and\ \citenamefont {Nagaosa}}]{nakosai2013majorana}%
		\BibitemOpen
		\bibfield  {author} {\bibinfo {author} {\bibfnamefont {S.}~\bibnamefont
				{Nakosai}}, \bibinfo {author} {\bibfnamefont {J.~C.}\ \bibnamefont {Budich}},
			\bibinfo {author} {\bibfnamefont {Y.}~\bibnamefont {Tanaka}}, \bibinfo
			{author} {\bibfnamefont {B.}~\bibnamefont {Trauzettel}},\ and\ \bibinfo
			{author} {\bibfnamefont {N.}~\bibnamefont {Nagaosa}},\ }\bibfield  {title}
		{\bibinfo {title} {{M}ajorana {Bound} {States} and {Nonlocal} {Spin}
				{Correlations} in a {Quantum} {Wire} on an {Unconventional}
				{Superconductor}},\ }\href {https://doi.org/10.1103/PhysRevLett.110.117002}
		{\bibfield  {journal} {\bibinfo  {journal} {Phys. Rev. Lett.}\ }\textbf
			{\bibinfo {volume} {110}},\ \bibinfo {pages} {117002} (\bibinfo {year}
			{2013})}\BibitemShut {NoStop}%
		\bibitem [{\citenamefont {Dmytruk}\ and\ \citenamefont
			{Schir\'o}(2021)}]{dmytruk2021gauge}%
		\BibitemOpen
		\bibfield  {author} {\bibinfo {author} {\bibfnamefont {O.}~\bibnamefont
				{Dmytruk}}\ and\ \bibinfo {author} {\bibfnamefont {M.}~\bibnamefont
				{Schir\'o}},\ }\bibfield  {title} {\bibinfo {title} {Gauge fixing for
				strongly correlated electrons coupled to quantum light},\ }\href
		{https://doi.org/10.1103/PhysRevB.103.075131} {\bibfield  {journal} {\bibinfo
				{journal} {Phys. Rev. B}\ }\textbf {\bibinfo {volume} {103}},\ \bibinfo
			{pages} {075131} (\bibinfo {year} {2021})}\BibitemShut {NoStop}%
		\bibitem [{\citenamefont {Sentef}\ \emph {et~al.}(2020)\citenamefont {Sentef},
			\citenamefont {Li}, \citenamefont {K\"unzel},\ and\ \citenamefont
			{Eckstein}}]{sentef2020quantum}%
		\BibitemOpen
		\bibfield  {author} {\bibinfo {author} {\bibfnamefont {M.~A.}\ \bibnamefont
				{Sentef}}, \bibinfo {author} {\bibfnamefont {J.}~\bibnamefont {Li}}, \bibinfo
			{author} {\bibfnamefont {F.}~\bibnamefont {K\"unzel}},\ and\ \bibinfo
			{author} {\bibfnamefont {M.}~\bibnamefont {Eckstein}},\ }\bibfield  {title}
		{\bibinfo {title} {Quantum to classical crossover of {F}loquet engineering in
				correlated quantum systems},\ }\href
		{https://doi.org/10.1103/PhysRevResearch.2.033033} {\bibfield  {journal}
			{\bibinfo  {journal} {Phys. Rev. Res.}\ }\textbf {\bibinfo {volume} {2}},\
			\bibinfo {pages} {033033} (\bibinfo {year} {2020})}\BibitemShut {NoStop}%
		\bibitem [{\citenamefont {Klinovaja}\ and\ \citenamefont
			{Loss}(2012)}]{klinovaja2012composite}%
		\BibitemOpen
		\bibfield  {author} {\bibinfo {author} {\bibfnamefont {J.}~\bibnamefont
				{Klinovaja}}\ and\ \bibinfo {author} {\bibfnamefont {D.}~\bibnamefont
				{Loss}},\ }\bibfield  {title} {\bibinfo {title} {Composite {M}ajorana fermion
				wave functions in nanowires},\ }\href
		{https://doi.org/10.1103/PhysRevB.86.085408} {\bibfield  {journal} {\bibinfo
				{journal} {Phys. Rev. B}\ }\textbf {\bibinfo {volume} {86}},\ \bibinfo
			{pages} {085408} (\bibinfo {year} {2012})}\BibitemShut {NoStop}%
		\bibitem [{\citenamefont {Buonemani}\ \emph {et~al.}(2026)\citenamefont
			{Buonemani}, \citenamefont {Gómez-León}, \citenamefont {Schirò},\ and\
			\citenamefont {Dmytruk}}]{buonemani_poor_2026}%
		\BibitemOpen
		\bibfield  {author} {\bibinfo {author} {\bibfnamefont {F.}~\bibnamefont
				{Buonemani}}, \bibinfo {author} {\bibfnamefont {A.}~\bibnamefont
				{Gómez-León}}, \bibinfo {author} {\bibfnamefont {M.}~\bibnamefont
				{Schirò}},\ and\ \bibinfo {author} {\bibfnamefont {O.}~\bibnamefont
				{Dmytruk}},\ }\href {https://doi.org/10.48550/arXiv.2604.15036} {\bibinfo
			{title} {Poor man's {{M}ajorana} bound states in quantum dot based {{K}itaev}
				chain coupled to a photonic cavity}} (\bibinfo {year} {2026}),\ \bibinfo
		{note} {arXiv:2604.15036 [cond-mat.mes-hall]}\BibitemShut {NoStop}%
		\bibitem [{\citenamefont {Mikami}\ \emph {et~al.}(2016)\citenamefont {Mikami},
			\citenamefont {Kitamura}, \citenamefont {Yasuda}, \citenamefont {Tsuji},
			\citenamefont {Oka},\ and\ \citenamefont {Aoki}}]{mikami2016brillouin}%
		\BibitemOpen
		\bibfield  {author} {\bibinfo {author} {\bibfnamefont {T.}~\bibnamefont
				{Mikami}}, \bibinfo {author} {\bibfnamefont {S.}~\bibnamefont {Kitamura}},
			\bibinfo {author} {\bibfnamefont {K.}~\bibnamefont {Yasuda}}, \bibinfo
			{author} {\bibfnamefont {N.}~\bibnamefont {Tsuji}}, \bibinfo {author}
			{\bibfnamefont {T.}~\bibnamefont {Oka}},\ and\ \bibinfo {author}
			{\bibfnamefont {H.}~\bibnamefont {Aoki}},\ }\bibfield  {title} {\bibinfo
			{title} {Brillouin-{W}igner theory for high-frequency expansion in
				periodically driven systems: {A}pplication to {F}loquet topological
				insulators},\ }\href {https://doi.org/10.1103/PhysRevB.93.144307} {\bibfield
			{journal} {\bibinfo  {journal} {Phys. Rev. B}\ }\textbf {\bibinfo {volume}
				{93}},\ \bibinfo {pages} {144307} (\bibinfo {year} {2016})}\BibitemShut
		{NoStop}%
		\bibitem [{\citenamefont {Li}\ and\ \citenamefont
			{Eckstein}(2020)}]{li2020manipulating}%
		\BibitemOpen
		\bibfield  {author} {\bibinfo {author} {\bibfnamefont {J.}~\bibnamefont
				{Li}}\ and\ \bibinfo {author} {\bibfnamefont {M.}~\bibnamefont {Eckstein}},\
		}\bibfield  {title} {\bibinfo {title} {Manipulating {I}ntertwined {O}rders in
				{S}olids with {Q}uantum {L}ight},\ }\href
		{https://doi.org/10.1103/PhysRevLett.125.217402} {\bibfield  {journal}
			{\bibinfo  {journal} {Phys. Rev. Lett.}\ }\textbf {\bibinfo {volume} {125}},\
			\bibinfo {pages} {217402} (\bibinfo {year} {2020})}\BibitemShut {NoStop}%
		\bibitem [{\citenamefont {Li}\ \emph {et~al.}(2022)\citenamefont {Li},
			\citenamefont {Schamri\ss{}},\ and\ \citenamefont
			{Eckstein}}]{li2022effective}%
		\BibitemOpen
		\bibfield  {author} {\bibinfo {author} {\bibfnamefont {J.}~\bibnamefont
				{Li}}, \bibinfo {author} {\bibfnamefont {L.}~\bibnamefont {Schamri\ss{}}},\
			and\ \bibinfo {author} {\bibfnamefont {M.}~\bibnamefont {Eckstein}},\
		}\bibfield  {title} {\bibinfo {title} {Effective theory of lattice electrons
				strongly coupled to quantum electromagnetic fields},\ }\href
		{https://doi.org/10.1103/PhysRevB.105.165121} {\bibfield  {journal} {\bibinfo
				{journal} {Phys. Rev. B}\ }\textbf {\bibinfo {volume} {105}},\ \bibinfo
			{pages} {165121} (\bibinfo {year} {2022})}\BibitemShut {NoStop}%
		\bibitem [{\citenamefont {Reeg}\ and\ \citenamefont
			{Maslov}(2017)}]{Reeg2017Transport}%
		\BibitemOpen
		\bibfield  {author} {\bibinfo {author} {\bibfnamefont {C.}~\bibnamefont
				{Reeg}}\ and\ \bibinfo {author} {\bibfnamefont {D.~L.}\ \bibnamefont
				{Maslov}},\ }\bibfield  {title} {\bibinfo {title} {Transport signatures of
				topological superconductivity in a proximity-coupled nanowire},\ }\href
		{https://doi.org/10.1103/PhysRevB.95.205439} {\bibfield  {journal} {\bibinfo
				{journal} {Phys. Rev. B}\ }\textbf {\bibinfo {volume} {95}},\ \bibinfo
			{pages} {205439} (\bibinfo {year} {2017})}\BibitemShut {NoStop}%
		\bibitem [{\citenamefont {Stanescu}\ \emph
			{et~al.}(2011{\natexlab{b}})\citenamefont {Stanescu}, \citenamefont
			{Lutchyn},\ and\ \citenamefont {Das~Sarma}}]{stanescu_majorana_2011}%
		\BibitemOpen
		\bibfield  {author} {\bibinfo {author} {\bibfnamefont {T.~D.}\ \bibnamefont
				{Stanescu}}, \bibinfo {author} {\bibfnamefont {R.~M.}\ \bibnamefont
				{Lutchyn}},\ and\ \bibinfo {author} {\bibfnamefont {S.}~\bibnamefont
				{Das~Sarma}},\ }\bibfield  {title} {\bibinfo {title} {{M}ajorana fermions in
				semiconductor nanowires},\ }\href
		{https://doi.org/10.1103/PhysRevB.84.144522} {\bibfield  {journal} {\bibinfo
				{journal} {Physical Review B}\ }\textbf {\bibinfo {volume} {84}},\ \bibinfo
			{pages} {144522} (\bibinfo {year} {2011}{\natexlab{b}})}\BibitemShut
		{NoStop}%
		\bibitem [{\citenamefont {Stanescu}\ \emph {et~al.}(2010)\citenamefont
			{Stanescu}, \citenamefont {Sau}, \citenamefont {Lutchyn},\ and\ \citenamefont
			{Das~Sarma}}]{stanescu_proximity_2010}%
		\BibitemOpen
		\bibfield  {author} {\bibinfo {author} {\bibfnamefont {T.~D.}\ \bibnamefont
				{Stanescu}}, \bibinfo {author} {\bibfnamefont {J.~D.}\ \bibnamefont {Sau}},
			\bibinfo {author} {\bibfnamefont {R.~M.}\ \bibnamefont {Lutchyn}},\ and\
			\bibinfo {author} {\bibfnamefont {S.}~\bibnamefont {Das~Sarma}},\ }\bibfield
		{title} {\bibinfo {title} {Proximity effect at the
				superconductor–topological insulator interface},\ }\href
		{https://doi.org/10.1103/PhysRevB.81.241310} {\bibfield  {journal} {\bibinfo
				{journal} {Physical Review B}\ }\textbf {\bibinfo {volume} {81}},\ \bibinfo
			{pages} {241310} (\bibinfo {year} {2010})}\BibitemShut {NoStop}%
\end{thebibliography}
\end{document}